\documentclass[11pt,a4paper]{article}
\pdfoutput=1
\usepackage[pdftex]{graphics}
\usepackage{jheppub}

\newcommand{\be}{\begin{eqnarray}}
\newcommand{\ee}{\end{eqnarray}}
\newcommand{\nn}{\nonumber}
\newcommand{\bn}{\begin{enumerate}}
\newcommand{\en}{\end{enumerate}}

\def\IC{\mathbb{C}}

\def\IR{\mathbb{R}}
\def\IZ{\mathbb{Z}}

\def\CA{{\cal A}}

\def\CL{{\cal L}}

\def\CN{{\cal N}}
\def\CO{{\cal O}}

\def\a{\alpha}
\def\b{\beta}
\def\g{\gamma}
\def\d{\delta}
\def\e{\epsilon}

\def\z{\zeta}

\def\l{\lambda}
\def\m{\mu}

\def\s{\sigma}

\def\t{\tau}

\def\L{\Lambda}

\def\half{\frac{1}{2}}
\def\thalf{{\textstyle \frac{1}{2}}}
\def\imp{\Longrightarrow}

\def\goto{\rightarrow}

\def\det{{\rm det}}

\def\jmath{{j}}

\newcommand{\eq}{\begin{equation}}
\newcommand{\eqe}{\end{equation}}

\newcommand{\eqa}{\begin{eqnarray}}
\newcommand{\eqae}{\end{eqnarray}}

\title{Tree-level Recursion Relation and Dual Superconformal Symmetry of the ABJM Theory}

\author[a]{Dongmin Gang,}
\author[b]{Yu-tin Huang,} 
\author[a]{Eunkyung Koh,} 
\author[c]{Sangmin Lee} 
\author[d]{and Arthur E. Lipstein}

\affiliation[a]{School of Physics, Korea Institute for Advanced Study, Seoul 130-722, Korea}
\affiliation[b]{Department of Physics and Astronomy, UCLA, Los Angeles, CA 90095-1547, USA}
\affiliation[c]{Department of Physics, University of Seoul, Seoul 130-743, Korea}
\affiliation[d]{Department of Physics, California Institute of Technology, Pasadena, CA 91125, USA}

\emailAdd{arima275@kias.re.kr}
\emailAdd{yhuang@physics.ucla.edu}
\emailAdd{ekoh@kias.re.kr}
\emailAdd{mphys@uos.ac.kr}
\emailAdd{arthur@theory.caltech.edu}

\abstract{
We propose a recursion relation for tree-level scattering amplitudes
in three-dimensional Chern-Simons-matter theories. The recursion relation involves
a complex deformation of momenta which generalizes
the BCFW-deformation used in higher dimensions. Using background field methods, we show that all tree-level superamplitudes of the ABJM theory vanish for large deformations, establishing the validity of the recursion formula.
Furthermore, we use the recursion relation to compute six-point and eight-point component amplitudes and
match them with independent computations based on
Feynman diagrams or the Grassmannian integral formula.
As an application of the recursion relation,
we prove that all tree-level amplitudes of the ABJM theory
have dual superconformal symmetry. Using generalized unitarity methods, we extend this symmetry to the cut-constructible parts of the loop amplitudes.
}

\arxivnumber{1012.5032}

\begin{document}
\begin{flushright} CALT 68-2813/UCLA-10-TEP-109/KIAS-P10046 \end{flushright}
\maketitle

\section{Introduction}
In the past year, there has been substantial progress in the study of scattering amplitudes in three-dimensional Chern-Simons-matter theories~\cite{Agarwal:2008pu,blm,hl1,lee,hl2}. Much of this work has focused on the $\mathcal{N}=6$ superconformal Chern-Simons theory developed by Aharony, Bergman, Jafferis and Maldacena (ABJM), since in a certain limit, this theory is dual to  type IIA string theory on AdS$_{4}\times \IC P^3$, and therefore provides a new example of the AdS/CFT correspondence ~\cite{ Aharony:2008ug}.
The AdS$_4$/CFT$_3$ correspondence has many features in common with the AdS$_5$/CFT$_4$ correspondence, which relates $\mathcal{N}=4$ super Yang-Mills theory (sYM) to type IIB string theory on AdS$_5 \times S^5$~\cite{Maldacena:1997re}. This suggests that many of the symmetries, structures, and dualities exhibited by the planar amplitudes of $\mathcal{N}=4$ sYM may also present in the ABJM theory.

The first hint of this possibility was the discovery that the four- and six-point tree-level amplitudes of the ABJM theory have Yangian symmetry~\cite{blm}.
In \cite{yangian0902}, it was demonstrated that the Yangian symmetry of  $\mathcal{N}=4$ sYM is equivalent to superconformal plus dual superconformal symmetry. Dual superconformal symmetry is hidden from the point of view of the action, and can only be seen by studying on-shell amplitudes~\cite{DualConformal, Drummond:2008vq}. The equivalence of Yangian symmetry with superconformal plus dual superconformal symmetry when acting on amplitudes was also demonstrated for the ABJM theory in ~\cite{hl2}.
Furthermore, an integral formula similar to the Grassmannian integral formula of $\mathcal{N}=4$ sYM~\cite{acck0907} was proposed for the ABJM theory in \cite{lee}. This formula has Yangian symmetry and is conjectured to generate all tree-level amplitudes. This was verified for the four-point amplitude and claimed to be verified for the six-point amplitude \cite{lee}.
In order to demonstrate dual superconformal symmetry and test the Grassmannian integral formula for amplitudes with more than six external legs, we need an efficient way to compute higher-point on-shell amplitudes in the ABJM theory. Unfortunately, explicit Feynman diagram calculations become tedious for amplitudes with more than six external legs.

The computation of amplitudes can be drastically simplified using linear identities, such as supersymmetric Ward identities, which  relate amplitudes with the same number of legs and helicity structure in four dimensions~\cite{Grisaru:1976vm,Grisaru:1976vm+}, or off-shell nonlinear recursion relations such as the Berends and Giele recursion relations~\cite{Berends:1987me}, which relate higher-point amplitudes to products of lower-point amplitudes. In recent years, twistor-inspired methods \cite{witten03} have led to new representations of tree-level amplitudes such as the MHV vertex expansion~\cite{csw1} and the Britto, Feng, Cachazo and Witten (BCFW)~\cite{bcfw,bcfw+} recursion relations, both of which give a systematic procedure for  constructing higher-point amplitudes from lower-point amplitudes.

The BCFW recursion relation is derived by analytically continuing the momenta into the complex plane. In particular, one shifts the momenta of two external particles by a complex parameter. The analytic continuation is such that the amplitudes become rational functions in the complex plane, whose only singularities arise from propagators becoming on-shell. The recursion relation then follows from the fact that a rational function can be reconstructed from the residues of its poles, provided that the function vanishes at infinity. The residues in this case are simply products of
lower-point on-shell amplitudes.

One of the crucial ingredients of the BCFW formalism is the complex deformation, which is defined such that  (super-)momentum conservation and on-shell properties of the external fields are preserved. As we describe in section 3, however, the standard deformation used in the BCFW approach is only valid for $D\geq4$.  In this paper, we will define a complex deformation such that all properties of the on-shell amplitudes are preserved in three dimensions. To construct a recursion relation from this deformation, one must demonstrate that the amplitudes vanish when the parameter of the deformation goes to infinity, which is not always true even for $D\geq4$. We argue that the superamplitudes of the ABJM theory indeed have vanishing large-$z$ behavior. We confirm the validity of our recursion relation by using it to reproduce six-point amplitudes computed in ref.~\cite{blm} using Feynman diagrams and an eight-point amplitude computed from the Grassmannian integral formula given in ref.~\cite{lee}.

While an on-shell recursion relation provides an efficient way to construct higher-point amplitudes from lower-point amplitudes, it also serves as a convenient tool for analyzing the symmetries of amplitudes. If one can demonstrate that the recursion relation preserves symmetries of the lowest-point tree-level amplitude, it follows by induction that the symmetry holds for all tree-level amplitudes. Indeed, the proof of dual superconformal symmetry of four-~\cite{bht}, six-~\cite{Dennen:2010dh}, and ten-dimensional~\cite{CaronHuot:2010rj} maximal sYM planar amplitudes was based on the fact that the BCFW recursion formula preserves dual superconformal symmetry. Equipped with a similar recursion relation in three dimensions, and the fact that the four- and six-point ABJM tree amplitudes are dual superconformal invariant~\cite{Agarwal:2008pu,hl2}, we will prove that the recursion relation also preserves the dual symmetry, and hence all tree-level amplitudes of the ABJM theory have dual superconformal symmetry. Using generalized unitarity methods~\cite{UnitarityMethod1,UnitarityMethod2,UnitarityMethod3,UnitarityMethod4}, we further extend this symmetry to the cut constructible part of the loop amplitude.

Dual superconformal symmetry in $\mathcal{N}=4$ sYM is related to the fact that type IIB string theory on AdS$_5 \times S^5$ is self-dual after T-dualizing certain directions corresponding to bosonic and fermionic isometries~\cite{Alday:2007hr,Berkovits:2008ic,Beisert:2008iq}. The dual superconformal symmetry of all tree-level amplitudes in the ABJM theory therefore strongly suggests that type IIA string theory on AdS$_4 \times \IC P^3$ is self-dual after T-dualizing certain bosonic and fermionic isometries, although various attempts to demonstrate this have encountered singularities~\cite{Adam:2009kt,Grassi:2009yj,Adam:2010hh,Bakhmatov:2010fp,Dekel:2011qw}.

For $\mathcal{N}=4$ sYM, the combination of ordinary and dual superconformal symmetry is sufficient to fix the tree-level amplitudes~\cite{Korchemsky:2010ut}. Since the Grassmannian formula for the ABJM theory has both of these symmetries, the fact that the ABJM tree-level amplitudes have dual conformal symmetry suggests that they are indeed generated by the Grassmannian formula.

The structure of this paper is as follows. In section 2, we review the spinor-helicity formalism for three-dimensions. In section 3, we explain the difficulties of extending the BCFW recursion relation to three-dimensions and derive a new recursion relation for three-dimensional theories. The novel feature of this recursion relation is that it requires deforming the momentum spinors of two external particles in a nonlinear way. In section 4, we use background field methods to argue that the superamplitudes of the ABJM theory vanish when the deformation parameter is taken to infinity. In section 5, we review the Grassmannian integral formula for the ABJM theory and show that it reproduces the six-point superamplitude computed in ref.~\cite{blm} using Feynman diagrams. We also use it to compute an eight-point amplitude for the first time. In section 6, we use our recursion relation to reproduce the six-point and eight-point amplitudes computed in section 5. In section 7, we use our recursion relation to prove for the first time that all tree-level amplitudes of the ABJM theory have dual superconformal symmetry. We extend this result to loop-level using generalized unitarity methods in section 8. We conclude with some discussion in section 9. In appendices A and B, we review our conventions and derive the Feynman rules for the ABJM theory in the background field approach. In appendix C, we describe an alternative recursion relation for three-dimensions which requires deforming the momentum spinors of four external particles in a linear way.
We also present a superconformally covariant form
of the recursion relation derived in section 3.
Appendix D provide additional computational details for various results in the paper.

{\it Note added}: Immediately after the first version of this paper
was submitted to the e-print archive, we were informed of
the paper~\cite{rey} which also studies the scattering amplitudes
of the ABJM theory and its mass deformation.

\section{Three dimensional spinor helicity and on-shell superspace}
Here we give a brief introduction to the three dimensional spinor helicity formalism~\cite{blm,Agarwal:2008pu,hl1}. A vector in three dimensions, when written in bi-spinor form, is given by a symmetric $2\times2$ matrix. Each spinor transforms under the three-dimensional Lorentz group ${\rm Spin}(1,2)={\rm SL(2,\IR)}$.  A null momentum in three dimensions can be written
in bi-spinor form using a single spinor:
\be
p^{\a\b} = p_\mu (\s^\mu)^{\a\b} = \l^\a \l^\b \,.
\ee
Our conventions for spinors and gamma matrices are
summarized in appendix \ref{AppA1}.
Here, we simply note that for real $p_\m$, the
gamma matrices are such that $p^{\a\b}$ is also real.
The spinor $\l^\a$ is real for outgoing ($p_0<0$) particles
and purely imaginary for incoming ($p_0>0$) particles.

The spinors are contracted via the SL$(2,\IR)$ metric
\eq
\langle ij\rangle \equiv \lambda_i^\alpha\lambda_{j\alpha}=\lambda_i^\alpha\epsilon_{\alpha\beta}\lambda_{j}^{\beta} \,,
\eqe
where $\epsilon_{12}=-\epsilon^{12}=1$. The vector and spinor Lorentz invariants are related by
\eq
2p_i\cdot p_j=-\langle ij\rangle^2\,.
\eqe

In later sections, we will make use of the ``$\l$-parity" symmetry
$\lambda \goto -\lambda$. It does not change the momentum
$p^{\a\b}$, but flips the sign of the fermion wave-function.
As a result, it acts on amplitudes by
\be
A(\l_1, \ldots, -\l_i,\ldots, \l_n) = (-1)^{F_i}
A(\l_1, \ldots, \l_i,\ldots, \l_n) \,,
\label{l-parity}
\ee
where $F_i$ is the fermion number of the particle on leg $i$. We note that for Chern-Simons matter theory, only amplitudes with even number of external states are non-vanishing. This is because for odd number of states, there will always be at least one Chern-Simons gauge field on the external line. Since the gauge field does not carry any physical degree of freedom, the amplitude vanishes.

The on-shell superspace for the ABJM theory is built upon
three fermionic coordinates $\eta^I$ in addition to $\l^\a$~\cite{blm}, which transform as a $\mathbf{3}$ under the U(3) subgroup of the SO(6) R-symmetry group. The particle/anti-particle superfields
take the form
\be
&&\Phi = \phi^4 + \eta^I \psi_I + \thalf \e_{IJK} \eta^I\eta^J \phi^K
+ \tfrac{1}{6} \e_{IJK} \eta^I \eta^J \eta^K \psi_4 \,,
\nn \\
&&\bar{\Phi} = \bar{\psi}^4 + \eta^I \bar{\phi}_I +
\thalf \e_{IJK} \eta^I\eta^J \bar{\psi}^K
+ \tfrac{1}{6} \e_{IJK} \eta^I \eta^J \eta^K \bar{\phi}_4 \,.
\label{sfield}
\ee
The SO(6) R-symmetry acting on the $\CN=6$ supercharges
are realized by $\eta^I$ and their conjugates $\z_I \equiv \partial/\partial \eta^I$
through the Clifford algebra,
\be
\{ \eta^I , \z_J \} = \d^I_J
\;\;\;\;\;
(I,J=1,2,3)  \,.
\ee
It is customary~\cite{blm,lee} to introduce a collective notation
$ \Lambda = ( \lambda ; \eta) $.
The color ordered amplitude, following the convention of ref.~\cite{lee}, is then a function of  $\L_{i}$ and we choose to associate $\L_{\rm odd/even}$ to $\bar{\Phi}$/$\Phi$ multiplet. Then, the supersymmetric extension of
eq.\eqref{l-parity} is
\be
{ \cal A} ( \Lambda_1, \cdots, - \Lambda_i , \cdots, \Lambda_n )  = (-1)^{i} { \cal A} ( \Lambda_1, \ldots , \Lambda_i , \cdots, \Lambda_n )  \,.
\label{super-l-parity}
\ee

The generators of the superconformal symmetry come in three types:
\be
\L \frac{\partial}{\partial \L}
\,,
\;\;\;
\L\L \,,
\;\;\;
\frac{\partial^2}{\partial \L\partial \L} \,.
\ee
The invariance of the amplitudes under the first type of generators
will be manifest from the notation. For the second type,
we will use the notation
\be
p^{\a\b} = \l^\a \l^\b ,
\;\;\;
q^{\a I} = \l^\a \eta^I ,
\;\;\;
r^{IJ} = \eta^I \eta^J .
\ee
The invariance under $p^{\a\b}$ and $q^{\a I}$ will be imposed by
the super-momentum conserving delta functions
\be
\d^3(P) \d^6(Q)
\;\;\; {\rm with} \;\;\;
P \equiv \sum_i p^{\a\b}_i , \;\; Q \equiv \sum_i q^{\a I}_i .
\ee
A similar delta function does not exist for $r^{IJ}$.
In ref.~\cite{blm}, it was shown that the $r^{IJ}$ invariance
introduces a coset ${\rm O}(2k-4)/{\rm U}(k-2)$
for the $2k$-point amplitude. As we will review in section 5,
the same coset appears in the Grassmannian integral formula
of ref.~\cite{lee}.


\section{Recursion relation} \label{sec:recursion}

\subsection{Momentum shift}
In the $D$-dimensional BCFW formalism (with $D>3$), one introduces a complex parameter $z$ into the amplitude by shifting the momenta of two of the external legs~\cite{bcfw+,ahk,cheung}. For legs $i$ and $j$, the shift is given by
\be
p_i\rightarrow p_i+zq,\;\; p_j\rightarrow p_j-zq\,,
\ee
for some vector $q$. For the external legs to remain on shell, the shift vector $q$ is required to satisfy
\be
q\cdot p_i=q\cdot p_j=q^2=0\,.
\label{Ddimshift}
\ee
Although these constraints admit nontrivial solutions in $D>3$ dimensions, for $D=3$ the only solution is $q=0$. To see this, recall that a null vector in three dimensions can be written as a bi-spinor product $\lambda^\alpha\lambda^\beta$. Since the spinor is two dimensional, without loss of generality one can write the spinor for the vector $q$ as
\be
\lambda_q=a\lambda_i+b\lambda_j\,.
\ee
One then immediately sees that the requirements in eq.(\ref{Ddimshift}) imply $a=b=0$.

The analysis above can be translated into the spinor language. Under a general two-line shift $\lambda \rightarrow \hat{\lambda}(z)$, momentum conservation implies that
\be
\lambda_i\lambda_i+\lambda_j\lambda_j=\hat{\lambda}_i(z) \hat{\lambda}_i(z)+\hat{\lambda}_j(z) \hat{\lambda}_j(z)\,.
\ee
Assuming that the shift is a linear transformation on $(\l_i,\l_j)$, one can write
\begin{align}
\begin{pmatrix} \hat \lambda_i(z)  \\ \hat \lambda_{j} (z)  \end{pmatrix} = R(z)  \begin{pmatrix} \lambda_i \\ \lambda_{j}
\end{pmatrix} \,.
\label{R}
\end{align}
Then momentum conservation requires that
\be
R^T(z)R(z)=I \,,
\label{Rcondition}
\ee
i.e. the shift is an element of SO$(2,\IC)$.
\footnote{Since we are dealing with
a continuous deformation, we cannot use O$(2,\IC)$ including orientation reversal.}

Note that the BCFW deformation in four dimensions can also be formulated in terms of a shift matrix. In four dimensions, the momentum conservation can be written as
\be
\sum_{i=1}^n p_i = \sum_{i=1}^n \bar{\l}_i \l_i= 0 \,,
\ee
which is invariant under an arbitrary GL$(n,\IC)$ transformation,
\be
\l_i \goto M_i{}^j \l_j \,, \;\;\;
\bar{\l}_i \goto \bar{\l}_j (M^{-1})^j{}_i \,.
\ee
The BCFW shift $(\l_i \goto \l_i + z \l_j, \;\; \bar{\l}_j \goto \bar{\l}_i - z \bar{\l}_i)$ can then be regarded as a particular element of GL$(2,\IC)\subset {\rm GL}(n,\IC)$:
\be
\begin{pmatrix}
\ \hat{\lambda}_i(z) \\ \hat{\l}_j(z)
\end{pmatrix}
=
\begin{pmatrix}
1 & z \\
0 & 1
\end{pmatrix}
\begin{pmatrix}
\l_i \\ \l_j
\end{pmatrix} \,.
\ee
{}Starting from the three dimensional momentum conservation
and following the same logic, one naturally arrives at
the SO$(2,\IC)$ shift as above.

Let us now return to the case of three-dimensions. If we assume that $R(z)$ in eq.(\ref{R}) is linear in $z$, then  it can be parameterized as
\eq
R(z)=\left(\begin{array}{cc}1+a_{11}z & a_{12}z \\ a_{21}z & 1+a_{22}z\end{array}\right)\,.
\eqe
It is not difficult, however, to see that eq.(\ref{Rcondition}) constrains all of $a$'s to be zero. A similar analysis shows that eq.(\ref{Rcondition}) also constrains the three-line shift to be trivial, i.e. if we take $R(z)$ to be an SO$(3,\IC)$ matrix that depends linearly on $z$ eq.(\ref{Rcondition}), constrains it to be the identity. The first non-trivial solution which depends linearly on $z$ appears at four-lines. Multi-line shifts, when compared to two-line shifts, have the disadvantage of introducing a larger number of diagrams for each recursion. Furthermore, for the purpose of analyzing the perseverance of symmetries throughout the recursion, one would prefer as few shifted variables as possible. However, since multi-line shifts generally have better large $z$ behavior, which may be necessary for lower supersymmetric theories, we give a detail account of the four-line shift in appendix \ref{multiline}.

Alternatively, we could introduce a mass deformation. In that case, the spinor helicity formalism is essentially four-dimensional, so one can define a linear two-line shift. For the purpose of proving dual superconformal symmetry of the amplitudes, however, it would be better to avoid introducing a mass-deformation.

In order to construct a two-line shift without introducing a mass deformation, we must relax the assumption that  $R(z)$ is linear in $z$. In the next section, we will construct such an SO$(2,\IC)$ matrix and use it to derive a recursion relation for tree-level amplitudes.

\subsection{Derivation of the recursion relation}

We parameterize the $R(z)$ matrix by
\begin{equation}
R(z) = \begin{pmatrix} \frac{z+z^{-1}}{2} & - \frac{ z - z^{-1} }{ 2 i }   \\ \frac{ z- z^{-1} }{2 i } & \frac{ z+ z^{-1} }{ 2} \end{pmatrix} .
\label{Rmat}
\end{equation}
It acts on two reference external legs, say, the $i$'th and the $j$'th.  Using the cyclic symmetry,
we can set $i=1$ and the $j=l$ with $2 \leq l \leq k+1$ without loss of generality. 
Since we are interested in a recursion relation for superamplitudes,
we also need to consider super-momentum conservation,
\eq
\sum^n_{i=1}q^{\alpha I}_i=\sum^n_{i=1}\lambda^\alpha_i\eta_i^I=0 \ .
\eqe
The conservation of both momentum and super-momentum can be maintained if we also deform the $\eta$'s~\cite{bht}. Thus we define the super-shift as:
\eqa
\begin{pmatrix} \hat \lambda_1(z)  \\ \hat \lambda_{l} (z)  \end{pmatrix} = R(z)  \begin{pmatrix} \lambda_1 \\ \lambda_{l} \end{pmatrix},\; \begin{pmatrix} \hat \eta_1(z)  \\ \hat \eta_{l} (z)  \end{pmatrix} = R(z)  \begin{pmatrix} \eta_1 \\ \eta_{l} \end{pmatrix}
\,, \label{deformation}
\eqae
or collectively we have
\eq
\begin{pmatrix} \hat \Lambda_1(z)  \\ \hat \Lambda_{l} (z)  \end{pmatrix} = {R}(z)  \begin{pmatrix} \Lambda_1 \\ \Lambda_{l} \end{pmatrix} \ .
\eqe
Note that, since $R^T  R=1$, the super-shift preserves not only the (super-)momentum but also all the other superconformal generators,
\be
\sum_i \L_i \frac{\partial}{\partial \L_i}
\,,
\;\;\;
\sum_i \L_i\L_i \,,
\;\;\;
\sum_i \frac{\partial^2}{\partial \L_i \partial \L_i} \,.
\ee

We will denote the tree-level superamplitude for given external particles and their momenta as
\be
\CA( \Lambda_1, \cdots, \Lambda_{2k}) = A( \Lambda_1, \cdots, \Lambda_{2k}) \delta^{3}(P).
\ee
After the deformation, we can regard it as a function of $z$.
Following the original proof of the BCFW recursion relation~\cite{bcfw},
we express the undeformed amplitude  at $z = 1$ as a contour
integral along a small circle around $z=1$,
\begin{equation}
A(z=1) =  \frac{1}{ 2 \pi i } \oint_{z=1} \frac{ A (z)}{z-1} \,.
\label{undeformed}
\end{equation}
The on-shell tree-level superamplitude has the following factorization property~\cite{blm},
\begin{align}
A|_{p_f^2 \to 0}  = \int d^3 \eta \frac{ A_{L} ( \Lambda_I, \cdots, \Lambda_{J}, \Lambda_f ) A_{R} ( i \Lambda_f, \Lambda_{J+1} , \cdots, \Lambda_{I-1 } ) }{ p_f^2 } + (\mbox{regular at } p_f^2 \to 0) \label{factorization}
\end{align}
in the limit $p_f^2 \to 0$,
where $p_f = p_I + p_{I+1} + \cdots + p_{J}$ is the momentum of the factorization channel $f$. The left/right sub-amplitudes $ A_L,  A_R$ are on-shell at $p_f^2 = 0$.  The variable $ \Lambda_f =( \hat \lambda^{ \alpha}_f ; \eta^I )$ is allocated to the internal propagator, where $\hat \l_f$ is determined by the momentum conservation.
The relation \eqref{factorization} holds up to an unspecified overall constant.

Now we apply the $SO(2)$ deformation $R(z)$ to eq.\eqref{factorization} and plug the resulting $A(z)$ into the integrand of eq.\eqref{undeformed}. Then,
we change the contour of eq.\eqref{undeformed} to encircle
the poles arising from the deformed internal propagator $1/\hat{p}_f(z)^2$.
As we will show in the next section,
the integrand vanishes sufficiently fast at $z \to \infty$ and
$z\to 0$.\footnote{
$A(1/z)$ is the same as $A(z)$ up to a sign because exchanging
$z$ with $1/z$ in eq.\eqref{Rmat} acts as an orientation reversal.
Sufficiently fast vanishing means $A(z\to \infty) \sim \CO(1/z)$
which implies $A(z\to 0) \sim \CO(z)$.
\label{ori-rev}
}
Thus we obtain
\begin{align}
A(z =1) & = - \frac{1}{2 \pi i}   \sum_f   \int d^3   \eta  \sum_{ i=1  }^2  \left( \oint_{z=+z^{ \ast}_{i,f}} + \oint_{z=-z^{ \ast}_{i,f}}  \right)  \frac{dz}{z-1} \frac{ A_{L} (z; \eta )  A_R (z ; i \eta)   }{\hat{p}^2_{f} (z) } .  \label{BCFW}
\end{align}
This is the three-dimensional $\CN=6$ supersymmetric recursion relation.
 The  sum over all possible channels is denoted by $ \sum_f$, and  $(\pm z^{ \ast}_{i,f})$ are zeroes of $\hat{p}_f^2 (z)$.
The recursion relation for component amplitudes can be obtained by applying the corresponding  $ \eta$-integration to both sides of eq.\eqref{BCFW}. We will see explicit examples in the next section. 

In our deformation, the on-shell condition $\hat{p}^2_f (z)=0$ takes the following form
\begin{equation}
\hat{p}^2_f (z) = a_f z^{-2} + b_f + c_f z^2 = 0, \label{pole}
\end{equation}
with $a_f, b_f, c_f$ independent of $z$.
To see this, let the deformed momenta be $ \hat p_1$ and $ \hat p_{l}$. They are in the following form,
\begin{align}
&\hat p_1 = \hat \lambda_1 \hat \lambda_1  =  \half ( p_1 + p_{l} ) +  z^2 q +  z^{-2} \tilde q,
\nn \\
&\hat p_{l}  =\hat \lambda_{l} \hat \lambda_{l} =  \half ( p_1 + p_{ l} ) -   z^2 q -  z^{-2} \tilde q,
\end{align}
where $q$ and $ \tilde q$ are given by
\begin{align}
q^{ \alpha \beta} = \frac{1}{4} ( \lambda_1 + i  \lambda_{ l} )^{ \alpha} ( \lambda_1 + i \lambda_{ l} )^{ \beta} , \quad \tilde q^{ \alpha \beta} = \frac{1}{4} ( \lambda_1 - i \lambda_{ l} )^{ \alpha} ( \lambda_1 - i \lambda_{l} )^{\beta}  .
\end{align}
Using the cyclicity, let us choose $I$ and $J$ to satisfy  $1 < I \leq l \leq J $, then $\hat{p}_f(z)$ is given by $ \hat p_f (z) = p_I + \cdots + \hat{p}_{l} (z) + \cdots + p_J$. {}From this, we obtain
\begin{align}
& a_f = - 2  \tilde {q} \cdot \left( p_f - p_l  \right) , \quad b_f = ( p_f +p_1 ) \cdot (p_f - p_l )  \quad c_f = -  2 q \cdot \left(  p_f - p_l  \right) ,
\label{pf-abc}
\end{align}
after some manipulation using $ q+ \tilde q = \frac{1}{2} (p_1 - p_l)$,  $ q \cdot (p_1 + p_l) = \tilde {q} \cdot (p_1 +p_l)=0$, etc. Using eq.\eqref{pf-abc}, one can explicitly write down the zeros  $\{  \pm z_{1,f}^{ \ast} ,  \pm z_{2, f}^{ \ast} \} $ of eq.\eqref{pole}:
\begin{align}
\{ (z_{1,f}^\ast)^2, (z_{2,f}^\ast)^2 \} =   \left\{ \frac{ ( p_f + p_1) \cdot (p_f - p_l ) \pm \sqrt{ ( p_f + p_1)^2 ( p_f - p_l )^2 } }{ 4 q \cdot (p_f - p_l ) }  \right\} \,, \label{zeros}
\end{align}
where we used a variation of Schouten's identity,
\begin{align}
\langle r \vert p \vert s \rangle^2 =  \langle r \vert p \vert r \rangle \langle s \vert p \vert s \rangle + p^2 \langle rs \rangle^2 , \quad (\mbox{for each } r,s).
\end{align}

Let us rewrite the sum over residues of eq.\eqref{BCFW} so that the factorization limit is transparent:
\begin{align}
A  (z=1) =
&  \sum_f    \int d^3 \eta  \frac{1}{ p_f^2 } \big(   H(z_{1,f}^{ \ast} , z_{2,f}^{ \ast} ) A_L ( z_{1,f}^*; \eta ) A_R (z_{1,f}^* ; i\eta)+(z_{1,f}^* \leftrightarrow z_{2,f}^* )  \big) \,,\label{fact-limit}
\end{align}
where the function $H(a,b)$ is defined by
\begin{align}
H(a,b)
& = \begin{cases}  \frac{a^2 (b^2 -1) }{a^2 -b^2} , & ( l \mbox{ odd}), \\
 \frac{a (b^2 -1)}{a^2 - b^2}  , &  ( l \mbox{ even}),
\end{cases} \ : = \frac{ h(a,b) }{a^2 - b^2}  \,.
\label{function F}
\end{align}
The second equality defines the function $h(a,b)$.
We used the fact that
\[
A_L (-z) A_R(-z) = (-1)^{l+1} A_L (z) A_R (z) \,,
\]
which follows from eq.\eqref{super-l-parity}. Since $ h(a,b) A_L(a) A_R (a)$ is invariant under both $ a \to -a $ and $b \to -b$, it is a function of $a^2, b^2 $. The numerator in eq.\eqref{fact-limit},
\[
h(z_{1,f}^*,z_{2,f}^*) A_L(z_{1,f}^*) A_R (z_{1,f}^*) - h(z_{2,f}^*,z_{1,f}^*) A_L(z_{2,f}^*) A_R (z_{2,f}^*),
\]
is antisymmetric in $z_{1,f}^{ \ast} \leftrightarrow z_{2,f}^{ \ast}$.
Given that it is a function of $(z_{1,f}^{ \ast})^2$ and $(z_{2,f}^{ \ast})^2$, it should be divisible by
$(z_{1,f}^{ \ast})^2 - (z_{2,f}^{ \ast})^2$. Moreover, since
the quotient must now be a symmetric function under $z_{1,f}^{ \ast} \leftrightarrow z_{2,f}^{ \ast}$, it can be written as a rational function containing
$(z^*_{1,f})^2 + (z^*_{2,f})^2 = - b_f/c_f$ and $(z^*_{1,f})^2 (z^*_{2,f})^2 = a_f/c_f$. Thus one concludes that the final result is free from
any square-root factors that $ \{  \pm z_{1,f}^{ \ast} ,  \pm z_{2, f}^{ \ast} \} $ may contain.

\section{Asymptotic behavior of the superamplitude} \label{z-count}
In this section, we study the asymptotic behavior of the deformed superamplitude  under the deformation in eq.\eqref{deformation}.
The change of integration contour from eq.\eqref{undeformed} to eq.\eqref{BCFW} is justified only when  $\mathcal{A}(z)$  vanishes as $z \to \infty$ or $0$, so that there is no pole at $z = \infty$ and $z = 0$. We will focus on the behavior as $z \to \infty$, as the behavior for $z \to 0$
can be derived in a similar fashion; see footnote \ref{ori-rev}.

Under a BCFW deformation, the large-$z$ behavior of a gluon amplitude in $\mathcal{N}=4$ sYM depends on the helicity of the shifted legs. Although certain helicity configurations lead to bad large-$z$ behavior, when all the component amplitudes are packaged into a superamplitude, the entire superamplitude vanishes at large $z$ if the supershift is appropriately defined.

Here, we work in the opposite direction. In particular, we first identify component amplitudes that share the same large-$z$ behavior as the superamplitude.
Then we show that these component amplitudes behave as $\frac{1}{z}$ as  $z \rightarrow\infty$, which implies that the superamplitude also vanishes at large $z$.

\paragraph{Choosing a component amplitude}

We need to clarify what it means to say
that a component amplitude has the same large-$z$ behavior
as the superamplitude. In principle, the large-$z$ behavior of the superamplitude receives contributions from different component amplitudes, each of which has different large-$z$ behavior, so it is not obvious that one can make a direct connection between the large-$z$ behavior of particular component amplitudes and the superamplitude.

To answer this question, consider shifting legs $i$ and $j$ of the superamplitude, and expand the superamplitude in $(\eta_i,\eta_j)$:
\be
\CA = \CA^{(0,0)} + \CA^{(1,0)}_I \eta_i^I + \CA^{(0,1)}_I \eta_j^I
+\ldots + \CA^{(3,3)} (\eta_i)^3 (\eta_j)^3 \,.
\ee
Each sub-amplitude $\CA^{(m,n)}$ depends on all $\l$'s and
all $\eta$'s except $(\eta_i,\eta_j)$.
After the super-shift, $(\L_i,\L_j)\goto (\hat{\L}_i(z),\hat{\L}_j(z))$,
the superamplitude becomes
\begin{align}
\CA(z) &= \CA^{(0,0)}(z) +
\CA^{(1,0)}_I(z) \hat{\eta}_i^I(z) + \CA^{(0,1)}_I(z) \hat{\eta}_j^I(z)
+\cdots
\nn \\
&= \tilde{\CA}^{(0,0)}(z) +
\tilde{\CA}^{(1,0)}_I(z) \eta_i^I + \tilde{\CA}^{(0,1)}_I(z) \eta_j^I +
\cdots \,.
\end{align}
On the first line, the $z$-dependence of $\CA^{m,n}(z)$ is entirely
due to the shift $(\l_i, \l_j)\goto (\hat{\l}_i(z), \hat{\l}_j(z))$. Thus, these sub-amplitudes contain the large $z$ behavior of the component amplitudes under the bosonic shift.
On the second line, the expansion variables are the undeformed $(\eta_i,\eta_j)$. The relation between $\tilde{\CA}^{(m,n)}(z)$
and $\CA^{(m,n)}(z)$ follows from the shift property of $(\eta_i,\eta_j)$. For example, $\CA^{(0,0)}$ and $\CA^{(3,3)}$ are singlets in the sense that $\tilde{\CA}^{(0,0)}= \CA^{(0,0)}$ and $\tilde{\CA}^{(3,3)}=\CA^{(3,3)}$.
On the other hand, we have
\be
\begin{pmatrix}
\tilde{\CA}^{(1,0)}(z) \\
\tilde{\CA}^{(0,1)}(z)
\end{pmatrix}
=
R(z)^{T}
\begin{pmatrix}
\CA^{(1,0)}(z) \\
\CA^{(0,1)}(z)
\end{pmatrix} \,.
\ee

Now, the key point is that
all $\tilde{\CA}^{(m,n)}(z)$ share the same large-$z$ behavior
even though $\CA^{(m,n)}(z)$ do not. To see this, recall that the super-shift preserves the super-momentum conservation condition, or equivalently the supersymmetry generator $Q^{\a I}=\left(\sum_l q^{\a I}_l \right)$.
On the other hand, the individual supersymmetry generators $q^{\a I}_l$ can relate different sub-amplitudes $\tilde{\CA}^{(m,n)}(z)$, which implies the supersymmetric Ward identity. In particular, applying super-momentum conservation
and collecting terms linear in $\eta_i^I$ and independent of $\eta_j^I$,
we find
\be
\left(\sum_l q^{\a I}_l \right) \CA(z) = 0
\;\;\; \imp \;\;\;
\tilde{\CA}^{(0,0)}(z) = \frac{\l_{j\a} \check{Q}^{\a I}}{\langle i j\rangle}
\tilde{\CA}^{(1,0)}_I(z)
\;\;\;\;\;
(\mbox{no sum in}\; I)\,,
\ee
where $\check{Q}^{\a I} = \sum_{l \neq i,j} q^{\a I}_l$.
Since the prefactor on the right-hand side has no $z$-dependence,
$\tilde{\CA}^{(0,0)}$ and $\tilde{\CA}^{(1,0)}$
share the same large-$z$ behavior.
Applying the same method iteratively, we can show that
all $\tilde{\CA}^{(m,n)}(z)$ share the same large-$z$ behavior,
which then defines the large-$z$ behavior of the full superamplitude $\CA(z)$.

Note that  $\CA^{(3,3)}$ only includes component amplitudes with legs $i,j$ corresponding to either $\bar{\phi}_4(\l_i)$ or $\psi_4(\l_j)$, since these are the lowest weighted components in the superfield expansion in eq.(\ref{sfield}).
Since $\CA^{(3,3)}(z) = \tilde{\CA}^{(3,3)}(z)$,
their large-$z$ behavior is identical. Hence, the large-$z$ behavior of the superamplitude
 is the same as the large-$z$ behavior of component amplitudes of the form  $\langle\cdots\left(\bar{\phi}_4(\hat{i})\;{\rm or}\;\psi_4(\hat{i})\right)\cdots\left(\bar{\phi}_4(\hat{j})\;{\rm or}\;\psi_4(\hat{j})\right)\cdots\rangle$.

\subsection{Background field method and naive large-$z$ behavior}
We will now demonstrate that component amplitudes of the form $\langle\cdots\psi_4(\hat{i})\cdots\bar{\phi}_4(\hat{j})\cdots\rangle$ vanish as $1/z$ in the large-$z$ limit. To simplify the analysis, we will take the shifted legs to be adjacent. This fixes the amplitudes under consideration to have the form $\langle\psi_4(\hat{1})\cdots\cdots\bar{\phi}_4(\hat{l})\rangle$. Note that the shifted legs do not have to be adjacent in order to have good large-$z$ behavior. In the next section, we will use the Grassmannian integral formula to show that non-adjacent shifts result in even better large-$z$ behavior.
Another consequence of the adjacent shift is that
$A(-z) = - A(z)$ as follows from the ``$\l$-parity" \eqref{super-l-parity}.
We will see that several apparently $\CO(z^0)$ terms all cancel out.

\begin{figure}
\begin{center}
\includegraphics[scale=0.75]{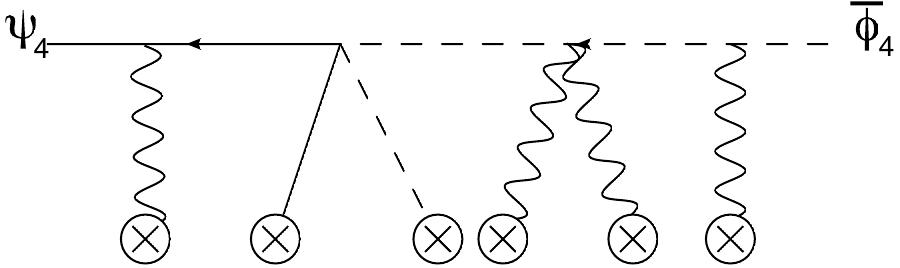}
\caption{Diagrammatic representation of a background field computation. 
 }
\label{Background}
\end{center}
\end{figure}

In the large-$z$ limit, it is convenient to analyze these amplitudes using a background field formulation which describes hard particles scattering through a soft background~\cite{ahk}.  Since we are shifting two external legs, the amplitudes under consideration reduce to diagrams of the form depicted in Fig.\ref{Background}, where the crosses represent background fields.

We proceed by separating each field into a background contribution plus a quantum fluctuation:
\eq
\nonumber A^\mu \rightarrow A^\mu+a^\mu , \qquad
\nonumber \Phi^I \rightarrow \Phi^I+\phi^I , \qquad
\Psi^I \rightarrow \Psi^I+\psi^I .
\eqe
There are similar equations for the hermitian conjugates of the matter fields.\footnote{Note that the ABJM theory has two gauge fields $A,\hat{A}$ for the ${\rm U}(N)\times {\rm U}(N)$ gauge group. In this section, we will not distinguish between them, since this does not play an important role in the analysis. For simplicity, we will also neglect color-indices and R-symmetry indices when they are not essential to the discussion.} While the background fields are solutions to the classical equations of motion, we take the fluctuations to describe fields whose momentum has been shifted and therefore approaches $z^2q$ as $z\rightarrow \infty$, where $q$ is the $z^2$ term in the deformation of external momenta $p_1$ and $p_l$,
\eq
q^{\alpha\beta}=\frac{1}{4}\left(\lambda_1^\alpha+i\lambda_l^\alpha\right) (\lambda_1^\beta+i\lambda_l^\beta )\,.
\label{defq}
\eqe
We define $ \lambda^{ \alpha}_q \equiv \frac{1}{2} ( \lambda^{\alpha}_1 + i \lambda^{\alpha}_l)$.

Since all the vertices in  Fig.\ref{Background} involve at most two fluctuation fields,  it is sufficient to expand the action to quadratic order in fluctuations for the purpose of analyzing diagrams of this type. Schematically, such terms take the form
\eqa
\nonumber \epsilon_{\mu\nu\rho}a^\mu\partial^\nu a^\rho,\;\;\;\phi\Box\bar{\phi}, \;\;\;\psi\displaystyle{\not}\partial_\mu\bar{\psi},
\\
\nonumber a^2A,\;\;\;a^2\Phi^2,\;\;\;a\phi\Phi A,\;\;\;A^2\phi^2,\;\;\; A\psi^2,\;\;\;\psi a\Psi,\;\;\;\Phi^2\psi^2,\;\;\;\Phi\Psi\psi\phi,\;\;\;\Psi^2\phi^2,\;\;\phi^2\Phi^4,
\\
a^\mu (\partial_\mu \phi) \Phi,\;\;\;A^\mu (\partial_\mu \phi) \phi,\;\;a^\mu (\partial_\mu \Phi) \phi.
\label{lego}
\eqae
Note that terms linear in the fluctuations vanish by the equations of motion for the background fields. Let us analyze their leading large-$z$ behavior individually. The first row contains the kinetic terms for the quantum fields. The corresponding propagators can be shown to behave as $(1,\frac{1}{z^2},1)$ respectively. Hence, only the scalar propagators improve the large-$z$ behavior. The second and third rows contain interaction terms. Note that terms in the second row do not contain any derivatives, so their vertices do not contribute any powers of $z$.

We now focus on the third row. The $a^\mu (\partial_\mu \Phi) \phi$ vertex does not introduce $z$-dependence since the derivative is on a background field. On the other hand, the $a^\mu (\partial_\mu \phi) \Phi$ vertex behaves as $z^2$. Since a gauge field fluctuation cannot appear as an external leg in the diagrams that we consider, any vertex involving gauge field fluctuations must be connected to a matter vertex via a vector propagator. The combination of the  $a^\mu (\partial_\mu \phi) \Phi$ vertex and a single vector propagator behaves as:
\eq
 \lim_{z\rightarrow \infty}\frac{\epsilon_{\mu\nu\rho}(p+z^2q)^\nu(z^2q)^\rho}{(p+z^2q)^2}=\frac{\epsilon_{\mu\nu\rho}(p)^\nu(z^2q)^\rho}{2z^2q\cdot p}\sim1.
\eqe
Hence, this combination approaches a constant as $z\rightarrow\infty$. Finally, the large-$z$ dependence of the $A^\mu (\partial_\mu \phi) \phi$ vertex can be set to zero if gauge-fix the background so that
\eq
q\cdot  A=0\,.
\label{gaugecondition}
\eqe
There are some circumstances for which this gauge choice cannot be implemented. As explained in ref.~\cite{ahk},
this problem will arise for diagrams in which the shifted external lines interact with a single background gauge field via a three-point contact term. This situation does not arise in our analysis, however, because we take the shifted external lines to be a scalar and a fermion, which cannot interact with a single background gauge field via a three-point contact term.

From the above analysis, we see that the propagators and vertices in eq.(\ref{lego}) will at worst give a constant in the large-$z$ limit. Combined with the fact that there is an external wavefunction for the fermion in $\langle\psi_4(-z^2q)\cdots \bar{\phi}_4(z^2q)\rangle$, which behaves as $\lim_{z\rightarrow \infty}\psi_4(z^2q)=z\lambda_q$, a naive analysis would imply that the amplitudes can have at most an $\mathcal{O}(z)$ divergence at large $z$. On the other hand, since the scalar propagator introduces a factor of $\frac{1}{z^2}$, diagrams containing at least one scalar propagator are $\mathcal{O}(1/z)$ in the large-$z$ limit, so the only diagrams that we need to worry about are those which do not contain scalar propagators. As we will see in the next subsection, however, the large-$z$ behavior of these diagrams is $\mathcal{O}(1/z)$ as well.
\subsection{Improved large-$z$ behavior}
To see the effect of the external wave function, we analyze all possible ways of connecting the external $\psi_4(z)$ line to the rest of the diagram. We focus on diagrams without scalar propagators, and we will show that each case behaves as $\frac{1}{z}$ asymptotically. Since from the previous discussion we've concluded that the building blocks of the amplitude behave at worst as a constant, the $\frac{1}{z}$ behavior of the sub diagram then implies the entire diagram can behave at worst as $\frac{1}{z}$.

\paragraph{Fermion propagator:} If the external hard fermion is connected to the rest of the diagram via $\bar{\psi}\displaystyle{\not}A\psi,\bar{\psi}\psi\Phi\bar{\Phi}$, there will be a fermion propagator. The asymptotic of the two scenarios is
  \eqa
  \nonumber && \lim_{z\rightarrow \infty}\langle \psi_4(\hat{p}_1)|\displaystyle{\not}A(k)\frac{(\displaystyle{\not}p+z^2\displaystyle{\not}q)}{(p+z^2q)^2}=\frac{z\lambda_q\displaystyle{\not}p}{2z^2q\cdot p}\sim\frac{1}{z}\,,\\
  &&\lim_{z\rightarrow \infty}\langle \psi_4(\hat{p}_1)|\frac{(\displaystyle{\not}p+z^2\displaystyle{\not}q)}{(p+z^2q)^2}=\frac{z\lambda_q\displaystyle{\not}p}{2z^2q\cdot p}\sim\frac{1}{z}\,,
  \eqae
where we used the notation $\langle \psi(p)|$ and $\langle \Psi(k)|$ to represent the external wave function and the background field respectively. For both cases we have used the fact that the leading large $z$ contribution in the external wave function is $\psi_4(\hat{p}_1)\rightarrow z\lambda_q$, and $q\cdot A(k)=q^{\alpha\beta}\lambda_{q\beta} =0$.

\paragraph{Gluon propagator:} If the external hard fermion is connected to the rest of the diagram via $\psi\displaystyle{\not}a\bar{\Psi}$, there will be a gluon propagator. Since the gluon will be required to turn into a scalar at some point along the line, the other side of the propagator will be required to connect to $a^\mu (\partial_\mu \bar{\phi})\Phi,\; a^\mu \bar{\phi}\Phi A_\mu$, \;$\epsilon_{\mu\nu\rho}a^\mu a^\nu A^\rho$ or $a^\mu\bar{\Phi}\Phi a_\mu$.
 \begin{figure}
\begin{center}
\includegraphics[scale=0.8]{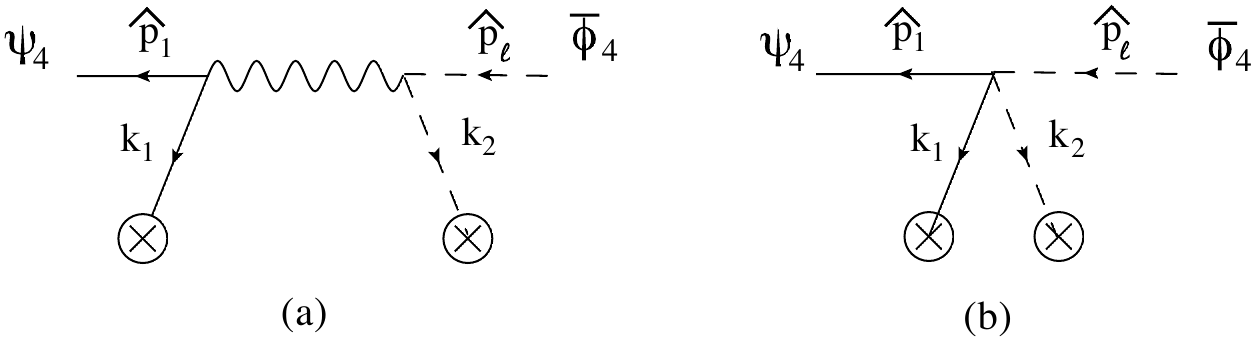}
\caption{Gluon propagator connecting to scalar vertex cancels against contact vertex in the leading order.}
\label{4ptcontact}
\end{center}
\end{figure}
\begin{figure}
\begin{center}
\includegraphics[scale=0.8]{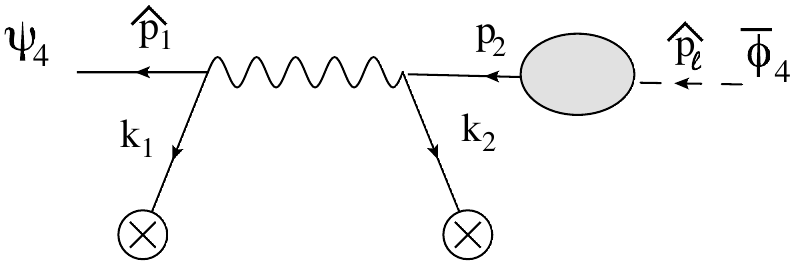}
\caption{Gluon propagator connecting to fermion vertex.}
\label{4ptfermi}
\end{center}
\end{figure}

\begin{enumerate}

\item
We first consider the case where the gluon propagator directly connects to the $a^\mu (\partial_\mu \bar{\phi}) \Phi$ vertex. Such diagrams generically take the form of Fig.\ref{4ptcontact}(a).\footnote{Note that we have denoted the scalar to be $\bar{\phi}_4$, even though this is not always the case. The justification is that if the scalar is of a different R-index, then one needs to connect to other scalar vertices to change the R-index. This will introduce scalar propagators.} We note that for every Fig.\ref{4ptcontact}(a) there will be complimentary diagram using the four-point contact vertex $\bar{\Psi}\psi\bar{\phi}\Phi$.  as in Fig.\ref{4ptcontact}(b). The asymptotic behavior is of order $z$ for both diagrams, and they cancel each other. The details of this cancelation is given in appendix \ref{AppB}. However, it is not difficult to be convinced of this result once one notes that if the background fields are replaced with on-shell external lines, it is simply the four-point amplitude which we know vanishes as $\frac{1}{z}$ at large $z$.

\item
The gluon propagator can also connect to a fermion vertex, and change to scalar later down the line. This correspond to the diagram in Fig.\ref{4ptfermi}.
The second fermion-vertex will have to connect to a fermion propagator. Using the Feynman rules given in appendix \ref{Feynman}, the two vertex two propagator combination gives
\eqa
\langle\psi( \hat{p}_1)|\sigma^\mu\bar{\Psi}(k_1)\frac{\epsilon_{\mu\nu\rho}(\hat{p}_1+k_1)^\nu}{(\hat{p}_1+k_1)^2}\bar{\Psi}(k_2)\frac{\sigma^\rho(\displaystyle{\not}\hat{p}_1+\displaystyle{\not}k_2+\displaystyle{\not}k_1)}{(\hat{p}_1+k_2+k_1)^2}\,.
\eqae
The leading large $z$ behavior takes the form:
\eqa
z\langle q|\sigma^\mu\bar{\Psi}(k_1)\frac{\epsilon_{\mu\nu\rho}q^\nu}{q\cdot k_1}\bar{\Psi}(k_2)\sigma^\rho \frac{\displaystyle{\not}q}{q\cdot(k_1+k_3)},
\eqae
which vanishes using the identity eq.(\ref{epsilon}).

\item
Next we consider the gluon propagator connecting to  $\epsilon_{\mu\nu\rho}a^\mu a^\nu A^\rho$. This corresponds to the diagram in Fig.\ref{5pt} and gives
\begin{figure}
\begin{center}
\includegraphics[scale=0.8]{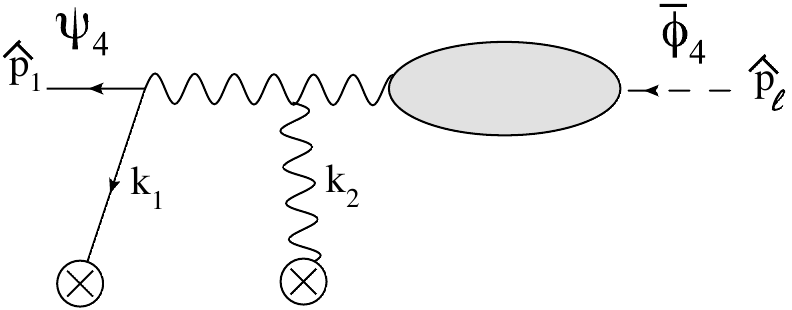}
\caption{Gluon propagator connecting to three gluon vertex.}
\label{5pt}
\end{center}
\end{figure}
\eq
-g^3\left\langle\psi(\hat{p}_{1})\right|\sigma^{\mu} \bar{\Psi}(k_{1}) \frac{\epsilon_{\mu\nu\lambda}\left(\hat{p}_{1}+k_{1}\right)^{\lambda}}{\left(\hat{p}_{1}+k_{1}\right)^{2}}\left(\epsilon^{\nu\rho\sigma}\right)A_{\rho}(k_{2})\frac{\epsilon_{\omega\sigma\delta}\left(k_{3}+p_{2}\right)^{\delta}}{\left(k_{3}+p_{2}\right)^{2}}{\normalcolor .}
\eqe
For the leading $z$ contribution, $\hat{p}_1=p_2=z^2q$, and the computation reduces to
\eqa
\nonumber&&-zg^{3}\left\langle q\right|\sigma^{\mu}\bar{\Psi}(k_{1}) \frac{\epsilon_{\mu\nu\lambda}q^{\lambda}\epsilon^{\nu\rho\sigma}A_{\rho}(k_{2})\epsilon_{\omega\sigma\delta}q^{\delta}}{2q\cdot k_{1}q\cdot k_{3}}+\mathcal{O}(1/z) \\
&=&zg^{3}\frac{q\cdot A(k_{2})}{2q\cdot k_{1}q\cdot k_{3}}\left\langle q\right|\sigma^{\mu}\bar{\Psi}(k_{1}) \epsilon_{\mu\omega\delta}q^{\delta}+\mathcal{O}(1/z){\normalcolor .}
\eqae
where we've used $\epsilon_{\mu\nu\lambda}\epsilon^{\nu\rho\sigma}=\frac{1}{2}\delta_{[\mu}^{\rho}\delta_{\lambda]}^{\sigma}$. Since the order $z$ piece is proportional to $q\cdot A(k_2)$, it vanishes for our background field gauge, and hence the diagrams start at $\frac{1}{z}$.

\item
For the case where the gluon propagator is connected to the vertex $ a^\mu \bar{\phi}\Phi A_\mu$, as indicated in Fig.\ref{5ptb}, there are two types of vertices depending on the type of gauge fields on the four-point vertex. The computation for the two are the same, so we present the result for the first diagram:
\begin{figure}
\begin{center}
\includegraphics[scale=0.8]{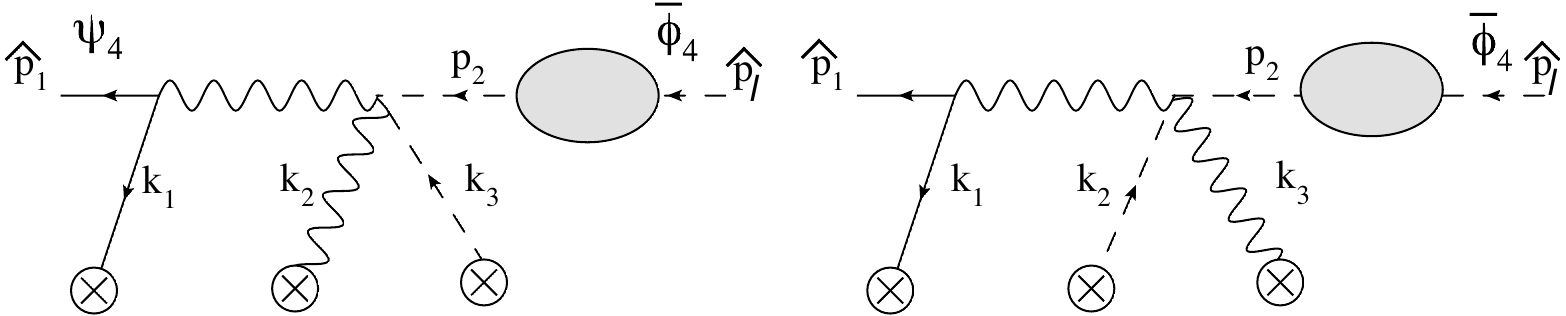}
\caption{Gluon propagator connecting to scalar propagator via quartic vertex.}
\label{5ptb}
\end{center}
\end{figure}
\eq
g^3\left\langle \psi(\hat{p}_{1})\right|\sigma^{\mu}\bar{\Psi}(k_{1}) \frac{\epsilon_{\mu\nu\lambda}\left(\hat{p}_{1}+k_{1}\right)^{\lambda}}{\left(\hat{p}_{1}+k_{1}\right)^{2}}A^{\nu}(k_{2}){\normalcolor .}
\eqe
In the large-$z$ limit this becomes
\eqa
\nonumber &&-zg^{3}\left\langle q\right|\sigma^{\mu}\bar{\Psi}(k_{1}) \frac{\epsilon_{\mu\lambda\nu}q^{\lambda}A^{\nu}(k_{2})}{2q\cdot k_{1}}+\mathcal{O}(1/z) \\
&=&-zg^{3}\frac{q\cdot A(k_{2})\left\langle q\right|\bar{\Psi}(k_{1}) }{2q\cdot k_{1}}+\mathcal{O}(1/z){\normalcolor .}
\eqae
where we have used eq.(\ref{epsilon}). Thus again the leading $z$ pieces vanish due to the background field gauge.
\begin{figure}
\begin{center}
\includegraphics[scale=0.8]{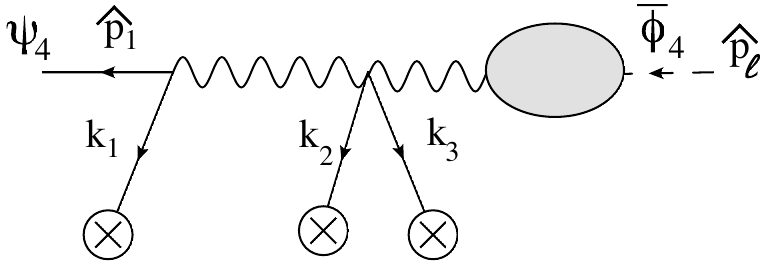}
\caption{Gluon propagator connecting to another gluon propagator via quartic vertex.}
\label{last}
\end{center}
\end{figure}

\item
Finally, we consider the gluon propagator connected to the vertex  $ a^\mu \bar{\Phi}\Phi a_\mu$. This is shown in Fig.\ref{last} and gives
\eqa
\langle\psi(\hat{p}_1)|\sigma^\mu\bar{\Psi}(k_1)\frac{\epsilon_{\mu\nu\rho}(\hat{p}_1+k_1)^\nu}{(\hat{p}_1+k_1)^2}\frac{\epsilon^{\rho\sigma\tau}(\hat{p}_1+k_1+k_2+k_3)^\sigma}{(\hat{p}_1+k_1+k_2+k_3)^2} \,.
\eqae
The leading large $z$ behavior takes the form:
\eqa
z\langle q|\sigma^\mu\bar{\Psi}(k_1)\frac{\epsilon_{\mu\nu\rho}q^\nu}{k_1\cdot q}\frac{\epsilon^{\rho\sigma\tau}q^\sigma}{q\cdot(k_1+k_3+k_4)} \,.
\eqae
Contracting the Levi-Cevita tensors, this term vanishes since $q^2=\displaystyle{\not\,}q\lambda_q=0$.

\end{enumerate}

Thus we see that for all cases, due to the fact that leading large $z$ behavior of the external fermion wave function is proportional to $\lambda_q$, the asymptotic behavior of amplitudes with adjacent $\bar{\phi}_4,\psi_4$ as shifted lines are improved to $\frac{1}{z}$. Since these amplitudes share the same large $z$ behavior as the superamplitude, our recursion relation is applicable. In the next section, the same large $z$ behavior will be reproduced by shifting the Grassmannian integral formula.
Finally, note that for the limit $z\to 0$, the same analysis would give that the amplitude behaves as $z$ and hence vanishes as well.


\section{Grassmannian integral}

In this section, we digress to discuss the matrix integral formula
for the tree level amplitudes of planar ABJM theory proposed in ref.~\cite{lee}:
\be
\CL_{2k}(\L) =
\int  \frac{d^{k\times 2k} C}{{\rm vol}[{\rm GL}(k)]}\frac{\d^{k(k+1)/2}(C\cdot  C^T)\, \d^{2k|3k}(C\cdot \L)}
{M_1 M_2 \cdots M_{k-1} M_k}
\,.
\label{3d-master}
\ee
The integration variable $C$ is a $(k\times 2k)$ matrix.
The dot products denote $(C\cdot C^T)_{mn} = C_{mi}C_{ni}$, $(C\cdot \L)_m = C_{mi}\L_i$.
The $i$-th minor $M_i$ of $C$ is defined by
\be
M_i = \e^{m_1 \cdots m_k} C_{m_1 (i)} C_{m_2 (i+1)} \cdots C_{m_k (i+k-1)} \,.
\ee
In ref.~\cite{lee}, this formula was shown to satisfy
the same cyclic symmetry and superconformal symmetry
as the tree-level $(2k)$-point superamplitude.
It was also shown to reproduce the known 4-point superamplitude
and satisfy Yangian invariance.

After a brief review of its defining properties, we will use the formula
to revisit the asymptotic $z$ behavior of the previous section and then to compute some amplitudes up to 8-point. For both purposes, the formula reveals remarkably simple structures that are not visible in a Feynman diagram analysis.

\paragraph{Orthogonal Grassmannian}
The integral (\ref{3d-master}) should be understood as a contour integral
on the moduli space of rank $k$, $(k\times 2k)$ matrices $C_{mi}$
$(m=1,\ldots,k;i=1,\ldots,2k)$
subject to the constraint $(C\cdot C^T)_{mn} = C_{m i} C_{n i} = 0$
and the equivalence
relation $C \sim g \,C$ with $g \in {\rm GL}(k)$.
This moduli space is known as the orthogonal Grassmannian OG$(k,2k)$.
The constraint and GL$(k)$ gauge symmetry determines the dimension of OG$(k,2k)$ as
\be
\dim_\IC\left[{\rm OG}(k,2k)\right] = 2k^2 - k^2 - \frac{k(k+1)}{2} = \frac{k(k-1)}{2} \,.
\ee
Here, we give an informal introduction to the orthogonal Grassmannian.\footnote{We thank Hoil Kim for discussions on basic facts about
Grassmannian.}
For more formal treatments, see, for instance, refs.~\cite{kresch,smt}.

In the mathematics literature, OG$(k,2k)$ is defined to
be the moduli space of maximal isotropic subspace of $\IC^{2k}$.
We assume that $\IC^{2k}$ is equipped with a non-degenerate
quadratic form $(V,W)$ for $V,W \in \IC^{2k}$ and choose to work in a basis of $\IC^{2k}$
such that $(V,W) = \sum_{i=1}^{2k} V_i W_i$.
A subspace $\mathcal{X} \subset \IC^{2k}$ is called isotropic
if $(V,W)=0$ for all $V,W\in \mathcal{X}$.
The maximal dimension of such $\mathcal{X}$ is known to be $k$.
If we interpret the $(k\times 2k)$ matrix $C$
as a collection of $k$ vectors in $\IC^{2k}$,
the constraint $C\cdot C^T=0$ and the equivalence relation
$C \sim h \,C$, $h \in {\rm GL}(k)$ dictate that
the $k$ vectors form a basis of a maximally isotropic subspace $\mathcal{X}$.

Alternatively, OG$(k,2k)$ can be defined to be the coset space:
\footnote{As we will see shortly, for the integral formula to work,
it is important to include both of the two disconnected SO$(2k)$
components of
O$(2k)$.}
\be
{\rm OG}(k,2k) = {\rm O}(2k)/{\rm U}(k) \,.
\ee
Geometrically, it is the moduli space of complex structures of $\IR^{2k}$
compatible with the standard metric.
To see this, we introduce complex orthonormal frames in $\IR^{2k}$
satisfying
\begin{align}
&ds^2 = dx^i dx^i = (E_{mi} dx^i)(\bar{E}^{m}_{j} dx^j) = E_m \otimes \bar{E}_{m} \,,
\;\;\; \bar{E}^m_i = (E_{mi})^* \,,
\nn \\
&
E_{m(i} \bar{E}^m_{j)} = \d_{ij} \,, \;\;\;\;
E_{mi} \bar{E}^{n}_i = 2 \d^n_m \,, \;\;\;\;
E_{mi} E_{ni} = 0 = \bar{E}^m_i \bar{E}^n_i \,.
\label{e-frame}
\end{align}
In terms of the orthonormal frames, the complex structure $J^k{}_j$
can be written as
\be
J^k{}_j = \d^{ki} J_{ij} \,,
\;\;\; \half\, J_{ij} \,dx^i \wedge dx^j  = \bar{E}^m \wedge E_m  \,.
\ee
The complex structure is compatible
with the metric by construction. It is also clear that the moduli space
of $J^k{}_j$ is ${\rm O}(2k)/{\rm U}(k)$, where ${\rm O}(2k)$ and ${\rm U}(k)$
acts on $E_{mi}$ from the right and left, respectively.

The two definitions of OG$(k,2k)$ can be related to each
other as follows. To go from the coset to the $C$ matrix,
we simply take $C_{mi} = E_{mi}$.
To define the inverse map, note that $C\cdot C^{\dagger}$ is a
positive definite hermitian matrix. By diagonalizing it, we can write
\be
C\cdot C^{\dagger} = U^\dagger D U
\;\;\; \imp \;\;\;
\left( \sqrt{\frac{2}{D}} U C \right)\cdot \left( \sqrt{\frac{2}{D}} U C \right)^{\dagger} = 2 I_{k}  \,.
\ee
So, $E = \sqrt{2/D}\, UC$ defines the inverse map.
It is possible to show that the map between (the equivalence classes of) $C$ and $E$
is one-to-one, but we will not give the details here.

We can discuss coordinate patches of OG$(k,2k)$ in the same way as for
the ordinary Grassmannian.
We choose $k$ out of $2k$ columns of $C$ and fill them with
the $(k\times k)$ identity matrix $I_k$. There are $(2k)!/(k!)^2$
different patches. For instance, we can have
\be
C =
\begin{pmatrix}
I_k
& c
\end{pmatrix} \,, \;\;\;
C\cdot C^T = 0
\;\;\; \imp \;\;\;
cc^T + I_{k} = 0 \,.
\ee
The $(k\times k)$ matrix $c$ is ($\pm i$ times) a complex-valued orthogonal
matrix.
Thus each coordinate patch is isomorphic to O$(k,\IC)$.

After taking account of the bosonic delta function $\d(C\cdot \l)$
and the overall momentum conservation,
the net number of integration variables of eq.\eqref{3d-master}
becomes~\cite{lee}
\be
\frac{k(k-1)}{2} - 2k + 3 = \frac{(k-2)(k-3)}{2} \,.
\label{var-counting}
\ee
The delta function constraint $C_{mi} \lambda_i^\a = 0$ means that
$(\l^{\a=1}_i,\l^{\a=2}_i)$ regarded as two vectors in $\IC^{2k}$ are contained
in the maximal isotropic subspace spanned by $C$.
Following steps similar to those of the four dimensional case~\cite{acck0907}, it is possible
to do a partial gauge fixing by choosing two other vectors in $\IC^{2k}$ such
that the remaining integration variables parameterize ${\rm OG}(k-2,2k-4)$
whose dimension is given by eq.\eqref{var-counting}.
\footnote{
The importance of this coset in constructing R-symmetry invariants
was noted in ref.~\cite{blm}.}

\subsection{Asymptotic $z$ revisited} \label{z-revisited}

If we consider $R(z)$ as an element of ${\rm O}(2k,\IC)$ by the natural embedding ${\rm O}(2,\IC)\subset {\rm O}(2k,\IC)$,
it acts on the $k\times 2k$ matrix integration variable $C$ from the right.
We can use this fact to rewrite the
deformed Grassmannian integral as
\begin{align}
\mathcal{L}_{2k} (\hat{\L}(z)) &=  \int \frac{d^{k\times 2k }C}{\textrm{vol}[{\rm GL}(k)]}\frac{\delta^{\frac{k(k+1)}2} (C \cdot C^T) \delta^{2k|3k} (C \cdot R(z) \cdot \Lambda)}{M_1 (C) \cdots M_k(C)}\;,  \nonumber
\\
&= \int \frac{d^{k\times 2k }\tilde{C}}{\textrm{vol}[{\rm GL}(k)]} \frac{\delta^{\frac{k(k+1)}2} ( \tilde{C} \cdot \tilde{C}^T) \delta^{2k|3k} (\tilde{C} \cdot \Lambda)}{M_1 (\tilde{C} \cdot R^{-1}(z))  \cdots M_k(\tilde{C}\cdot R^{-1}(z))}\;,
\quad (\tilde{C} \equiv C\cdot R(z)) \nonumber
\\
&= \int \frac{d^{k\times 2k }C}{\textrm{vol}[{\rm GL}(k)]} \frac{\delta^{\frac{k(k+1)}2} (C \cdot C^T) \delta^{2k|3k} (C \cdot \Lambda)}{M_1 (\hat{C}(z)) \cdots M_k(\hat{C}(z))} \, ,
\quad (\hat{C}(z) \equiv C \cdot R^{-1}(z)) \;.
\label{deform-grass}
\end{align}
On the second line, we used the GL$(k)$ gauge invariance.
On the third line, we removed the tilde to avoid clutter.
Note that all $z$-dependence has been absorbed into the minors.
This is in accord with the discussion in section 4 that the full superamplitude has a well-defined large $z$-behavior.

As in section 3.1, suppose the legs $1$ and $l$ are deformed.
{}From the $z$-dependence of the minors, we now argue that
the asymptotic behavior of the superamplitude is
\begin{align}
A(z)  \stackrel{z \to \infty}{\longrightarrow} {\cal O} (z^{-(l-1)}) \; ,  \qquad( 2 \leq l \leq k ) \,.
  \label{scale}
\end{align}
To see this, first note that
\begin{align}
C_\pm \equiv C_1 \pm i\, C_l
\;\;\;
\imp
\;\;\;
\hat{C}_1 (z) = \frac{z} 2 C_+ + \frac{z^{-1}}2 C_- \,,
\;\;
\hat{C}_l (z) = \frac{z}{2i} C_+ - \frac{z^{-1}}{2 i }C_- \,.
\label{hat C_1 , C_l}
\end{align}
Here, $C_i$ ($1\leq i\leq 2k$) denote the $i$-th column in $C$. $M_1$ is the determinant of
\[
\left(  \hat{C}_1(z), C_2, \cdots,    \hat{C}_{l}(z), \cdots, C_{k} \right) ,
\]
thus $M_1$ is independent of $z$.  $M_{i >l}$ are also independent of $z$ since the determinant includes neither of $\hat{C}_1(z)$, $\hat{C}_{l}(z)$. On the other hand, each $M_{2 \leq i \leq l } (z)$ are linear in $z$ at the leading order as $z \to \infty$, thereby
confirming the $z^{-(l-1)}$ behavior claimed in eq.\eqref{scale}.

For the case $l=k+1$, there is a subtlety in finding the large $z$ behavior from the Grassmannian integral. Recall that the integral \eqref{deform-grass} is actually a contour integral. After solving the
bosonic delta function constraints, we can write the contour integral
schematically as
\begin{align}
\mathcal{L}_{2k} (\hat{\Lambda}(z)) = \oint d\tau \frac{J(\tau) \delta (C(\tau)\cdot \eta)}{M_1 (\tau ;z) M_2 (\tau;z)\ldots M_k (\tau;z)}\;,\quad  M(\tau ;z) \equiv M (\hat{C}(\tau;z) ) ,
\end{align}
where $\tau$ parameterize the $(k-2)(k-3)/2$ holomorphic coordinates
of OG$(k-2,2k-4)$. The function $J(\tau)$ denotes the Jacobian factor
arising from the delta function constraints.  From eq.\eqref{hat C_1 , C_l}, it is easy to see that ($\tau$ dependence is suppressed for simplicity)
\begin{align}
&M_1(z) = \frac{z}2 \big{(} \det (C_+ ,C_2 ,  \ldots , C_k) + \mathcal{O}(z^{-2}) \big{)} \equiv z m_1 +\mathcal{O}(z^{-1})\;, \nonumber
\\
&M_2(z) = \frac{z}{2i} \big{(} \det (C_2 , \ldots , C_k, C_+) + \mathcal{O}(z^{-2}) \big{)} \equiv z m_2 +\mathcal{O}(z^{-1})\;, \nonumber
\\
&M_i(z) = \frac{z}{2i}   \big{(} \det (C_i , \ldots , C_k, C_+ , \ldots ,C_{i+k-1}) + \mathcal{O}(z^{-2}) \big{)}\;, \quad \textrm{for }2<i \leq k.
\end{align}
One peculiarity of  the $l=k+1$ case is that the zeros of the two minors $M_1 (\tau;z), M_2 (\tau;z)$ tend to collapse at large $z$.
Since $\hat{C}_1(z)$ and $i\hat{C}_l(z)$ both approaches $(z/2)(C_+ +\CO(z^{-2}))$, if we denote the zeros of $M_i(\tau;z)$ by $\tau^*_i$, then
\begin{align}
\tau^*_1 = \tau^*_2 + \mathcal{O}(z^{-2}).
\end{align}
At $\tau^*_1$, all minors $M_{j \neq 1}$  are $\mathcal{O}(z)$ at large $z$, except $M_2$ which is  $\mathcal{O}(z^{-1})$:
\begin{align}
M_2 (\tau_1^*;z) =  z m_2 (\tau_2^* + \mathcal{O}(z^{-2})) +\mathcal{O}(z^{-1}) = \mathcal{O}(z^{-1}).
\end{align}
Thus residue at $\tau_1^*$, denoted by $\{1\}$, asymptotically behaves as
\begin{align}
\{1\} \stackrel{z \to \infty}{\longrightarrow} \mathcal{O}(z^{-1} \times \frac{1}{ z^{-1}\times z \ldots z}) = \mathcal{O}(z^{-(k-2)}).
\end{align}
First $z^{-1}$ factor comes from the residue of $1/M_1$ at $\tau_1^*$. The residue at $\tau_2^*$, $\{2\}$,
shares the same property. Since at $\tau^*_i$ ($i>2$) all minors $\hat{M}_{j\neq i}$ are $\mathcal{O}(z)$, we conclude that
\begin{align}
&\{i \} =\mathcal{O}(z^{-k}), \quad 2<i \leq k.
\end{align}
Thus depending on the choice of contour, the asymptotic behavior of superamplitude could be $\mathcal{O}(z^{-(k-2)})$ or $\mathcal{O} (z^{-k})$ for $l=k+1$. In any case, it vanishes as $z$ goes to infinity.
For $k=2,3$, there is no integration, so the superamplitudes behave as $\mathcal{O}(z^{-k})$ at large $z$.

If the Grassmannian integral is indeed capable
of producing all tree-level amplitudes,
the above observation shows that
the supersymmetric recursion relation should work for any choice of reference
legs. This is consistent with (and stronger than)
the analysis of section \ref{z-count} based
on Feynman diagrams.

We can also consider a non-supersymmetric recursion relation
by deforming $ \lambda$ only and leaving $ \eta$ intact.
The asymptotic behavior of the amplitudes can become worse, since the fermionic delta function can contribute up to $z^3$ order
to the numerator of ${ \cal L}_{2k} (z)$.
Similar supersymmetric improvements have been noted when applying the BCFW deformation to four dimensional ${ \cal N}=4$ sYM~\cite{bht}, and ${ \cal N}=8$ supergravity~\cite{ArkaniHamed:2008gz}.

Finally, eq.\eqref{scale} tells us that the asymptotic behavior of the amplitude becomes better when we deform two legs further apart. This tendency is also observed in four dimensions, which can be used to derive the generalized BCJ relations~\cite{Bern:2008qj} as demonstrated in~\cite{Feng:2010my}.

\subsection{Computing amplitudes up to 8-point}

\paragraph{4-point amplitude}
The 4-point superamplitude was first computed by Feynman diagrams
in ref.~\cite{Agarwal:2008pu}. In our convention, the amplitude is given by
\be
\CA_4 = \frac{\d^3(P) \d^6(Q)}{\langle 14 \rangle \langle 34 \rangle} \,.
\label{4pt-known}
\ee
It is rather trivial to reproduce this amplitude from the Grassmannian integral.
One can fix the GL$(2)$ gauge by setting, for instance,
\be
C = \begin{pmatrix}
c_{21} & 1 & c_{23} & 0 \\
c_{41} & 0 & c_{43} & 1
\end{pmatrix} \,.
\ee
The integral reduces to finding the zeroes of
the bosonic delta function $\d(C\cdot \l)$.
\be
\d^4(C\cdot \l) = \frac{1}{\langle 13 \rangle^2}\prod_{\bar{r}, s} \d^4( c_{\bar{r}s} - c^*_{\bar{r}s})   \,, \qquad
\begin{pmatrix}
c^*_{21} & c^*_{23} \\
c^*_{41} & c^*_{43}
\end{pmatrix}
=
-\frac{1}{\langle 13 \rangle}
\begin{pmatrix}
\langle 2 3 \rangle & \langle 1 2 \rangle \\
\langle 4 3 \rangle & \langle 1 4 \rangle
\end{pmatrix} \,.
\label{4pt-1}
\ee
Inserting $c^*_{\bar{r}s}$ to the remaining delta functions and the minors, one finds
\be
\d^{3}(C\cdot C^T) =  \frac{\langle 13 \rangle^6}{\langle 24 \rangle^3} \d^3(P) \,,
\;\;
\d^6(C\cdot \eta) = \frac{\langle 24 \rangle^3}{\langle 13 \rangle^6}\d^6(Q) \,,
\;\;
\frac{1}{M_1 M_2} 
= \frac{\langle 13 \rangle^2}{\langle 1 4 \rangle \langle 34 \rangle} \,.
\label{4pt-2}
\ee
Combining eq.\eqref{4pt-1} and eq.\eqref{4pt-2}, we recover the superamplitude eq.\eqref{4pt-known}.
For later convenience, we list some explicit examples of component amplitudes here.
\begin{align}
&  \langle \bar{\phi} (\lambda_1)\phi (\lambda_2) \bar{\phi} (\lambda_3)\phi(\lambda_4)\rangle =  \frac{\langle 13 \rangle^3}{\langle 14 \rangle \langle 34 \rangle}\,,  \;\;
\langle \bar{\phi} (\lambda_1) \phi (\lambda_2) \bar{\psi} (\lambda_3) \psi(\lambda_4) \rangle = -\frac{\langle 14 \rangle^3}{\langle 14 \rangle \langle 34 \rangle}\,,
\label{4pt-comp1}
\\
&\langle \bar{\phi} (\lambda_1)\psi (\lambda_2) \bar{\psi} (\lambda_3)\phi(\lambda_4) \rangle = \frac{\langle 12 \rangle^3}{\langle 14 \rangle \langle 34 \rangle} \,, \quad
\langle \bar{\psi} (\lambda_1)\phi (\lambda_2) \bar{\phi}(\lambda_3)\psi (\lambda_4) \rangle =  \frac{\langle 34 \rangle^3}{\langle 14 \rangle \langle 34 \rangle} \,.
\label{4pt-comp2}
\end{align}
Throughout this section, $(\phi,\bar{\phi},\psi,\bar{\psi})$
without superscripts will denote $(\phi^4,\bar{\phi}_4,\psi_4,\bar{\psi}^4)$.

\subsubsection{6-point amplitude}

\paragraph{Generality}
The counting \eqref{var-counting} for the number of integration variables
shows that there is no integral to do
other than solving the delta function constraints.
But, since ${\rm OG}(1,2) = O(2)/U(1)=\IZ_2$,
the amplitude is expected to be a sum of two terms.
In fact, the general structure of the 6-point superamplitude was
analyzed in ref.~\cite{blm}.  There,
based on the invariance under the $SO(6)$ $R$-symmetry,
it was shown to take
the following general form,
\be
\CA_{6} = \d^3(P) \d^6(Q) \left( \d^3(\a_+) f_+(\l) + \d^3(\a_-) f_-(\l) \right) \,.
\label{6pt-general}
\ee
Here, $\d^{3}(\a) = \frac{1}{6} \e_{IJK} \a^I \a^J \a^K$ and
$\a^I_{\pm} = \a_{\pm}^i (\l) \eta^I_i$
is some linear combination of $\eta_i^I$
that is linearly independent of the $Q$-directions.
The two terms are related to each other by the ``$\l$-parity" symmetry.

As noted in ref.~\cite{blm}, there is some ambiguity in
the precise definition of $\a_\pm$.
{}From our perspective, the ambiguity is nothing but
the GL$(k)$ gauge symmetry of the Grassmannian integral.
It was also shown in ref.~\cite{blm}
that superconformal invariance imposes severe constraints on $f_{\pm}(\l)$
but does not fully determine them. To remedy this problem,
two component amplitudes were computed by Feynman diagrams.
We will see below that
the Grassmannian integral is capable of reproducing the
full 6-point superamplitude without any extra input.

\paragraph{Cyclic gauge}
Let us begin by taking what we may call a ``cyclic" gauge.
\be
C =
\begin{pmatrix}
c_{21} & 1 & c_{23} & 0 & c_{25} & 0 \\
c_{41} & 0 & c_{43} & 1 & c_{45} & 0 \\
c_{61} & 0 & c_{63} & 0 & c_{65} & 1 \\
\end{pmatrix} .
\ee
The bosonic delta functions can be ``integrated" by using the following relation,
\be
\int d^{3\times 3} c \, \d^6(\l_{\bar{r}} + c_{\bar{r} s} \l_s)
\d^6(c c^T + 1) F(c) = \d^3(P) \left( F(c^*_+) + F(c^*_-) \right) ,
\ee
where $c^*_{\pm}$ are the two solutions
of the delta-function constraints for the ($3\times 3$) matrix $c_{\bar{r}s}$.
The barred (unbarred) indices label even (odd)-numbered particles, respectively.
Both types of indices are defined modulo 6.
Concretely, the solutions $c^*_{\pm}$ take the form
\be
(c^*_{\pm})_{\bar{r}s} = \frac{-\langle \bar{r}|p_{135}|s\rangle\pm i \langle \bar{r}+ 2, \bar{r}-2\rangle \langle s-2 , s+2\rangle}{(p_{135})^2}.
\ee
Here, we defined $p_{135}\equiv p_1 + p_3 + p_5$.
Substituting these solutions to the fermionic delta function
determines $\a_{\pm}$:
\be
\left.\d^9(C\cdot \eta)\right|_{c=c^*_\pm} = \d^6(Q) \d^3(\a_{\pm})  \,,
\;\;\;\;\;
\a_{\pm}^I = - \frac{1}{2(p_{135})^2} \left( \e_{\bar{p}\bar{q}\bar{r}} \langle \bar{p} \bar{q} \rangle \eta_{\bar{r}}^I \pm i \,\e_{pqr} \langle p q \rangle \eta_r^I  \right) \,.
\ee
The minors in the denominator determine the bosonic functions $f_{\pm}(\l)$,
\be
f_{\pm} = \left.\frac{1}{M_1 M_2 M_3}\right|_{c=c^*_\pm}
=  \frac{1}{(c^*_{25} c^*_{41} c^*_{63})_\pm} \,.
\ee
The component amplitudes can be easily read off from the superamplitude,
for example,
\begin{align}
&\langle \bar{\psi}(\lambda_1) \psi (\lambda_2) \bar{\psi} (\lambda_3) \psi (\lambda_4)\bar{\psi} (\lambda_5) \psi (\lambda_6)  \rangle  = \frac{-1}{(c^*_{63}c^*_{25}c^*_{41})_+}+\frac{-1}{(c^*_{63}c^*_{25}c^*_{41})_-}
\equiv A_{6\psi} \,, \nonumber
\\
&\langle \bar{\phi} (\lambda_1) \phi(\lambda_2) \bar{\phi}(\lambda_3) \phi (\lambda_4)\bar{\phi} (\lambda_5) \phi (\lambda_6) \rangle
= \frac{-i}{(c^*_{63}c^*_{25}c^*_{41})_+}
+\frac{+i}{(c^*_{63}c^*_{25}c^*_{41})_-}  \equiv A_{6\phi} \,, \nonumber
\\
&\langle \bar{\phi} (\lambda_1)\phi (\lambda_2) \bar{\phi}(\lambda_3)\phi (\lambda_4) \bar{\psi} (\lambda_5) \psi (\lambda_6) \rangle  = \frac{-i(c^*_{65})^3_+}{(c^*_{63}c^*_{25}c^*_{41})_+}+\frac{-i(c^*_{65})^3_-}{(c^*_{63}c^*_{25}c^*_{41})_-}   \equiv A_{\bar{\phi} \phi \bar{ \phi} \phi \bar{\psi} \psi }
\,.
\end{align}

In ref.~\cite{blm}, $A_{6\psi}$ and $A_{6\phi}$ were computed by Feynman diagrams. It was shown that the two component amplitudes
and superconformal symmetry are sufficient to determine
the full superamplitude.
Even after judicious use of color-ordered Feynman rules
and efforts to simplify the result, the final expressions
for $A_{6\psi}$ and $A_{6\phi}$ given in ref.~\cite{blm}
are somewhat lengthy. There, the cyclic symmetry
of the amplitudes results from summing over diagrams
related to each other by cyclic shifts.

In contrast, our expressions for $A_{6\psi}$ and $A_{6\phi}$
take an extremely simple form.
The component amplitudes contain two terms related by ``$\l$-parity",
and each term exhibits manifest cyclic symmetry.
Remarkably, our results exactly match those of ref.~\cite{blm}
evaluated at on-shell momenta.
Technically, the comparison was done only numerically.
But, since we are comparing two rational functions of the $\l$-variables
with quadratic constraints, numerical checks for sufficiently
many $\l$ configurations amount to a complete proof.

\paragraph{Factorization gauge}

Next, consider using the ``factorization" gauge,
\be
C =
\begin{pmatrix}
1 & 0 & 0 & c_{14} & c_{15} & c_{16} \\
0 & 1 & 0 & c_{24} & c_{25} & c_{26} \\
0 & 0 & 1 & c_{34} & c_{35} & c_{36}  \\
\end{pmatrix} .
\label{factor-gauge}
\ee
The computation is almost the same as in the cyclic gauge.
The full superamplitude still takes the general form \eqref{6pt-general}:
\be
\CA_{6} = \d^3(P) \d^6(Q) \left( \d^3(\a_+) f_+(\l) + \d^3(\a_-) f_-(\l) \right) . \nn
\ee
The fermions $\a_\pm$ are now given by $(\bar{r}=1,2,3;s=4,5,6)$
\be
\a_{\pm}^I = - \frac{1}{2(p_{123})^2} \left( \e_{\bar{p}\bar{q}\bar{r}} \langle \bar{p} \bar{q} \rangle \eta_{\bar{r}}^I \pm i \,\e_{pqr} \langle p q \rangle \eta_r^I  \right) .
\ee
The matrices $c^*_\pm$ can be expressed as
\begin{align}
(c^*_{\pm})_{\bar{r}s} = \frac{\langle \bar{r}|p_{123}|s\rangle \mp i \langle \bar{r}+ 1, \bar{r}+2\rangle \langle s+1 , s+2\rangle}{(p_{123})^2}
\,,
\end{align}
where cyclic identification is understood separately for the barred and unbarred indices. Because $M_1=1$ in this gauge,
the functions $f_\pm(\l)$ take a simpler form
\begin{align}
[f_\pm(\l)]^{-1} &= \left. M_1 M_2 M_3\right|_{c=c^*_{\pm}} = ( \mp i )  (c_{14}^* c_{36}^*)_{\pm}
\nn \\
&= ( \mp i )   \frac{\left( \langle  1 | p_{123} | 4 \rangle \mp i \langle 23\rangle \langle 56 \rangle  \right)
\left(  \langle  3 | p_{123} | 6 \rangle  \mp i \langle 12\rangle \langle 45 \rangle  \right)}{[(p_{123})^2]^{2}} \,.
\label{6pt-factor-minor}
\end{align}
As a result, the component amplitudes also become simpler, for instance,
\begin{align}
&A_{\bar{ \phi} \phi \bar{ \phi} \phi \bar{\psi} \psi } = \frac{-i}{(p_{123})^2} \left[(-1) \frac{( \langle 2 \vert p_{123} \vert 6 \rangle- i \langle 31 \rangle \langle 45 \rangle )^3}
{(\langle 1 \vert p_{123} \vert 4 \rangle - i \langle 23 \rangle \langle 56 \rangle )( \langle 3 \vert p_{123} \vert 6 \rangle - i \langle 12 \rangle \langle 45 \rangle )} \right. \nonumber
\\
&\qquad \qquad \qquad \qquad
\left.+\frac{ (\langle 2 \vert p_{123} \vert 6 \rangle + i \langle 31 \rangle \langle 45 \rangle )^3}
{(\langle 1 \vert p_{123} \vert  4 \rangle +i \langle 23 \rangle \langle 56 \rangle )(  \langle 3 \vert p_{123} \vert 6 \rangle  + i \langle 12 \rangle \langle 45 \rangle )}\right]. \label{6pt-4b2f}
\end{align}

We have confirmed that the component amplitudes obtained in the two different gauges are indeed the same.
The manifest appearance of the internal propagator $1/(p_{123})^2$
in eq.\eqref{6pt-4b2f} suggests that the factorization gauge is more suitable
for the study of the recursion relation.
In fact, we will recover eq.\eqref{6pt-4b2f}
from the recursion relation in the next section.
For a later convenience, we show two explicit examples in this gauge
\begin{align}
& \langle \bar \phi ( \lambda_1)  \phi ( \lambda_2 ) \bar{ \psi} ( \lambda_3 ) \phi ( \lambda_4 ) \bar{ \phi} ( \lambda_5 ) \psi ( \lambda_6 ) \rangle  = \left( \frac{ (c^{*}_{14})^3 }{ c^{*}_{14} c^{*}_{36} }  \right)_{+} + \left(  \frac{ (c^{*}_{14})^3}{ c^*_{14} c^*_{36} }  \right)_{-} \equiv A_{ \bar{ \phi} \phi \bar{\psi} \phi \bar{ \phi} \psi} \,,
\nn \\
& \langle \bar \phi ( \lambda_1)  \psi ( \lambda_2 ) \bar{ \psi} ( \lambda_3 ) \phi ( \lambda_4 ) \bar{ \psi} ( \lambda_5 ) \psi ( \lambda_6 ) \rangle =\left(  \frac{ (c^*_{36})^3}{c^*_{14 }c^*_{36} } \right)_{+} - \left( \frac{ (c^*_{36})^3 }{ c^{*}_{ 14} c^{*}_{ 36} } \right)_{-}  \label{6pt-for-8pt} \equiv  A_{ \bar{\phi}  \psi \bar{\psi} \phi \bar{ \psi}   \psi } \,.
\end{align}

\subsubsection{8-point amplitude}

\paragraph{Generality} \label{8pt-generality}

The counting \eqref{var-counting} indicates that
there is a one parameter family of solutions to delta function constraints.
Following the general discussion above, if we fix the GL$(4)$ gauge by filling up four of the columns of the $C$ matrix
by the identity matrix,
the remaining ($4\times 4$) matrix $c_{\bar{r}s}$ 
defines ($\pm i$ times) an element of O$(4,\IC)$.
Consider the condition
\be
\l_{\bar{r}}^\a + c_{\bar{r}s} \l_{s}^\a = 0\,.
\label{8pt-pcons}
\ee
We can understand it geometrically as follows.
We construct two sets of orthonormal basis for $\IC^4$,
one associated to $(-i) \l_{\bar{r}}$ (call it $\bar{e}_{i}$) and
the other associated to $\l_{s}$ (call it $e_{i}$).
We define $e_1$ and $e_2$ such that $e_1$ is parallel to $\l^{\a=1}_s$
and $e_2$ is parallel to $\l^{\a=2}_s$ minus its projection onto $e_1$. Then, $e_3$ and $e_4$ are defined to span the 2-plane orthogonal
to $e_{1,2}$.
The other set $\{ \bar{e}_i \}$ is defined in a similar way
with $(-i)\l_{\bar{r}}$ replacing $\l_{s}$.

The momentum conservation (\ref{8pt-pcons}) requires that
$i c_{\bar{r}s}$ map $(e_1,e_2)$ onto $(\bar{e}_1,\bar{e}_2)$.
After imposing this condition, we need to map the 2-plane spanned by
$\bar{e}_{3,4}$ to the 2-plane spanned by $e_{3,4}$.
There is precisely an O$(2)$ freedom in such a map, and that explains
the origin of the single integration variable.

The group O$(2)$ contains two copies of SO$(2)$ related to
each other by orientation reversal.
Our experience with the integral formula in the 6-point example suggests
that we should include both copies to find the correct answer.
For a given fixed orientation, we parameterize the SO$(2)$ rotation
by $\tau = e^{i\theta}$. Since we are working with a contour integral, $\theta$
and $\tau$ should be understood as complex variables.

Concretely, we can write the pair of one-parameter solutions $c_{\pm}(\tau)$
as follows.
\begin{align}
&i c_+(\tau) = \bar{e}_1 e_1^T +\bar{e}_2 e_2^T
+ \tau ( \bar{e}_+ e_-^T) + \tau^{-1} (\bar{e}_- e_+^T) \,,
\nn \\
& i c_-(\tau) = \bar{e}_1 e_1^T +\bar{e}_2 e_2^T
+ \tau ( \bar{e}_+ e_+^T) + \tau^{-1} (\bar{e}_- e_-^T) \, ,
\label{c-pm}
\end{align}
where $e_{ \pm} = \frac{1}{ \sqrt{2} } ( e_3 \pm i e_4), \bar{e}_{ \pm} = \frac{1}{ \sqrt{2} }  ( \bar{e}_3 \pm i \bar{e}_4 ) $.
The orthogonality condition $cc^T=-1$ ensures that
the minors $M_i$ are at most quadratic in $c_{\bar{r}s}$.
Note that even for the minors quadratic in $c_{\bar{r}s}$,
due to the determinant structure,
\be
M(\tau) = c_{\bar{r}_1 s_1}(\tau) c_{\bar{r}_2 s_2}(\tau) -
c_{\bar{r}_1 s_2}(\tau) c_{\bar{r}_2 s_1}(\tau)\,,
\ee
the $\tau^{\pm 2}$ terms all cancel out, so that the zeroes of the minors
can be computed explicitly:
\be
M(\tau) = \alpha \tau^{-1} + \b + \g \tau = 0
\;\;\; \imp \;\;\;
\tau  = \frac{-\b \pm \sqrt{\b^2 -4 \a \g}}{2\g} \,.
\ee
The unpleasant square-root factors $\sqrt{\b^2 -4 \a \g}$
will all disappear once we sum over residues,
and the final result will be a rational function of
$\langle i j \rangle$ $(i,j=1,\cdots 8)$.

Since we reduce the matrix integral to an ordinary contour integral
in the $\tau$-variable by solving delta function constraints,
it is important to compute the Jacobian factors.
As explained in appendix \ref{details}, the Jacobian factor is fairly simple
$(p_0 \equiv \sum_{\bar{r}} p_{\bar{r}})$:
\begin{align}
\CA_8 (\Lambda) &= \frac{\d^{3}(P)}{\sqrt{-p_0^2}} \frac{1}{ 2 \pi i} \left( \oint_{C_+} + \oint_{C_-}\right) \frac{d \tau}{\tau} \frac{\delta^{(12)}(c_{\bar{r}s}(\tau)\eta^I_s +\eta^I_{\bar{r}})}{M_1(\t) M_2(\t) M_3(\t) M_4(\t) } \,,
\nn \\
& =  \frac{\d^{3}(P) \d^{6}(Q) }{ \sqrt{ - p_0^2 } }  \frac{1}{ 2 \pi i}  \left( \oint_{C_+} + \oint_{C_-}\right) \frac{d \tau}{\tau} \frac{   \prod_{ I = 1}^3  a^I  }{M_1(\t) M_2(\t) M_3(\t) M_4(\t) }  \,,
\label{8pt-master}
\end{align}
where $a^I$ is given by
\begin{align}
a^I= \frac{1}{- (p_0)^2 } \frac{1}{4} \epsilon_{ \bar{p} \bar{q} \bar{r} \bar{s} } \langle \bar{p} \bar{q} \rangle ( \eta_{ \bar{r} }^I + c_{ \bar{r} t  } \eta_{t}^I )( \eta^I_{ \bar{s} } + c_{ \bar{s}u  } \eta_{u}^I ) .
\qquad \mbox{(no sum in} \; I)
\end{align}
The equality in eq.\eqref{8pt-master} holds up to an overall constant factor that cannot be determined by the Grassmannian integral.

\paragraph{Factorization gauge}
Our earlier discussion of the 6-point amplitude
shows that the amplitude in the factorization gauge
is the simplest and most suitable for comparison with the recursion relation.
For simplicity,
we present the result only in the factorization gauge,
($\bar{r}=1,2,3,4$, $s=5,6,7,8$)
\begin{equation}
C = \begin{pmatrix} 1 & 0 & 0 & 0 & c_{ 1 5} & c_{ 16} & c_{17} & c_{18} \\
0& 1&0&0& c_{25} & c_{ 26} & c_{ 27} & c_{ 28} \\
0&0& 1& 0 & c_{ 35} & c_{ 36} & c_{ 37} & c_{ 38}   \\
0& 0 & 0 & 1 & c_{ 45} & c_{ 46} & c_{ 47} & c_{ 48}
\end{pmatrix} .
\label{8pt-fact-gauge}
\end{equation}
The minors appearing in the denominator are given by $M_1=1$ and
 \be
 M_{2; \pm } = - c_{15 ; \pm } ( \tau ) \,, \;\;\;
 & M_{3 ; \pm} = \det \begin{pmatrix} c_{15}(\t) & c_{16}(\t) \\ c_{25}(\t) & c_{ 26}(\t) \end{pmatrix}_{ \pm} \,, \;\;\;
 & M_{4;\pm} = \pm c_{48 ; \pm }( \tau) \,, \label{8pt-minors}
 \ee
with $c_{\pm}(\t)$ defined in eq.\eqref{c-pm}.

For concreteness, we focus on a particular component amplitude
$A_{ \bar{\phi} \psi \bar{ \psi} \phi \bar{ \psi}  \phi \bar{ \phi} \psi} $.
This choice simplifies the evaluation of the contour integral
considerably because the fermionic delta function in the numerator
produces
 \be
 \left[ \det{}
 \begin{pmatrix}
 c_{ 37} & c_{ 38} \\ c_{ 47} & c_{ 48}
 \end{pmatrix}_{ \pm}
 \right]^3
 =  \left( \pm  M_{3 \pm} (\t)  \right)^3  \,,
 \nn
 \ee
which removes poles corresponding to $M_3(\tau)$. Note that the removal of the poles from $M_3(\tau)$ is gauge independent. The gauge choice \eqref{8pt-fact-gauge} further simplifies the calculation since $M_1( \tau)=1$.
We should stress that the net number of poles must be gauge independent;
the poles for $M_1(\tau)$ are merely pushed away to zero and infinity.

The amplitude thus becomes
\begin{align}
A_{  \bar{\phi} \psi \bar{ \psi} \phi \bar{ \psi}  \phi \bar{ \phi} \psi } & = \frac{1}{ \sqrt{ - p^2_{1234}} }   \frac{1}{ 2 \pi i}    \left( \oint_{C_+} \frac{ d \tau} { \tau} \frac{ M_{ 3 +}^2 (\tau) }{ M_{2+}(\tau) M_{4+} (\tau) } +\oint_{ C_{-} }  \frac{ d \tau} { \tau} \frac{ (-1) M_{3-}^2 (\tau) }{ M_{2-}(\tau) M_{4-}(\tau) } \right) .   
\end{align}
This looks much simpler than the generic integral \eqref{8pt-master}.
Let us introduce the notation
\begin{equation}
\{2\}_{ \pm},   \quad \{ 4 \}_{ \pm} , \quad \{ 0 \}_{ \pm} , \quad \{ \infty \}_{ \pm} ,
\end{equation}
where  $\{0 \}_{ \pm}, \{ \infty \}_{ \pm} $ denote residues evaluated at $\t  = 0$ and $ \t = \infty$, $\{n\}_{ \pm}$ is the sum of the two residues
for the $n$-th minor.
The sum of all residues should of course vanish,
\begin{equation}
\{2\}_+ + \{ 4 \}_+ + \{ 0 \}_+ + \{ \infty \}_+  = 0
= \{2\}_- + \{ 4 \}_- + \{ 0 \}_- + \{ \infty \}_- \,.
\end{equation}

To determine which linear combinations of the residues
yield the correct 8-point amplitude, we make use of
discrete symmetries of the amplitude.
First, consider a particular ``$\l$-parity" $\pi$ which acts as
\be
\pi  :  \lambda_1 \to - \lambda_1  \,.
\ee
Second, we consider the following permutation,
\begin{align}
& \sigma : \{ 1, 2, 3, 4, 5, 6, 7, 8 \} \to \{ 4,3,2,1,8,7,6,5 \} \,.
\end{align}
The amplitude at hand,
$A_{\bar{\phi}\psi \bar{\psi} \phi \bar{\psi}\phi\bar{\phi}\psi}$,
is invariant under both transformations;
$\pi$-invariance due to the fact that $\bar{\phi}(\l_1)$ is
a boson, and $\s$-invariance due to the charge conjugation symmetry.
On the other hand, the residues transform as
\begin{align}
\pi  \ : \ &  \{2\}_{ \pm} \leftrightarrow \{ 2\}_{ \mp} , \quad  \{ 4 \}_{ \pm} \leftrightarrow  \{ 4 \}_{ \mp} , \quad   \{0 \}_{\pm}  \leftrightarrow   \{ \infty\}_{ \mp}   \ ,
\\
\sigma \ : \ &  \{2\}_{\pm} \leftrightarrow \{4 \}_{\pm}, \quad    \{0 \}_{ \pm} \leftrightarrow \{ 0 \}_{ \pm} , \quad  \{ \infty \}_{ \pm} \leftrightarrow  \{ \infty \}_{ \pm}  \ .
\end{align}
These transformation properties show
that there is a unique combination of residues
to give the correct amplitude, namely,
\begin{align}
A_{\bar{\phi}\psi \bar{\psi} \phi \bar{\psi}\phi\bar{\phi}\psi}
& = \{ 0 \}_+ + \{ 0 \}_- + \{ \infty \}_+ +\{ \infty \}_-  \ \\
& = - \left(\{ 2 \}_+ + \{ 2 \}_- + \{ 4 \}_+ +\{ 4 \}_- \right) .
\label{8pt-residues}
\end{align}
To emphasize the discrete symmetries, we can also write
\begin{align}
A_{\bar{\phi}\psi \bar{\psi} \phi \bar{\psi}\phi\bar{\phi}\psi}
 =( 1+\pi) \cdot ( \{ 0 \}_+ +  \{ 0 \}_-)  = - ( 1+ \pi) \cdot ( 1 + \sigma) \cdot \{2 \}_{+} \,.
\end{align}

After some algebra, we can write down the residues explicitly,
\begin{align}
\{0 \}_{+} + \{0\}_{-}
= \frac{1}{ \sqrt{ - p_{1234}^2} } \frac{1}{p^2_{123} p^2_{678} p^2_{234} p^2_{567} }  \left( n_{32+} n_{34+} + n_{32-} n_{34-} \right) ,
\label{zero-residue}
\end{align}
where $n_{32 \pm}, n_{34 \pm}$ are given by
\be
n_{32\pm}  &=&   -i \sqrt{ -p^2_{1234} }  (\langle 34 \rangle \langle 6 | p_{78} | 1 \rangle   \mp \langle 78 \rangle \langle 2 | p_{34} | 5 \rangle )
\nn \\
&&
-( \langle 12 \rangle \langle 56 \rangle \pm   \langle 34 \rangle \langle 78 \rangle )  \langle 1 | p_{234} | 5 \rangle  +p_{1234}^2 \langle 6|p_{781} | 2 \rangle  \, ,  \nn \\
n_{34 \pm}  &=&  i \sqrt{ - p_{1234}^2 } ( \langle 56 \rangle \langle 3 | p_{12} | 8 \rangle \mp \langle 12 \rangle  \langle 4 | p_{56} | 7 \rangle )
\nn \\
&&-( \langle 12 \rangle \langle 56 \rangle  \pm \langle 34 \rangle \langle 78\rangle)  \langle 4 | p_{567} | 8 \rangle \mp p_{1234}^2  \langle 3 | p_{456} | 7 \rangle \, .
\nn \label{8pt-semifinal}
\ee
The numerator in eq.(\ref{8pt-semifinal}) has the expansion
\begin{align}
 \left( n_{32+} n_{34+} + n_{32-} n_{34-} \right)  & = \sqrt{ - p^2_{1234} } ( \cdots ) + ( \cdots). \nn
 \end{align}
The second term has eigenvalue $(-1)$ under the $\pi $-transformation, thus it is projected out by $(1+ \pi)$. To summarize,
we obtain a remarkably simple final answer
\begin{align}
& A_{  \bar{\phi} \psi \bar{ \psi} \phi \bar{ \psi}  \phi \bar{ \phi} \psi} ( \lambda_1, \lambda_2, \lambda_3, \lambda_4, \lambda_5, \lambda_6, \lambda_7, \lambda_8 )
\nn \\
& =  \frac{1 }{p^2_{123} p^2_{678} p^2_{234} p^2_{567} }  \big\{ p^2_{1234}  \left( - \langle 2 | p_{34} | 5 \rangle \langle 3 | p_{456 } | p_7 | 8 \rangle + \langle 3 | p_{12} | 8 \rangle \langle 5 | p_6  | p_{781} | 2 \rangle  \right) \nn \\
&   \qquad \qquad \qquad \qquad + \langle 12 \rangle \langle 34 \rangle \left( \langle 78 \rangle \langle 1 | p_{234 } | 5 \rangle \langle 4 | p_{56} | 7 \rangle - \langle 56 \rangle \langle 8 | p_{123} | 4 \rangle \langle 6 | p_{78} | 1 \rangle \right) \nn \\
&  \qquad \qquad \qquad \qquad- \langle 12 \rangle \langle 56 \rangle^2  \langle 1 | p_{234 } | 5 \rangle \langle 8 | p_{12} | 3 \rangle + \langle 34 \rangle \langle 78 \rangle ^2  \langle 8 | p_{123} | 4 \rangle \langle 2 | p_{34} | 5 \rangle
\big\} \label{8pt-final},
\end{align}
up to an overall constant. 
The overall factor $ 1 /p^2_{123} p^2_{678} p^2_{234} p^2_{567}  $ in eq.\eqref{8pt-final} suggests that the 8-point amplitude can be understood as a sum over off-shell 4- and 6-point sub-diagrams joined with an internal propagator, $1/p^2_{123}, 1/p^2_{678} , 1/p^2_{234}$ and $1/p^2_{567}$. The absence of factors  such as $1/p^2_{345}$ or $1/p^2_{456}$ also can be expected, since the corresponding decomposition to the sub-diagrams does not exist. In passing, we also note that under the deformation given in eq.\eqref{deformation}, the large $z$ behavior of  $A_{ \bar{ \phi} \psi \bar{ \psi} \phi \bar{ \psi} \phi \bar{ \phi} \psi}(z)$ agrees not only with the Feynman diagram analysis in section \ref{z-count}, but also with the Grassmannian formula analysis in section \ref{z-revisited}.
\section{Examples of recursion relation}

In this section, we  give a few examples of the recursion relation derived in section \ref{sec:recursion}. We will match the recursion result with that derived in the previous sections. The checks are made up to an overall constant for the $n$-point superamplitude. In principle, by carefully defining
the notion of color-ordered amplitudes, one could determine
the overall coefficients without any ambiguity,
both in the recursion relation and in the Feynman diagram analysis.
As for the
Grassmannian integral formula, there is no known way
to determine the overall coefficients.
We will not keep track of these coefficients in this paper.

\subsection*{Example 1: $(6)=(4)\star(4)$}

\begin{figure}
\begin{center}
\includegraphics[width=9cm]{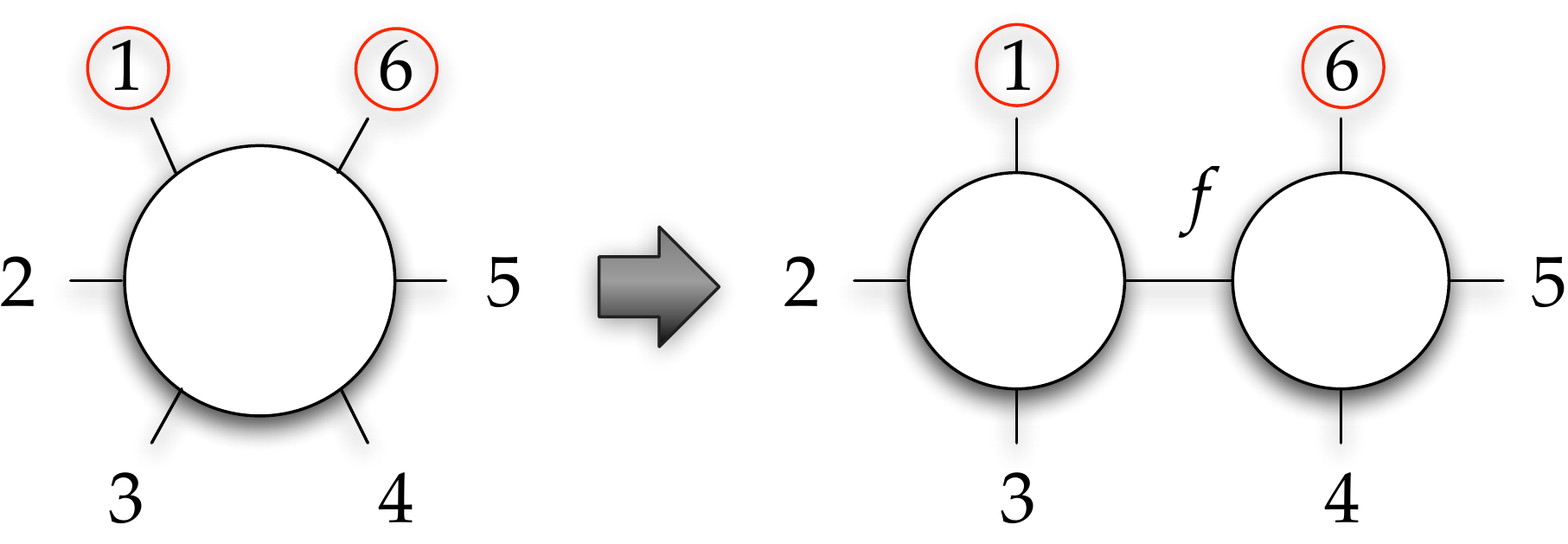}
\caption{Recursion relation applied to the 6-point amplitude.}
\label{6pt-example}
\end{center}
\end{figure}

To be specific, consider a component amplitude \eqref{6pt-4b2f}
computed in the factorization gauge, which
we reproduce here for the reader's convenience
\begin{align}
&A_{\bar{ \phi} \phi \bar{ \phi} \phi \bar{\psi} \psi } = \frac{i}{(p_{123})^2} \left[ \frac{( \langle 2 \vert p_{123} \vert 6 \rangle- i \langle 31 \rangle \langle 45 \rangle )^3}
{(\langle 1 \vert p_{123} \vert 4 \rangle - i \langle 23 \rangle \langle 56 \rangle )( \langle 3 \vert p_{123} \vert 6 \rangle - i \langle 12 \rangle \langle 45 \rangle )} \right. \nonumber
\\
&\qquad \qquad \qquad  \qquad \;\;\;
\left.- \frac{ (\langle 2 \vert p_{123} \vert 6 \rangle + i \langle 31 \rangle \langle 45 \rangle )^3}
{(\langle 1 \vert p_{123} \vert  4 \rangle +i \langle 23 \rangle \langle 56 \rangle )(  \langle 3 \vert p_{123} \vert 6 \rangle  + i \langle 12 \rangle \langle 45 \rangle )}\right].
\label{re:6pt-4b2f}
\end{align}
To compare this with the recursion relation, we choose to deform
$(\Lambda_1,\Lambda_6)$. This deformation admits only one factorization
channel (123:456) as demonstrated in Fig.\ref{6pt-example}. The recursion relation \eqref{fact-limit} reads
\begin{align}
&A_6 (\Lambda_1, \Lambda_2, \Lambda_3, \Lambda_4, \Lambda_5, \Lambda_6) \nonumber
\\
& \quad
= \frac{i}{(p_{123})^2} \int d^3 \eta  \left( H( z^*_1 , z^*_{2}  ) A_4 (\hat\Lambda_1 (z_{1}^*), \Lambda_2, \lambda_3, \Lambda_f) A_4 (i \Lambda_f , \Lambda_4 , \Lambda_5, \hat\Lambda_6 (z_1^*))+( z^*_1 \leftrightarrow z^*_2 ) \right) \,, 
\end{align}
where $H$ is the function defined in eq.\eqref{function F} for even $l$.
The component amplitude $A_{\bar{ \phi} \phi \bar{ \phi} \phi \bar{\psi} \psi}$ can be obtained from the superamplitude.
Noting that $\hat\eta_1 (z) \hat\eta_6 (z) = \eta_1 \eta_6$,
we find
\begin{align}
&A_{\bar{ \phi} \phi \bar{ \phi} \phi \bar{\psi} \psi }
(\l_1,\l_2,\l_3,\l_4,\l_5,\l_6)
\nn \\
& = \frac{i}{(p_{123})^2} \left( H(z^*_1 ,z^*_2) A_{ \bar{\phi} \phi \bar{ \phi} \phi } ( \hat{ \lambda}_1 (z_1^*) , \lambda_2, \lambda_3, \lambda_f ) A_{ \bar{ \phi} \phi \bar{ \psi} \psi } ( i \lambda_f , \lambda_4 , \lambda_5 , \hat{ \lambda}_6 ( z_1^* ))  + ( z^*_1  \leftrightarrow z^*_2) \right)  \,.
\label{bcfw analytic}
\end{align}

To obtain the explicit form of the amplitude,
we need the expression for 4-point component amplitudes given in eq.\eqref{4pt-comp1}.
Due to the small number of external legs,
the positions of poles take a simple form,
\begin{align}
{z^*_1}^2
= \frac{\langle 16\rangle^2 - (\langle 23 \rangle - \langle 45\rangle)^2}{
(\langle 1| + i \langle 6|) p_{45}(|1\rangle + i | 6 \rangle) } \,, \nonumber
\qquad
{z^*_2}^2 =  \frac{\langle 16\rangle^2 - (\langle 23 \rangle + \langle 45\rangle)^2}{
(\langle 1| + i \langle 6|) p_{45}(|1\rangle + i | 6 \rangle) } \,.
\end{align}
After repeated use of momentum conservation and Schouten's identity, we obtain
\begin{align}
& H(z_1^*, z_2^* ) A_{ \bar{\phi} \phi \bar{ \phi} \phi } (z_1^* ) A_{ \bar{ \phi} \phi \bar{ \psi} \psi} (z_1^*)
=- \frac{ (\langle 2 \vert p_{123} \vert 6 \rangle + i \langle 31 \rangle \langle 45 \rangle )^3}
{(\langle 1 \vert p_{123} \vert  4 \rangle +i \langle 23 \rangle \langle 56 \rangle )(  \langle 3 \vert p_{123} \vert 6 \rangle  + i \langle 12 \rangle \langle 45 \rangle )} \,,
\nn
\\
&H(z_2^*, z_1^* ) A_{ \bar{\phi} \phi \bar{ \phi} \phi } (z_2^* ) A_{ \bar{ \phi} \phi \bar{ \psi} \psi} (z_2^*)
= \frac{( \langle 2 \vert p_{123} \vert 6 \rangle- i \langle 31 \rangle \langle 45 \rangle )^3}
{(\langle 1 \vert p_{123} \vert 4 \rangle - i \langle 23 \rangle \langle 56 \rangle )( \langle 3 \vert p_{123} \vert 6 \rangle - i \langle 12 \rangle \langle 45 \rangle )} \,.
\end{align}
Thus, we have verified that
the amplitude obtained from the recursion relation \eqref{bcfw analytic} is
the same as the amplitude computed
directly from the Grassmannian integral \eqref{re:6pt-4b2f}. 

Obviously, we can check the recursion relation
against the
Grassmannian integral for many other component amplitudes
and different factorization channels.
For instance, if we deform  $(\L_1 ,\L_3 )$,
there are two channels labeled by $(561)$, $(345)$.
Schematically, the results can be expressed as
\begin{align}
& A_{ 6 \psi} ( \lambda_1, \lambda_2, \lambda_3, \lambda_4, \lambda_5, \lambda_6 ) = A_{ 4 \psi} ( \lambda_5, \lambda_6, \hat \lambda_1, \lambda_f ) \star A_{ 4 \psi} ( i \lambda_f , \lambda_2 , \hat{ \lambda}_3 , \lambda_4 ) \nn \\
& \qquad\qquad  \qquad\qquad  \qquad \quad \;\;\;\;  + A_{ 4 \psi} ( \hat{ \lambda}_3 , \lambda_4,  \lambda_5 , \lambda_f ) \star A_{ 4 \psi} ( i \lambda_f , \lambda_6 , \hat{ \lambda}_1 , \lambda_2 ) \,,
\label{psi-bcfw}
\\
& A_{ 6 \phi} ( \lambda_1, \lambda_2, \lambda_3, \lambda_4, \lambda_5, \lambda_6 ) = A_{ 4 \phi} ( \lambda_5, \lambda_6, \hat \lambda_1, \lambda_f ) \star A_{ 4 \psi} ( i \lambda_f , \lambda_2 , \hat{ \lambda}_3 , \lambda_4 ) \nn \\
& \qquad\qquad  \qquad\qquad  \qquad \quad \;\;\;\;  + A_{ 4 \phi} ( \hat{ \lambda}_3 , \lambda_4, \lambda_5 , \lambda_f ) \star A_{ 4 \phi} ( i \lambda_f , \lambda_6 , \hat{ \lambda}_1 , \lambda_2 )  \,.
\label{phi-bcfw}
\end{align}
To check these relations in practice, care should be taken to assign
the correct factors of $(\pm i)$ on the right-hand side.

Some component recursion relations can mix external particles. Let us show an explicit example.  We are still deforming $( \Lambda_1, \Lambda_3)$. The recursion relation for $A_{\bar{ \psi} \psi \bar{ \phi} \phi \bar{ \phi} \phi}$ can be read from the general recursion relation \eqref{BCFW} by integrating out $\eta_2^3 \eta_3^3 \eta_5^3$ of $A_6$. On the right hand side of the recursion relation, the component amplitude of $A_4$ which depends on $ \hat{\eta}_1$ in turn depends on $ \eta_3$, thus it leads to
\begin{align}
& A_{ \bar{ \psi} \psi \bar{ \phi} \phi \bar{ \phi} \phi } ( \lambda_1, \lambda_2 , \lambda_3 , \lambda_4 , \lambda_5 , \lambda_6 )  =  -  A_6 |_{\eta_2^3 \eta_3^3 \eta_5^3} \nn \\
&\qquad=s^3( z^{ \ast} )  A_{ \bar{ \phi } \phi \bar{ \phi} \phi } ( \lambda_5 , \lambda_6 , \hat{ \lambda}_1 , \lambda_f )   \star A_{ \bar{\phi} \psi \bar{ \psi} \phi } ( i \lambda_f , \lambda_2 , \hat{ \lambda}_3 , \lambda_4 )  \nn \\
& \qquad\quad + c^3 ( z^{ \ast} ) A_{ \bar{ \phi} \phi \bar{ \psi} \psi} ( \lambda_5 , \lambda_6 , \hat{ \lambda}_1 , \lambda_f ) \star A_{ \bar{ \psi} \psi \bar{ \phi} \phi } ( i \lambda_f , \lambda_2 , \hat{ \lambda}_3 , \lambda_4 ) + \cdots  .
\label{changing-ptcl}
\end{align}
where $c( z^{ \ast}), s( z^{ \ast})$ are defined by $ c( z^{ \ast} ) = \frac{ z^{ \ast} + (z^*)^{-1} }{2}, s( z^{ \ast} ) = \frac{ z^{ \ast} - (z^{\ast} )^{-1} }{2i}$, for $ z^{ \ast}$ the zeros of the on-shell condition of the channel $p_f^2 ( z^{ \ast} ) = 0$. The first line of eq.\eqref{changing-ptcl} comes from the expansion of $ \eta_2^3 \hat{ \eta}_1^3 \eta_5^3$, while the second line from $ \eta_2^3 \hat{ \eta}_3^3 \eta_5^3$. Note that in the first line, the external particles are changed from $ \bar \psi ( \lambda_1) \cdots\bar{\phi} ( \lambda_3) \cdots $ to $ \bar \phi ( \hat{\lambda}_1)\cdots  \bar \psi ( \hat{ \lambda}_3 ) \cdots  $.

\subsection*{Example 2: $(8)=(4)\star(6)+(6)\star(4)$}

\begin{figure}
\begin{center}
\includegraphics[width=13cm]{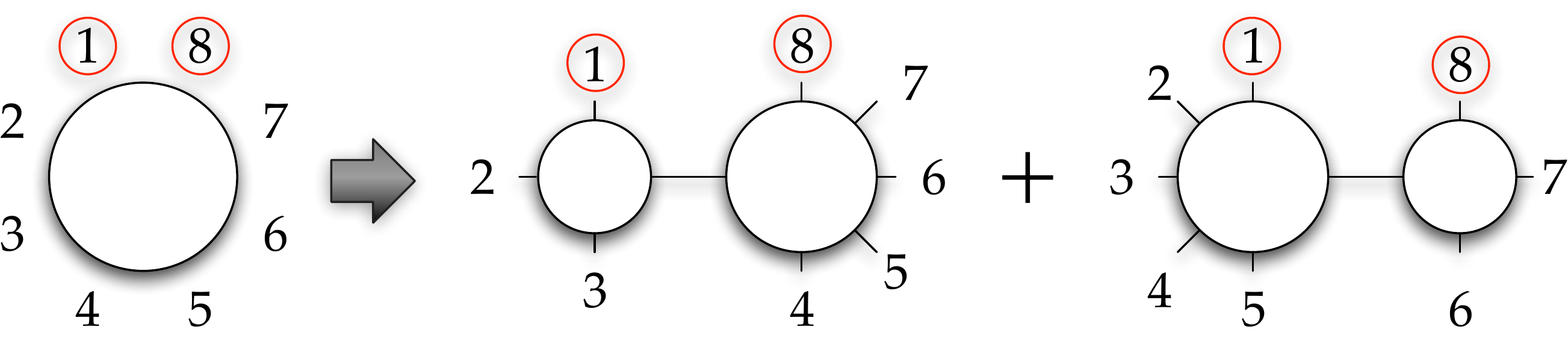}
\caption{Recursion relation applied to the 8-point amplitude.}
\label{8pt-example}
\end{center}
\end{figure}

We apply the recursion relation
to the 8-point component amplitude $A_{\bar{\phi}\psi \bar{\psi} \phi \bar{\psi}\phi\bar{\phi}\psi}$ \eqref{8pt-final}.
If we choose to deform $(\Lambda_1,\Lambda_8)$,
only two factorization channels contribute to the recursion relation
as demonstrated in Fig.\ref{8pt-example}.
Schematically, the recursion relation takes the form
\begin{align}
& A_{ \bar{\phi} \psi \bar{ \psi} \phi \bar{ \psi} \phi \bar{ \phi} \psi } ( \lambda_1, \lambda_2, \lambda_3, \lambda_4, \lambda_5 , \lambda_6, \lambda_7 , \lambda_8 )
\nn \\
& \qquad =A_{ \bar \phi \psi \bar{ \psi} \phi } ( \hat{ \lambda}_1 , \lambda_2, \lambda_3, \lambda_f ) \star A_{ \bar{ \phi} \phi \bar{ \psi} \phi \bar{ \phi} \psi} ( i \lambda_f , \lambda_4 , \lambda_5 , \lambda_6 , \lambda_7 , \hat{ \lambda}_8 ) \nn \\
& \qquad \quad +  A_{ \bar{ \phi} \psi \bar{ \psi} \phi \bar{ \psi} \psi}  ( \hat{ \lambda}_1 , \lambda_2 , \lambda_3 , \lambda_4 , \lambda_5 , \lambda_f ) \star A_{ \bar{ \psi} \phi \bar{ \phi} \psi} ( i \lambda_f , \lambda_6 , \lambda_7 , \hat{ \lambda}_8 ) \,.
\label{recursion-8pt}
\end{align}
%
To evaluate the right-hand side,
we use the 4-point amplitudes in eq.\eqref{4pt-comp2}
and the 6-point amplitudes in eq.\eqref{6pt-for-8pt}.

By numerically computing both sides of eq.\eqref{recursion-8pt} for
a large number of $\{\l_i\}$ configurations satisfying
the overall momentum conservation, we have confirmed
that the recursion relation holds.
Since we are comparing two rational functions of complex variables,
the numerical check amounts to a complete proof of the on-shell equivalence.

It is worth noting that, if we express the 8-point amplitude  as the sum of residues \eqref{8pt-residues}, the
two channels of the recursion relation in eq.\eqref{recursion-8pt} separately match the residues in the following way,
\begin{align}
&- ( \{4 \}_{+} + \{4 \}_{-} ) = A_{ \bar \phi \psi \bar{ \psi} \phi } ( \hat{ \lambda}_1 , \lambda_2, \lambda_3, \lambda_f ) \star A_{ \bar{ \phi} \phi \bar{ \psi} \phi \bar{ \phi} \psi} ( i \lambda_f , \lambda_4 , \lambda_5 , \lambda_6 , \lambda_7 , \hat{ \lambda}_8 )  \,, \nn \\
& - ( \{ 2 \}_{+} + \{ 2 \}_{-} ) =   A_{ \bar{ \phi} \psi \bar{ \psi} \phi \bar{ \psi} \psi}  ( \hat{ \lambda}_1 , \lambda_2 , \lambda_3 , \lambda_4 , \lambda_5 , \lambda_f ) \star A_{ \bar{ \psi} \phi \bar{ \phi} \psi} ( i \lambda_f , \lambda_6 , \lambda_7 , \hat{ \lambda}_8 )   \,. \label{residue-channel}
\end{align}
One may expect it from the form of minors in the factorization gauge,  $M_2 \sim c_{15} , M_4 \sim c_{ 48} $, in eq.\eqref{8pt-minors}.


\section{Dual superconformal symmetry for all tree amplitudes}

In this section, we will use the recursion relation described in section 3 to demonstrate that the dual superconformal symmetry of the four-point amplitude can be extended to all on-shell tree-level amplitudes in the ABJM theory. As we will explain in section 7.2, this boils down to showing that the amplitudes are covariant under dual inversion.

The proof is based on induction: assuming that $\mathcal{A}_n$ with $n<2k$ transform covariantly under dual inversion, we demonstrate that each term in the recursive construction of $\CA_{2k}$ will transform in a universal covariant matter, and thus so will  $\CA_{2k}$. Since it was already shown that the four-point amplitude transforms covariantly under dual inversion~\cite{hl2}, this proves covariance for all ABJM tree amplitudes. Similar proofs were given for the maximal super Yang-Mills theories in four-, six- and ten-dimensions~\cite{bht,Dennen:2010dh,CaronHuot:2010rj}.

In the next subsection, we will review the construction of the dual superspace for the ABJM theory, and explain how to perform inversion in this space. We present the proof of dual superconformal symmetry in section 7.2.

\subsection{Dual space and dual inversion properties }
In ref~\cite{blm}, it was demonstrated that the four and six-point amplitudes of the ABJM theory have Yangian symmetry. Yangian symmetry of the scattering amplitudes was then shown to be equivalent to superconformal symmetry plus dual superconformal symmetry~\cite{hl2}. In particular, the nontrivial level-one Yangian generators were matched with generators of dual superconformal symmetry, which are defined in a dual space parameterized by the coordinates $(x,y,\theta)$.
The dual space coordinates are related to the on-shell space coordinates $(\l,\eta)$ as follows:
\be
x_{i,i+1}^{\a \b} &\equiv  x_i^{\a \b} - x_{i+1}^{\a \b} = p_i^{\a \b } = \l_i^\a \l_i^\b \,, \nn
\\
\theta_{i,i+1}^{I \a} &\equiv \theta_i^{I \a} - \theta_{i+1}^{I \a} = q_i^{I\a} = \l_i^\a \eta_i^I \,,\nn
\\
y_{i,i+1}^{IJ} &\equiv y_i^{IJ} - y_{i+1}^{IJ} = r_i^{IJ} = \eta_i^I \eta_i^J \,
 \label{dual coordinate}
\ee
where $x_{2k+1} \equiv x_1, \; \theta_{2k+1}\equiv \theta_1,\;y_{2k+1}\equiv y_1$. Note that the $y$ coordinates are Grassmann-even. The dual space coordinates are defined such that supermomentum and R-symmetry are automatically conserved:
\be
\sum_i p_i=\sum_i q_i=\sum_i r_i=0.
\ee
Inversion $I$ acts on the dual space as
\be
I[x_i^{\a\b}] =\frac{x_i^{\a\b}}{x_i^2}, \; I[\theta_i^{I\alpha}] = \frac{x_i^{\a\b}}{x_i^2} \theta^I_{i\b},\; I\left[y_{i}^{AB}\right]=\frac{\theta_{i}^{A\alpha}\theta_{i}^{B\beta}x_{i\alpha\beta}}{x_{i}^{2}}+y_i^{AB}.
\ee
The spinor indices $(\a,\b)$ are raised and lowered using the antisymmetric 2-index $\epsilon$ tensor.

As noted in ref.~\cite{hl2}, the coordinates ($x_i,\theta_i$) are sufficient to define ($\lambda_i,\eta_i$). Given the $x$'s, one can determine $\lambda$'s using the first line in eq.(\ref{dual coordinate}) (up to sign ambiguities). These sign ambiguities also appear when translating from spinors to dual coordinates via
\begin{equation}
\langle i i+1\rangle=\pm\sqrt{-x_{i,i+2}^2}.
\label{ambiguous}
\end{equation}
Given the $\theta$'s, one can then determine the $\eta$'s using the second line in eq.(\ref{dual coordinate}):
\be
\eta_i^I=\frac{(\theta_{i,i+1})^{I\beta}\lambda_{i+1\beta}}{\sqrt{-x_{i,i+2}^2}}.
\ee
Finally, given the $\eta$ coordinates, the $y$ coordinates can be determined using the third line in eq.(\ref{dual coordinate}) up to some reference point $y_0$, which the amplitudes do not depend on since they are invariant under translations in the dual superspace.
Hence, the amplitudes can be parameterized using only $x$ and $\theta$ coordinates. Using this parametrization, only half of the dual supersymmetry is manifest. In principle, it should be possible to write amplitudes using all three coordinates $(x,y,\theta)$ in such a way that all of the dual supersymmetry is manifest, but for simplicity, we will only use $x$ and $\theta$ coordinates to write the amplitudes.

Note that the dual coordinates in eq.\eqref{dual coordinate} are defined up to a constant shift.  For later convenience, we fix this ambiguity by choosing\footnote{Note that this choice can always be made by taking the origin of the dual space to be at the midpoint of the line $(x_2,x_{2k})$.}
\be
x_2 +x_{2k}=0 , \; \theta_1 =q_1. \label{ambiguity fixing}
\ee
For this choice,
\begin{align}
&x_1 = \frac{p_1 - p_{2k}}2 , \; x_2 = - x_{ 2k} =  - \frac{p_1 +p_{2k}}2 , \ldots , x_{2k} = \frac{p_1 +p_{2k}}2, \nonumber
\\
&\theta_1 = q_1 , \; \theta_2 = 0, \ldots, \theta_{2k} = q_1 +q_{2k}. \label{x-lambda relation}
\end{align}
Furthermore, if we apply the deformation in eq.\eqref{deformation} to legs 1 and $2k$, this only shifts the point $(x_1, \theta_1)$ in the dual space:
\be
&&\hat{x}_1 (z) = \frac{\hat p_1 (z) - \hat p_{2k}(z)}2, \; \hat{\theta}_1 (z) = \hat q_1 (z), \nn
\\
&&\hat x_{i} (z) = x_i, \; \hat \theta_i (z) = \theta_i , \; \textrm{for $i>1$.}
\ee
Although $x_1$ is deformed to $ \hat{x}_1 (z)$, its norm does not change, i.e. $\hat{x}_1^2(z)= x^2_1$.  This is one of the advantages of the choice in eq.\eqref{ambiguity fixing}. Another implication of this choice is
\be
x_2^2 = x_{2k}^2 = - x_1^2.
\ee

The inversion properties of $(\lambda,\eta)$  are given by~\cite{hl2}:
\be
&&I [\l_i^\a]  = \e_i \frac{(x_i)^{\a\b} \l_{i\b}}{\sqrt{(x_{i+1})^2(x_i)^2}}= \e_i \frac{(x_{i+1})^{\a\b} \l_{i\b}}{\sqrt{(x_{i+1})^2(x_i)^2}},
\;\;\;\;\; (\e_i=\pm 1)  \ \nonumber
\\
&&I [\eta_i^I] = - \e_i \frac{x^2_i}{\sqrt{x_i^2 x_{i+1}^2}} [\eta_i^I +(x^{-1}_{i+1})^{\a \b} \theta_{i \b}\l_{i \a}]. \label{inversion on lambda}
\ee
where $\e_i$ represents the sign ambiguity of the inversion rules, which is related to the sign ambiguity in eq.(\ref{ambiguous}). The  shifted dual coordinates obey these transformation rules if we choose $x_2 + x_{2k} =0$ and $\epsilon_1 \epsilon_{2k} = -1$:
\be
&&I [\hat x_1 (z)] = \frac{\hat x_1 (z)}{\hat x^2_1 (z)} = \frac{\hat{x}_1(z)}{x_1^2}, \; I [\hat\theta_1 (z)] = \frac{\hat x_1 (z) \cdot \hat\theta_1 (z)}{x_i^2}, \nonumber
\label{inversion identity}
\ee
where $(x \cdot \theta)^\a = x^{\a\b}\theta_\b$. This can be demonstrated straightforwardly:
\be
I [\hat x_1 (z)] &&=  I \left[\frac{(\l_1 - i \l_{2k})^2 z^{-2} +(\l_1 + i \l_{2k})^2 z^{2}}4\right]  \nonumber
\\
&&=  \frac{(x_1\cdot \l_1-i \e_1 \e_{2k}x_{2k}\cdot \l_{2k})^2}{4 x_1^2 x_2^2 }z^{-2}+\frac{(x_1\cdot \l_1+ i \e_1 \e_{2k}x_{2k}\cdot \l_{2k})^2}{4 x_1^2 x_2^2 }z^2    \nonumber
\\
&&=\frac{(\l_1-i  \l_{2k})^2}{4 x_1^2 }z^{-2}+\frac{(\l_1+ i  \l_{2k})^2}{4 x_1^2  }z^2  \nonumber\\
 &&= \frac{\hat x_1(z)}{x_1^2}.
\ee
In obtaining this result, we noted that $x^2_2 = x_{2k}^2 $ and used the explicit form of $x_1, x_2, x_{2k}$ in eq.\eqref{x-lambda relation}. If we had chosen $\epsilon_1 \epsilon_{2k}=1$, then eq.\eqref{inversion identity} would read
\be
I [\hat x_1 (z)] = \frac{\hat x_1 (1/z)}{x_1^2}.
\ee
For simplicity, we will choose $\epsilon_1 \epsilon_{2k}=-1$. Choosing $\epsilon_1 \epsilon_{2k}=1$ does not change our conclusions.

Having defined our conventions for the dual space, we will now proceed to the proof of dual superconformal invariance.

\subsection{Proof of dual superconformal invariance}

We would like to show that when the on-shell amplitudes are written in terms of the dual superspace, they transform as follows under dual inversion:
\be
&&I[\CA_{2k}] = \sqrt {\prod_{i=1}^{2k} x^2_i}\CA_{2k} \,,
\label{cov}
\ee
or equivalently,
\be
&&I[f_{2k}] = \sqrt {\prod_{i=1}^{2k} x^2_i} f_{2k}, \;\CA_{2k} = f_{2k}\delta^3(P)\delta^6(Q) ,
\label{dual conformal invariance}
\ee
where we use the identity $I[\delta^3(P)\delta^6(Q)] = \delta^3(P)\delta^6 (Q)$ which is proven in ref.~\cite{hl2}.
It is easy to verify eq.(\ref{dual conformal invariance}) for the case $2k=4$:\footnote{Note that there may be sign ambiguities when one combines square roots, since generically $\sqrt{A}\sqrt{B}=\pm\sqrt{AB}$. This ambiguity can be removed by keeping track of the phase for each factor in the square root. In the end, terms with the same factors appearing in the square root will have the same phase. }
\be
I[f_4] = I [ \frac{1}{\sqrt{x^2_{1,3}} \sqrt{x_{2,4}^2}}] = \sqrt{\prod_{i=1}^4 x_i^2 } \ f_4.
\ee

Under a dual inversion-translation-inversion, eq.(\ref{cov}) implies that the superamplitude transforms as follows:
\be
K_{\alpha\beta}\left[\mathcal{A}_{2k}\right]=IP_{\alpha\beta}I\left[\mathcal{A}_{2k}\right]=-\frac{1}{2}\left(\sum_{i=1}^{2k}x_{i\alpha\beta}\right)\mathcal{A}_{2k} \ .
\ee
Hence, if we define the dual conformal boost generator to be
\be
\tilde{K}_{\alpha\beta}=K_{\alpha\beta}+\frac{1}{2}\sum_{i=1}^{2k}x_{i\alpha\beta} \ ,
\ee
then this will be a symmetry of the amplitudes. Furthermore, it can be shown that $\tilde{K}$ matches a level-1 Yangian generator and that all the other nontrivial dual superconformal generators can be obtained by commuting $\tilde{K}$ with the ordinary superconformal generators (when acting on on-shell amplitudes)~\cite{hl2}. In summary, if we can prove that eq.(\ref{dual conformal invariance}) holds for all tree-level amplitudes, this implies that the amplitudes enjoy dual superconformal symmetry.

The proof is based on induction. Assuming that
\be
I[f_{2l}(x_i)] = f_{2l}(\frac{x_i}{x_i^2})= \sqrt{\prod_{i=1}^{2l} x_i^2} f_{2l} (x_i), \label{induction assumption}
\ee
for all $2l$, with $l<k$, we will use the recursion relation defined in section 3 to show that $f_{2k}$ inverts the same way.
Recall the recursion relation 
given in eq.\eqref{fact-limit}:
\be
A_{2k} = \sum_f \int d^3 \eta \frac{1}{p_f^2} [H(z_{1,f}^*,z_{2,f}^*)A_{L} (z^*_{1,f};\eta)A_R (z^*_{1,f};i\eta)+(z^*_{1,f}\leftrightarrow z^*_{2,f})],
\nn
\ee
where $f$ labels the different channels and  $\CA_L$ and $\CA_R$ are lower-point amplitudes. We have $p_f = p_{j+1} + \ldots +p_{2k} = - (p_1 +\ldots + p_j)$ and $\{\pm z^*_{1}, \pm z^*_2\}$ are solutions of $\hat{p}_f (z)^2 =(\hat p_1 (z)+p_2 +\ldots +p_j)^2 =0$. The function $H(a,b)$ is given in eq.\eqref{function F}.
Combining $\delta^6(Q_L)\delta^6(Q_R) = \delta^6(Q_{2k})\delta^6 (Q_R)$, one can extract $f_{2k}$:
\be
f_{2k} = \sum_f \int d^3 \eta \frac{1}{p_f^2} [\delta^6(Q_R)H(z^*_{1;f},z^*_{2;f})f_L (z^*_{1,f};\eta) f_R (z^*_{1,f};i\eta) +(z^*_{1,f}\leftrightarrow z^*_{2,f})].
\label{BCFW for dualconformal}
\ee

\begin{figure}
\begin{center}
\includegraphics[scale=0.8]{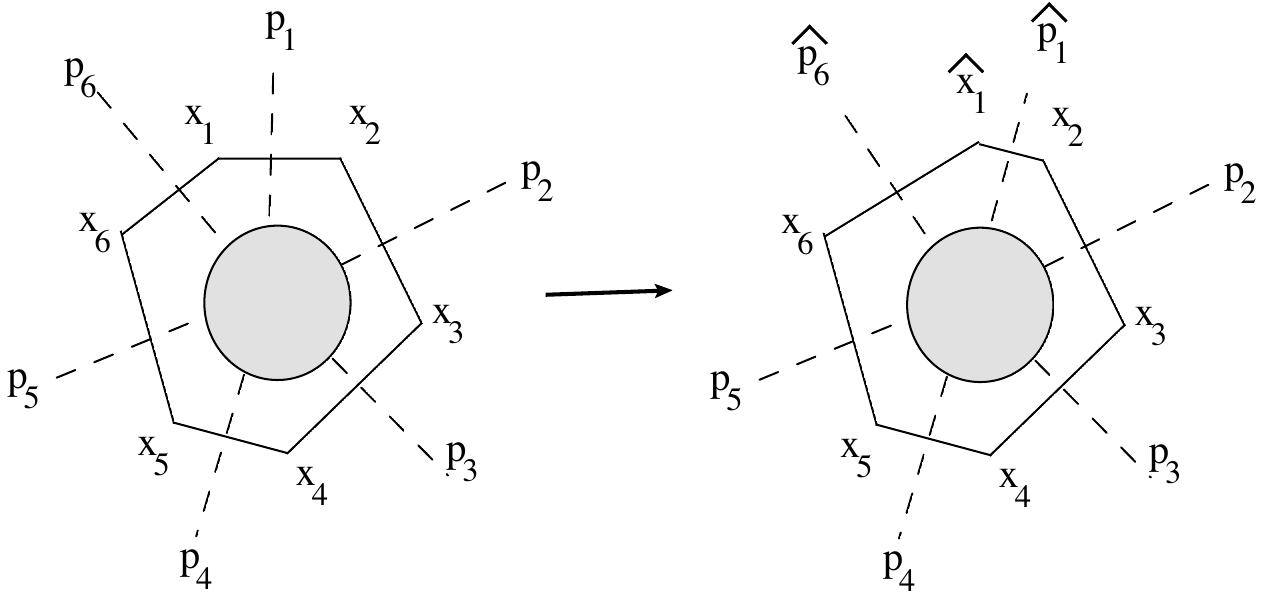}
\caption{The recursion in dual coordinates for $2k=6$. Note that the amplitude is deformed by shifting $x_1$ to $\hat{x}_{1}(z)$. The shifts creates poles in the $p_{123}$ channel and the residue is proportional to the product of four-point tree amplitudes at shifted kinematics.   }
\label{bcfwdual}
\end{center}
\end{figure}

We will now deduce the inversion property of $f_{2k}$ by applying a dual inversion to eq.(\ref{BCFW for dualconformal}). As explained in section 7.1, the deformation corresponds to a shift in the dual coordinate $x_1$. We illustrate this in Fig.\ref{bcfwdual} for the case $2k=6$.
In terms of the shifted dual coordinate,
\be
f_L (z^*) = f_L (\hat x_1 (z^*), x_2,\ldots,x_{j+1}), \; f_R (z^*) = f_R (\hat x_1 (z^*),x_{j+1},x_{j+2},\ldots, x_{2k}),
\ee
where we suppress the $\theta_i$ coordinates for simplicity. By assumption, the functions $f_L$ and $f_R$ invert as 
\be
&&I[f_L (\hat x_1 (z^*), \ldots, x_{j+1})] = \sqrt{(\hat x_1^2(z^*)\ldots x_{j+1}^2)} f_L (\hat x_1(z^*),\ldots, x_{j+1}), \nonumber
\\
&&I[f_R (\hat x_1 (z^*), \ldots, x_{2k})] = \sqrt{(\hat x_1^2(z^*)\ldots x_{2k}^2)} f_L (\hat x_1(z^*),\ldots, x_{2k}). \label{inversion bcfw 1}
\ee
The inversion of the propagator term in eq.(\ref{BCFW for dualconformal})  is given by
\be
I[\frac{1}{p_f^2}]  = I [\frac{1}{x_{1,j+1}^2}] = \frac{x_1^2 x_{j+1}^2}{x_{1,j+1}^2}
= \frac{x_1^2 x_{j+1}^2}{p_f^2  }.  \label{inversion bcfw 2}
\ee
A simple calculation shows that
\be
\int d^3 \eta \delta^6 (Q_R) && = \int d^3 \eta \delta^6 (q^{I \a}_{j+1}+ \ldots + \hat q^{I\a}_{2k}- \eta^I  \hat \l^\a_{f}) \nonumber
\\
&&= \int d^3 \eta \delta^6 (\theta^{I\a}_{j+1} - \hat \theta^{I\a}_{1} - \eta^I \hat{\l}^\a_{f}) \nonumber
\\
&& = \frac{1}{6} \epsilon_{IJK} (\hat \theta_1 - \theta_{j+1})^{I \a}\hat{\l}_{f,\a}(\hat \theta_1 - \theta_{j+1})^{J \b}\hat{\l}_{f,\b} (\hat \theta_1 - \theta_{j+1})^{K \g}\hat{\l}_{f,\g}\,.
\ee
To see how this expression inverts, we note that
\be
I [(\hat \theta_1 - \theta_{j+1})^{I \a} \hat \l_{f,\a}] && = [(\hat{x}^{-1}_1)^{\a \g} \hat \theta^I_{1,\g}-(x^{-1}_{j+1})^{\a\g}\theta^I_{j+1,\g}]\frac{(x_{j+1})_{\alpha}^{\phantom{\alpha} \b}\hat\l_{f,\b}}{\sqrt{x_{j+1}^2 \hat x_1^2}} \nonumber
\\
&&= (\hat{x}^{-1}_1)^{\a \g} \hat \theta^I_{1,\g}  \frac{(\hat x_{1})^{\phantom{\a} \b}_{\alpha} \hat\l_{f,\b}}{\sqrt{x_{j+1}^2 \hat x_1^2}}  -(x^{-1}_{j+1})^{\a\g}\theta^I_{j+1,\g} \frac{(x_{j+1})^{\phantom{\a} \b}_{\a} \hat\l_{f,\b}}{\sqrt{x_{j+1}^2 \hat x_1^2}} \nonumber
\\
&&= \frac{1}{\sqrt{\hat x_1^2(z^*) x_{j+1}^2}}(\hat \theta_1 - \theta_{j+1})^{I \a} \hat \l_{f,\a} \,.
\ee
In the second line, we used the identity $x_{i} \cdot \l_i = x_{i+1}\cdot \l_i$ with $i=j+1, i+1 =\hat 1,\l_i = \hat \l_f$. Hence, we find that
\begin{align}
I[\int d^3 \eta \delta^6(Q_R)] = \frac{1}{\hat x_1^2(z^*) x^2_{j+1}\sqrt{\hat x_1^2 (z^*) x^2_{j+1}}} \int d^3 \eta \delta^6(Q_R). \label{inversion bcfw 3}
\end{align}
The only remaining piece to invert in eq.\eqref{BCFW for dualconformal} is $H(z_1^*,z_2^*)$. From the observation that
\begin{align}
I [ \hat p_f (z)^2] & 
= \frac{\hat p_f (z)^2}{x_1^2 x_{j+1}^2}, \label{inversion-pf}
\end{align}
one can easily see that
$
\hat p_f (z)^2 =0 $ is equivalent to $ I [ \hat p_f (z)^2] =0 $. Thus
\be
I [z^*]= z^* \nn \ee
so
\be
I[H(z_1^*, z_2^*)]= H(z_1^*,z_2^*). \label{inversion bcfw 4}
\ee
Combining equations \eqref{BCFW for dualconformal},  \eqref{inversion bcfw 1}, \eqref{inversion bcfw 2}, \eqref{inversion bcfw 3} and \eqref{inversion bcfw 4}, and using $\hat x_1^2 = x_1^2$, we conclude that
\be
I [f_{2k}] =\sqrt{\prod_{i=1}^{2k} x_i^2 } f_{2k}. \label{inversion-amp}
\ee

Hence, if $f_{2l}$ inverts according to eq.(\ref{dual conformal invariance}) for all $l<k$, then so does $f_{2k}$. Since $f_4$ satisfies this property, this completes the proof that all tree-level ABJM amplitudes are dual superconformal invariant.
\section{Dual superconformal symmetry of loop amplitudes}
In this section, we will demonstrate that the dual conformal properties of the tree-level amplitudes can be extended to the cut-constructible parts of loop amplitudes via generalized unitarity methods~\cite{UnitarityMethod1,UnitarityMethod2,UnitarityMethod3,UnitarityMethod4}. To make a statement about the complete loop-level amplitudes, one needs to understand how the regulator modifies, or in some cases breaks, the symmetry. This is discussed in greater detail in the conclusion.

We will follow the argument which was used to establish dual conformal symmetry for six-dimensional super Yang-Mills~\cite{Dennen:2010dh}. The proof involves showing that all non-vanishing unitarity cuts of a given loop amplitude, with cut propagators restored, invert in a universal fashion. Since all cuts invert in the same way, so must the amplitude, and this leads to the following loop-level statement:
\begin{equation}
I\left[\mathcal{A}^L_n\right]=\sqrt{\left(\prod^n_{i=1}x_i^2\right)}\mathcal{A}^L_n\,.
\end{equation}
where $\mathcal{A}^L_n$ is understood to be the $L$-loop amplitude prior to integration, which is defined without a regulator.

\begin{figure}
\begin{center}
\includegraphics[scale=0.9]{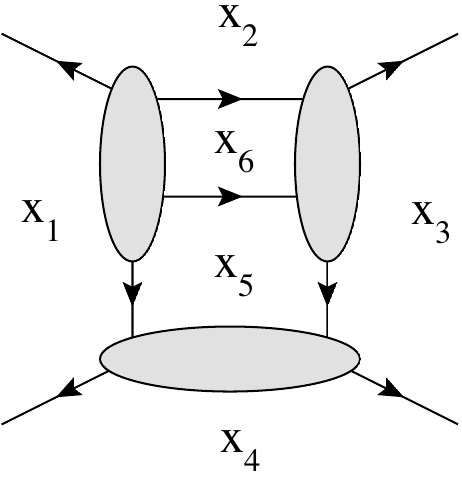}
\caption{A sample unitarity cut of four-point two-loop amplitude. Note that the dual points $x_2$ and $x_6$, while adjacent, are not labeled successively.    }
\label{dualloop}
\end{center}
\end{figure}

For loop amplitudes, the labeling of the dual space cannot always be done such that adjacent regions are labeled successively. For example, $x_6$ and $x_2$ in Fig.\ref{dualloop} are non-successive yet they are adjacent regions. Thus we utilize the following more general notation:
\begin{equation}
x^{\alpha\beta}_{i}-x^{\alpha\beta}_{j}=p^{\alpha\beta}_{\{ij\}},\;\;\theta^{\alpha I}_{i}-\theta^{\alpha I}_{j}=q^{\alpha I}_{\{ij\}}\,.
\end{equation}
In previous tree-level discussions, we had $j=i+1$. 

We begin by converting integrals over $\eta$'s to integrals over $\theta$'s. The conversion is done using the following identity:
\begin{equation}
 \prod_{\alpha} \delta^6\left(Q_{\alpha}\right)=\int \left[\prod_k d^6\theta_k\right]\left[ \prod_{\{rs\}}\delta^6\left(\theta_r-\theta_s-\lambda_{\{rs\}}\eta_{\{rs\}}\right)\right]
 \label{transformer}
\end{equation}
where $\alpha$ runs over the tree diagrams in the cut, $k$ runs over regions in the dual space, and $\{rs\}$ runs over all cut lines and external lines. The number of regions for a loop amplitude is $F=n+L$, where $L$ is the loop level. Since the translation from supermomentum to dual $\theta$ coordinates has an overall shift symmetry, the integration measure $d^6\theta_k$ is understood to include only $F-1$ of them. To see the equality in eq.(\ref{transformer}), one notes that the delta functions can be used to localize the $6(F-1)$ integrals. Denoting the total number of lines, cut or external, as $P$, there are then $6(P-F+1)$ delta functions left. For planar diagrams, $P-V=F-1$, where $V$ is the number of vertices, or in our case, the number of tree amplitudes in the cut. Thus, we are left with a supermomentum delta function for each tree amplitude. In other words, by converting to the dual $\theta$ representation, one automatically solves all supermomentum conservation constraints.

Using eq.(\ref{transformer}), one can rewrite the cut equation as:
\begin{eqnarray}
  \mathcal{A}_n^{L}\Bigr|_{\hbox{\footnotesize{cut}}} &=& \int
    \prod_{\{ij\}} d^3\eta_{\{ij\}} \times
    \mathcal{A}^{{\rm tree}}_{(1)}\mathcal{A}^{\rm tree}_{(2)}\mathcal{A}^{\rm tree}_{(3)}\ldots\mathcal{A}^{\rm tree}_{(m)} \nn\\
  &=& \delta^3(P) \int
   \prod_{\{ij\}} d^3\eta_{\{ij\}} \times
    \prod_{\alpha} \delta^6\left(Q_{\alpha}\right) f_{\alpha}  \nn\\
  &=&  \delta^3(P)\int
    \prod_{\{ij\}} d^3\eta_{\{ij\}} \times
    \prod_k d^6\theta_k \times
    \prod_\alpha f_\alpha \nn\\
  &&\times\prod_{\{rs\}}\delta^6\left(\theta_r-\theta_s-\lambda_{\{rs\}}\eta_{\{rs\}}\right)
     \,,
\label{loopamplitude}
\end{eqnarray}
where $\{ij\}$ runs over cut lines and $\{rs\}$ runs over cut lines and external lines. 
After using the delta functions to eliminate the $\eta_{\{ij\}}$-dependence from each $f_\alpha$, the $\eta_{\{ij\}}$-dependence comes solely from the delta functions. The integral over $\eta_{\{ij\}}$ then simplifies to
\begin{eqnarray}
  \int d^3\eta_{\{ij\}} \delta^6\left(\theta_i-\theta_j-\lambda_{\{ij\}}\eta_{\{ij\}}\right)=\delta^3\left(\theta_{ij}\lambda_{\{ij\}}\right) ,\,
\end{eqnarray}
where $\theta_{ij}\lambda_{\{ij\}}=(\theta_i-\theta_j)^{\alpha}\lambda_{\alpha\{ij\}}$, and we have suppressed the SU(3) R-index. Plugging this result into eq.(\ref{loopamplitude})  gives 
\begin{eqnarray}
  \mathcal{A}_n^{L}\Bigr|_{\hbox{\footnotesize{cut}}} &=& \delta^3(P) \int
    \prod_k d^6\theta_k\times
    \prod_\alpha f_\alpha \times
    \prod_{\{ij\}} \delta^3\left(\theta_{ij}\lambda_{\{ij\}}\right)\nn\\
  &&\times\prod_{\{rs\}}\delta^6\left(\theta_r-\theta_s-\lambda_{\{rs\}}\eta_{\{rs\}}\right) \,,
\end{eqnarray}
where $\{rs\}$ now only runs over the external lines, i.e. there are $n$ of them. Furthermore, one can pull out an overall supermomentum delta function, leaving $(n-1)$ delta functions, which completely saturate the $\theta$ integrations over the external regions. We are finally left with
\begin{eqnarray}
  \mathcal{A}_n^{L}\Bigr|_{\hbox{\footnotesize{cut}}}=\delta^3(P)\delta^6(Q) \int \left(\prod_k d^6 \theta_k \right)
  \prod_{\{ij\}} \delta^3\left(\theta_{ij}\lambda_{\{ij\}}\right)
  \prod_\alpha f_\alpha \,,
\label{cutintegrand}
\end{eqnarray}
where now $k$ now runs over the loop regions, which are regions 5 and 6 in the example shown in Fig.\ref{dualloop}.

Let us consider the inversion weight of each term in eq.(\ref{cutintegrand}):
\begin{itemize}
  \item For each loop region $k$, the $\theta_k$ measure contributes a factor $(x_k^2)^3$.
  \item Each cut leg $\{ij\}$ contributes $(x_i^2x_j^2)^{-\frac{3}{2}}$, which comes from $\delta^3\left(\theta_{ij}\lambda_{\{ij\}}\right)$.
  \item Each tree-level sub-amplitude contributes $\sqrt{\prod_ix_i^2}$, where $i$ runs over all regions adjacent to the tree.
\end{itemize}
After restoring the cut propagators, which invert as
\begin{equation}
I\left[\frac{1}{p^2_{\{ij\}}}\right]=I\left[\frac{1}{x^2_{ij}}\right]=\frac{x_i^2x_j^2}{x^2_{ij}}\,,
\end{equation}
the resulting object obtains the following factor after a dual inversion:
\begin{equation}
  \sqrt{\left(\prod_{i\in\varepsilon} x_i^2\right)}\left(\prod_{i=1}^{L} (x_{l_i}^2)^3\right)\,,
  \label{InvertNew}
\end{equation}
where $x_{l_i}$ are loop regions.

As a concrete example, consider the diagram in Fig.\ref{dualloop}. One has
\begin{eqnarray}
\nonumber I\left[ \mathcal{A}^{\rm Fig.10}\Bigr|_{\hbox{\footnotesize{cut}}}\right]&=&\frac{(x_5^2x_6^2)^3\sqrt{x_1^2x_2^2x^2_6x^2_5}\sqrt{x_1^2x_3^2x^2_4x^2_5}\sqrt{x_3^2x_2^2x^2_6x^2_5}}{(x_2^2x_6^2x_6^2x_5^2x_5^2x_1^2x_5^2x_3^2)^{3/2}}\mathcal{A}^{\rm Fig.10}\Bigr|_{\hbox{\footnotesize{cut}}}\\
&=&\frac{x^2_6\sqrt{x^2_4}}{\sqrt{x_3^2x_1^2x_2^2}}\mathcal{A}^{\rm Fig.10}\Bigr|_{\hbox{\footnotesize{cut}}}.
\end{eqnarray}
Furthermore, there are four propagators in Fig.\ref{dualloop}, which invert as follows:
\begin{equation}
I\left[\frac{1}{x^2_{15}}\right]=\frac{x_1^2x^2_5}{x^2_{15}},\;\;I\left[\frac{1}{x^2_{53}}\right]=\frac{x_3^2x^2_5}{x^2_{53}},\;\;I\left[\frac{1}{x^2_{56}}\right]=\frac{x_6^2x^2_5}{x^2_{56}},\;\;I\left[\frac{1}{x^2_{62}}\right]=\frac{x_2^2x^2_6}{x^2_{62}}.
\end{equation}
Thus, when $\mathcal{A}^{\rm Fig.10}\Bigr|_{\hbox{\footnotesize{cut}}}$ is combined with the cut propagators, the resulting object has the following inversion weight:
\begin{equation}
(x^2_6)^3(x^2_5)^3\sqrt{x_1^2x_2^2x_3^2x^2_4}\,,
\end{equation}
which matches the result in eq.(\ref{InvertNew})

There is one more ingredient missing, notably the loop integration measure $\int d^3l_i$, which is written in terms of dual coordinates as $\int d^3x_{l_i}$. In three dimensions, the loop integration measure will provide an extra inversion weight of $\prod_i (x_{l_i}^2)^{-3}$, which exactly cancels the extra weight of loop regions coming from the integrand. Hence, we find that all cuts, with the cut propagators and loop integration restored, 
invert the same way as the tree amplitudes do.
We conclude that the cut-constructible part of the $L$-loop amplitude inverts as
\begin{equation}
I\left[\mathcal{A}^L_n\right]=\sqrt{\left(\prod^n_{i=1}x_i^2\right)}\mathcal{A}^L_n\,.
\end{equation}
From the discussion in section 7, it follows that the cut-constructible parts of the loop-level amplitudes enjoy dual superconformal symmetry.

\section{Discussion}
In this paper, we constructed a BCFW-like recursion relation for three-dimensional Chern-Simons matter theories. This recursion relation involves shifting the momenta of two external particles. Unlike the usual BCFW approach, the shift is nonlinear in the complex deformation parameter $z$. 
Furthermore, the poles of the shifted amplitudes are computed by solving a  quadratic equation, rather than a linear equation. 

Using background field methods and the Grassmannian integral formula, we have argued that the superamplitudes of the ABJM theory vanish when the deformation parameter goes to infinity, which is required for the recursion relation to be applicable. We have explicitly checked the recursion relation for six-point and eight-point tree-level amplitudes of the ABJM theory by reproducing the results of Feynman diagram calculations and the Grassmannian formula. Finally, we have used this recursion relation to prove that all tree-level amplitudes of the ABJM theory enjoy dual superconformal symmetry. Using generalized unitarity methods~\cite{UnitarityMethod1,UnitarityMethod2,UnitarityMethod3,UnitarityMethod4}, we have
further extended the dual conformal symmetry to the cut constructible part of loop amplitudes.

For $\mathcal{N}=4$ sYM, the combination of ordinary and dual superconformal  symmetries, along with invariance under helicity rescalings, uniquely fixes the amplitudes~\cite{Korchemsky:2010ut}. Given that all the tree-level amplitudes of the ABJM theory have dual superconformal symmetry, it is natural to ask whether the amplitudes of the ABJM theory can also be fixed by the dual and ordinary superconformal symmetries. If this turns out to be true, it would probably imply that the Grassmannian formula proposed in ref.~\cite{lee} indeed generates all of the tree-level amplitudes.

Another important implication of dual superconformal symmetry is that the ABJM theory may have an amplitude/Wilson-loop duality. Since the one-loop correction to the scattering amplitudes vanishes, it would be useful to have an explicit calculation of the two-loop correction to the four-point scattering amplitude to compare with the Wilson loop computation~\cite{Henn:2010ps}.

Our discussion of dual superconformal symmetry at loop level is restricted to the cut-constructible parts of the amplitudes, or more precisely, prior to evaluating loop integrals. Due to infrared singularities, the integrals are ill-defined in the absence of a regulator. Introducing a regulator will generally render the symmetries anomalous. On the other hand, if one can choose a regulator and modify the symmetry generator in such a way that the regulator becomes symmetry preserving, this implies that the symmetry is preserved at the quantum level. 
This was done for $\mathcal{N}=4$ sYM by considering the dual symmetry as five-dimensional, with the extra dimension giving rise to a massive regulator~\cite{Alday:2009zm,henn1,henn2}. Alternatively, one can perform the unitarity cuts in a higher dimensional theory and perform dimensional reduction.
This approach was applied to $\mathcal{N}=4$ sYM, for example, by first demonstrating dual conformal symmetry in 6d maximal sYM
and then taking the higher dimensional components of the loop momenta to be regulators for the four-dimensional integrals ~\cite{Dennen:2010dh}. Unfortunately, Chern-Simons matter theories do not arise from dimensional reduction of any higher-dimensional theory, so it seems that the best approach is to introduce a mass deformation~\cite{Agarwal:2008pu,rey}. A first step would be to see if the unitarity cuts of mass-deformed amplitudes preserve an ``extended" dual conformal symmetry.

It also would be interesting to investigate if the recursion relation we proposed is applicable to three-dimensional theories other than the ABJM theory. The next candidate to consider is the BLG theory~\cite{Bagger:2007jr,Bagger:2007vi,Gustavsson:2007vu}, which has maximal supersymmetry and a similar structure to the ABJM theory. Similarly, one could ask if this recursion relation is applicable to theories with less supersymmetry. Although various component amplitudes in the ABJM theory have bad large-$z$ behavior, the theory has enough supersymmetry to guarantee that all the superamplitudes have good large-$z$ behavior. In other words, it is possible to relate all the badly-behaved component amplitudes of the ABJM theory to well-behaved component amplitudes using the supersymmetric Ward identities. This may not be possible for theories with less supersymmetry. It would therefore be interesting to determine the minimal amount of supersymmetry that is required to have good large-$z$ behavior of the superamplitudes.

\section*{Acknowledgments}
We are grateful to Nima Arkani-Hamed, Zvi Bern, Bo Feng, Yoon Pyo Hong, Hoil Kim, Ki-Myeong Lee, and John H. Schwarz for discussions.
The work of YH is supported by the US Department of Energy under contract DE--FG03--91ER40662 and the work of AEL is supported in part by the US DOE grant DE-FG02-92ER40701.
The work of SL is supported in part by the National Research Foundation of Korea Grants NRF--2009--0072755 and NRF--2009--0084601. YH would like to thank Mark Wise for the invitation as visiting scholar at Caltech.
The work of SL at the initial stage was done during the workshop ``QFT, String Theory and Mathematical Physics" held at Kavli Institute for Theoretical Physics China, CAS, Beijing 100190, China.
EK thanks Korea Institute for Advanced Study for providing KIAS Symbolic Cluster Systems.

\newpage

\centerline{\large \bf Appendix}

\appendix
\section{Conventions  \label{AppA1}}
We follow the conventions used in ref.~\cite{blm}. The SL$(2,\IR)$ metric is
\eq
\epsilon_{\alpha\beta}=\left(\begin{array}{cc}0 & 1 \\-1 & 0\end{array}\right),\;\epsilon^{\alpha\beta}=\left(\begin{array}{cc}0 & -1 \\1 & 0\end{array}\right).
\eqe
The spinor contraction is implemented as
\eq
\psi^\alpha \chi_\alpha =-\psi_\alpha \chi^\alpha, \quad \epsilon_{\beta\alpha}A^\alpha=A_\beta,\quad \epsilon^{\alpha\beta}A_{\beta}=A^{\alpha},\quad \epsilon^{\alpha\beta}\epsilon_{\beta\gamma}=\delta^\alpha_\gamma.
\eqe
The vector notation is translated to the bi-spinor notation
and vice versa through three dimensional gamma matrices,
\eq
x^{\alpha\beta}=x^\mu (\sigma_\mu)^{\alpha\beta},\;x^\mu=-\frac{1}{2}(\sigma^\mu)_{\alpha\beta}x^{\alpha\beta},
\eqe
with
\eq
\sigma^0=\left(\begin{array}{cc}-1 & 0 \\0 & -1\end{array}\right),\;\sigma^1=\left(\begin{array}{cc}-1 & 0 \\0 & 1\end{array}\right),\;\sigma^2=\left(\begin{array}{cc}0 & 1 \\1 & 0\end{array}\right),
\eqe
We list some useful identities
\begin{align}
&(\sigma^\mu)_{\alpha\beta}(\sigma^\nu)^{\alpha\beta}=-2\eta^{\mu\nu}\,,
\\
&(\sigma^\mu)_{\alpha\beta}(\sigma_\mu)_{\gamma\delta}=\epsilon_{\alpha\gamma}\epsilon_{\beta\delta}+\epsilon_{\beta\gamma}\epsilon_{\alpha\delta}\,,
\\
&\epsilon^{\mu\nu\rho}(\sigma_{\mu})^{ab}(\sigma_{\nu})^{cd}(\sigma_{\rho})^{ef}
=\frac{1}{2}(\epsilon^{ac}\epsilon^{be}\epsilon^{df}+\epsilon^{bc}\epsilon^{ae}\epsilon^{df}+\epsilon^{ad}\epsilon^{be}\epsilon^{cf}+\epsilon^{bd}\epsilon^{ae}\epsilon^{cf}
\nn \\
&\qquad\qquad\qquad\qquad\qquad\qquad
+\epsilon^{ac}\epsilon^{bf}\epsilon^{de}+\epsilon^{bc}\epsilon^{af}\epsilon^{de}+\epsilon^{ad}\epsilon^{bf}\epsilon^{ce}+\epsilon^{bd}\epsilon^{af}\epsilon^{ce}) \,,
\label{epsilon}
\\
& A^{[\alpha\beta]}=A^{\alpha\beta}-A^{\beta\alpha}=-\epsilon^{\alpha\beta}A^\gamma\,_\gamma,\\
& A_{[\alpha\beta]}=A_{\alpha\beta}-A_{\beta\alpha}=\epsilon_{\alpha\beta}A^\gamma\,_\gamma,\\
& x^{\alpha\beta}x_{\beta\gamma}=-x^2\delta^\alpha_\gamma,
\end{align}
where $x^2=x^\mu x_\mu$.

\section{Feynman rules for ABJM\label{Feynman}}
We start with the Lagrangian for the ABJM theory~\cite{Benna:2008zy,scs6,Bandres:2008ry}:
\begin{align}
\mathcal{L}=&\mathcal{L}_{2}+\mathcal{L}_{CS}+\mathcal{L}_{4}+\mathcal{L}_{6}\,, \nn \\
\mathcal{L}_{2}=&{\normalcolor tr}\left(D_{\mu}\phi^{A}D_{\mu}\overline{\phi}_{A}+i\psi_{A}\sigma^{\mu}D_{\mu}\overline{\psi}^{A}\right) \,,
\nn \\
\mathcal{L}_{CS}=&\epsilon^{\mu\nu\lambda}{\normalcolor tr}\left(\frac{1}{2}A_{\mu}\partial_{\nu}A_{\lambda}+\frac{i}{3}gA_{\mu}A_{\nu}A_{\lambda}-\frac{1}{2}\hat{A}_{\mu}\partial_{\nu}\hat{A}_{\lambda}-\frac{i}{3}g\hat{A}_{\mu}\hat{A}_{\nu}\hat{A}_{\lambda}\right) \,,
\nn \\
\mathcal{L}_{4}=&ig^{2}\epsilon^{ABCD}{\normalcolor tr}\left(\psi_{A}\overline{\phi}_{B}\psi_{C}\overline{\phi}_{D}\right)-ig^{2}\epsilon_{ABCD}{\normalcolor tr}\left(\overline{\psi}^{A}\phi^{B}\overline{\psi}^{C}\phi^{D}\right)+ig^{2}{\normalcolor tr}\left(\overline{\psi}^{A}\psi_{A}\overline{\phi}_{B}\phi^{B}\right) \,,
\nn \\
&-2ig^{2}{\normalcolor tr}\left(\overline{\psi}^{B}\psi_{A}\overline{\phi}_{B}\phi^{A}\right)-ig^{2}{\normalcolor tr}\left(\psi_{A}\overline{\psi}^{A}\phi^{B}\overline{\phi}_{B}\right)+2ig^{2}{\normalcolor tr}\left(\psi_{A}\overline{\psi}^{B}\phi^{A}\overline{\phi}_{B}\right) \,,
\nn
\end{align}
where $g=\sqrt{2\pi/k}$, $\phi^{A}=\overline{\phi}_{A}^{\dagger}$, $\psi_{A}=\overline{\psi}^{A\dagger}$,
and
\[
D_{\mu}\overline{\phi}_{A}=\partial_{\mu}\overline{\phi}_{A}+ig\left(A_{\mu}\overline{\phi}_{A}-\overline{\phi}_{A}\hat{A}_{\mu}\right) \,,
\;\; D_{\mu}\overline{\psi}^{A}=\partial_{\mu}\overline{\psi}^{A}+ig\left(A_{\mu}\overline{\psi}^{A}-\overline{\psi}^{A}\hat{A}_{\mu}\right)\,.
\]
Note
that the first two terms in $\mathcal{L}_{4}$ break the R-symmetry
from $SO(8)$ to $SU(4)$ because they contain $\epsilon^{ABCD}$.
The bosonic potential term
$\mathcal{L}_{6}$ is $\mathcal{O}\left(\phi^{6}\right)$ and will
not play an essential role in the analysis of this paper.

The gauge group is ${\rm U}(N)\times {\rm U}(N)$, where $A_{\mu}$
and $\hat{A}_{\mu}$ are the associated gauge fields, and $\overline{\phi}_{A}$
and $\overline{\psi}^{A}$ transform in the $(\mathbf{N},\bar{\mathbf{N}})$ representation. The
fields can be expanded in terms of matrices as follows:
\[
A_{\mu}=A_{\mu}^{a}T^{a}\,,\;\;\; \hat{A}_{\mu}=\hat{A}_{\mu}^{a}T^{a} \,,
\qquad
\overline{\phi}_{A}=\overline{\phi}_{A}^{a}\tilde{T}^{a}\,,
\;\;\;
\overline{\psi}^{A}=\overline{\psi}^{Aa}\tilde{T}^{a} \,,
\]
where $T^{a}$, $a=1,...,N^{2}-1$, are hermitian generators of ${\rm SU}(N)$
satisfying ${\normalcolor tr}\left(T^{a}T^{b}\right)=\delta^{ab}$,
$T^{N^{2}}$ is $1/\sqrt{N}$ times an $N\times N$ matrix, $\tilde{T}^{a}=\frac{i}{\sqrt{2}}T^{a}$
for $a=1,...,N^{2}-1$, and $\tilde{T}^{N^{2}}=\frac{1}{\sqrt{2}}T^{N^{2}}$.
Hence, at the level of matrix representations, the main difference
between the gauge fields and the matter fields is that the matter
fields are not hermitian. It should be emphasized that the matter fields
carry two different indices, one for each ${\rm U}(N)$.

\begin{figure}
\begin{center}
\includegraphics[scale=0.2]{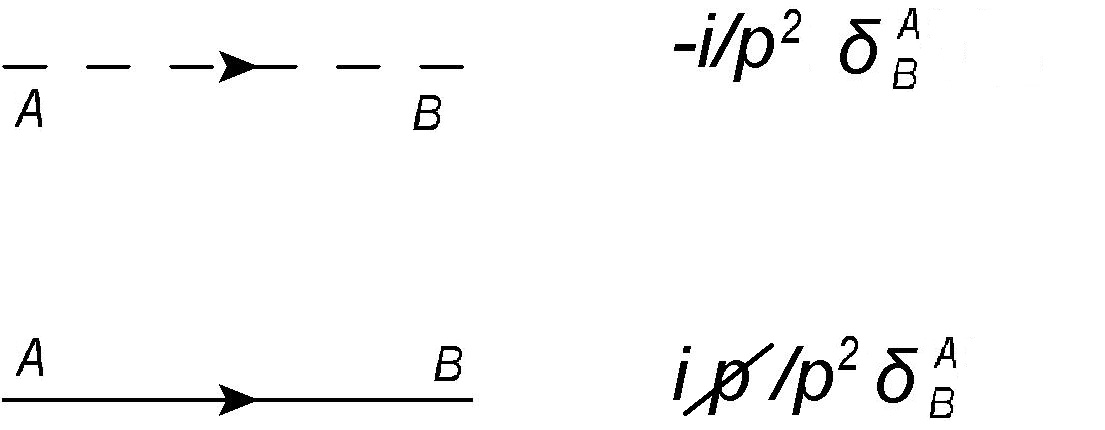}
\caption{Matter propagators }
\label{matprop}
\end{center}
\end{figure}

Plugging these expansions into the kinetic terms gives\[
-\frac{1}{2}\partial_{\mu}\phi^{A\, a}\partial^{\mu}\overline{\phi}_{A}^{a}+\frac{i}{2}\psi_{A}^{a}\sigma^{\mu}\partial_{\mu}\overline{\psi}^{Aa}+\frac{1}{2}\epsilon^{\mu\nu\lambda}\left(A_{\mu}^{a}\partial_{\nu}A_{\lambda}^{a}-\hat{A}_{\mu}^{a}\partial_{\nu}\hat{A}_{\lambda}^{a}\right){\normalcolor .}\]
The propagators for the matter
fields are depicted in Fig.\ref{matprop}.
\begin{figure}
\begin{center}
\includegraphics[scale=0.15]{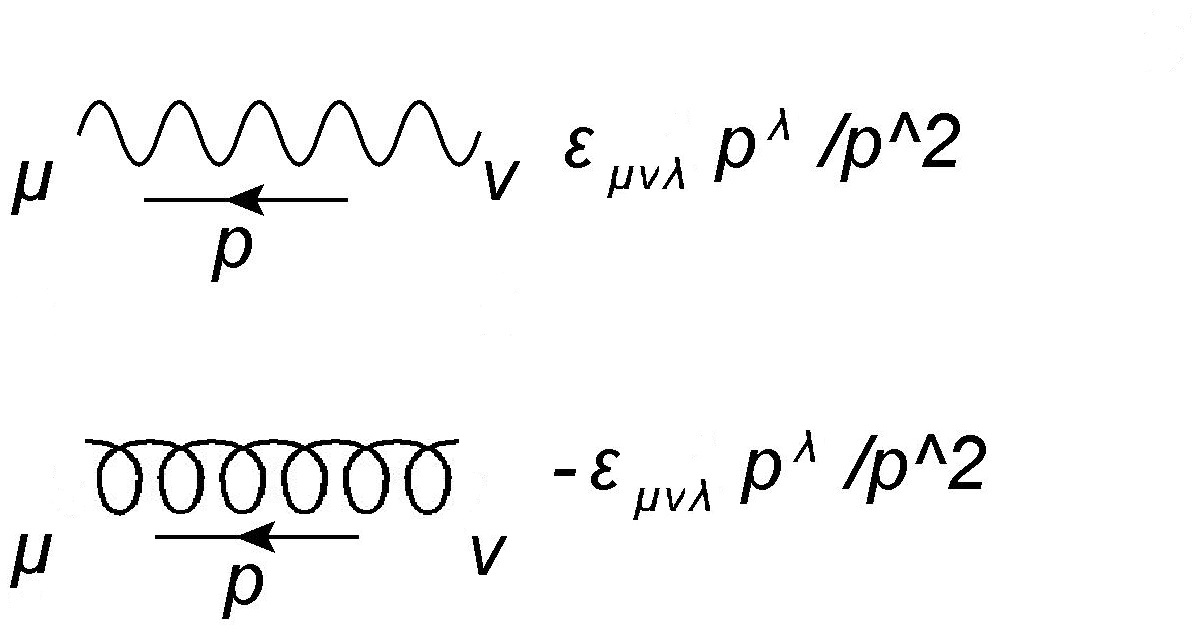}
\caption{Gauge propagators }
\label{gagprop}
\end{center}
\end{figure}
\begin{figure}
\begin{center}
\includegraphics[scale=0.15]{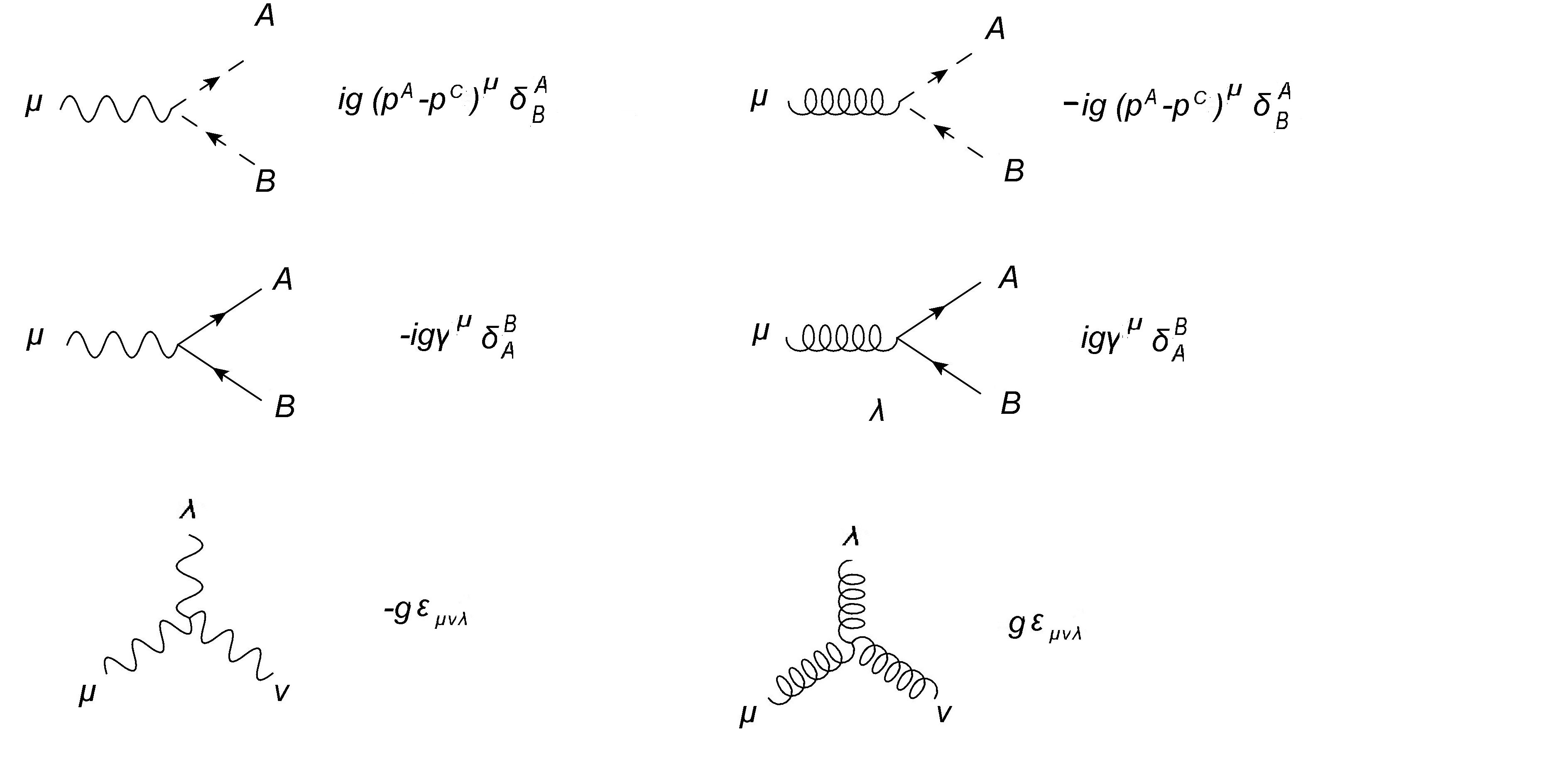}
\caption{Cubic vertices }
\label{3pt}
\end{center}
\end{figure}
\begin{figure}
\begin{center}
\includegraphics[scale=0.15]{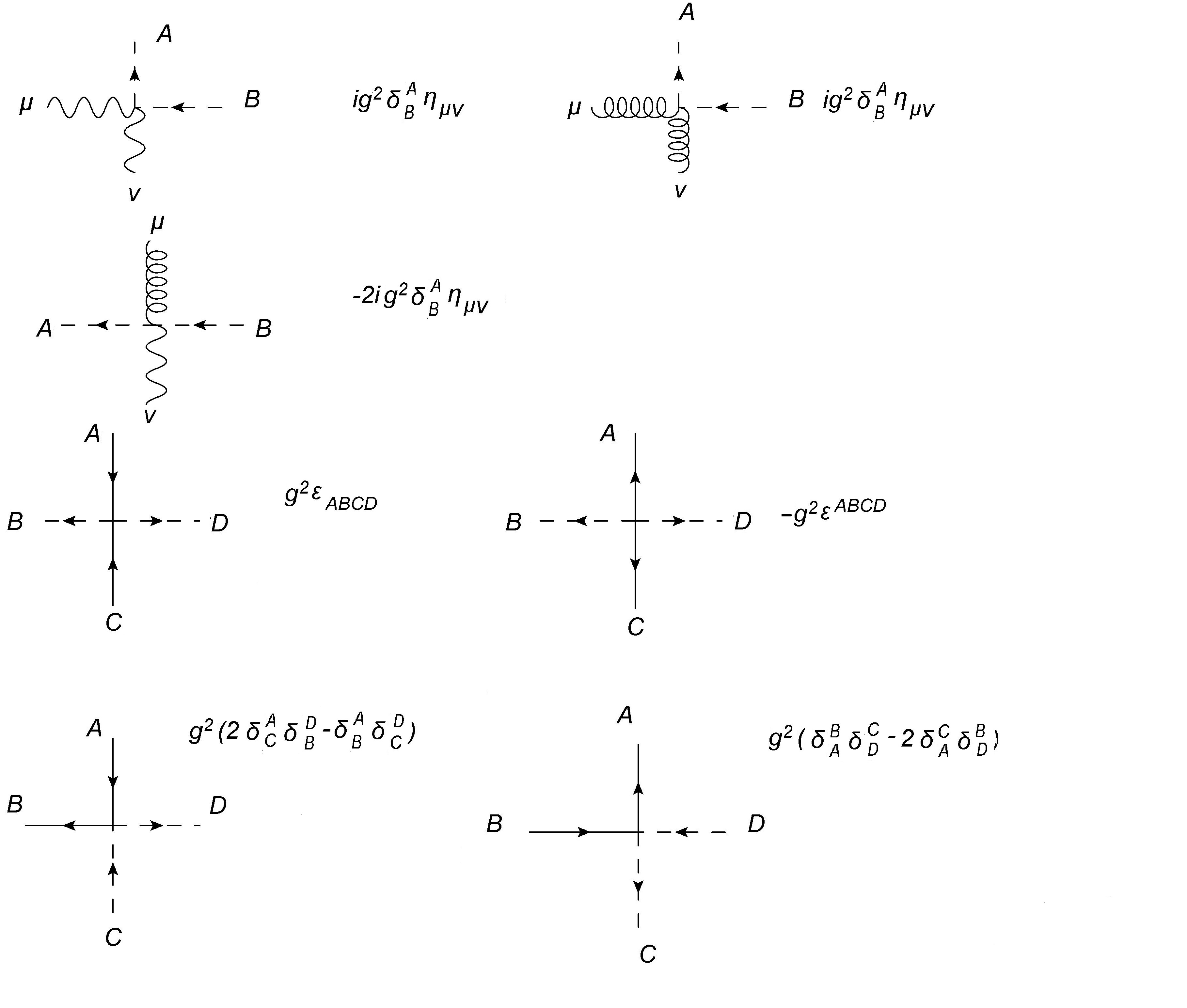}
\caption{Quartic vertices }
\label{4ptvertex}
\end{center}
\end{figure}

In the background field formalism, there are two kinds of gauge symmetry, notably gauge symmetry of the background and gauge symmetry of the fluctuations.
We use the background gauge symmetry to impose the following conditions on the background gauge fields:
\eq
q\cdot A^a=q\cdot \hat{A}^a=0\,.
\label{choice}
\eqe
After gauge-fixing the background, we still have the gauge symmetry of the fluctuations (under which the background is inert). We can therefore use the Faddeev-Popov procedure to introduce the following gauge-fixing terms for the gauge-field fluctuations:
\eq
\mathcal{L}_{gf}=tr\left[\frac{1}{\xi_1}(\partial \cdot a)^2-\frac{1}{\xi_2}(\partial\cdot \hat{a})^2\right]\,
\eqe
where $a$ and $\hat{a}$ represent the fluctuations of the gauge fields and $\xi_i$ ($ i=1,2$) are gauge-fixing parameters.
Note that these gauge-fixing terms don't  preserve the background gauge symmetry. On the other hand, the background gauge symmetry is already broken by eq. (\ref{choice}), so there's no need to choose gauge-fixing terms which preserve the background gauge symmetry.

The propagators for the gauge field fluctuations are then given by
 \eq
\pm \frac{\delta^{ab}}{p^{2}}\left(\epsilon_{\mu\nu\lambda}p^{\lambda}+i\xi_i\frac{p_\mu p_\nu}{p^2}\right) \,,
\eqe
where $\pm$ refers to the $a/\hat{a}$ fields and $a,b$ in the above equation are adjoint indices. Taking $\xi_i=0$ gives Landau gauge propagators:
 \eq
\pm\frac{\epsilon_{\mu\nu\lambda}p^{\lambda}}{p^{2}}\delta^{ab} \,.
\eqe
The resulting propagators are depicted in Fig.\ref{gagprop}.

The color-ordered Feynman rules for the cubic and quartic vertices are depicted
in Figs. \ref{3pt} and \ref{4ptvertex}. They can easily be adapted to the background field approach by choosing at least two legs in each diagram to be fluctuation fields. The symmetry factors are unchanged.

\section{Alternative forms of recursion relation}

\subsection*{Linear multi-line shift \label{multiline}}
Consider four external particles labeled $i,j,k,l$ and define
\[
\mu_{\alpha}=\left(\lambda_{i}+i\lambda_{j}\right)_{\alpha} ,
\;\;\;
\rho_{\alpha}=\left(\lambda_{k}+i\lambda_{l}\right)_{\alpha} .
\]
Also define $\bar{\mu}\equiv\lambda_{i}-i\lambda_{j}$ and $\bar{\rho}\equiv\lambda_{k}-i\lambda_{l}$.
Note that $\mu_{(\alpha}\bar{\mu}_{\beta)}=\left(p_{i}+p_{j}\right)_{\alpha\beta}$ and $\rho_{(\alpha}\bar{\rho}_{\beta)}=\left(p_{k}+p_{l}\right)_{\alpha\beta}$.
Now define the shift
\[
\lambda_{i}\rightarrow\lambda_{i}+\frac{1}{2}z\bar{\rho},\;\;\;
\lambda_{j}\rightarrow\lambda_{j}+\frac{i}{2}z\bar{\rho} ,
\;\;\;\;\;
\lambda_{k}\rightarrow\lambda_{k}-\frac{1}{2}z\mu,\,\,\,\lambda_{l}\rightarrow\lambda_{l}+\frac{i}{2}z\mu .
\]
Under this deformation, the momenta shift
as follows
\[
p_{i}+p_{j}\rightarrow p_{i}+p_{j}+zq \,, \;\;\;
p_{k}+p_{l}\rightarrow p_{k}+p_{l}-zq \,,
\]
where $q_{\alpha\beta}=\mu_{(\alpha}\bar{\rho}_{\beta)}$. Note
that $p_{i}+p_{j}+p_{k}+p_{l}$ is invariant under the shift. Also
note that each momentum remains on-shell after the shift. It is straight forward to define an analogous shift of the fermionic coordinates.

Near each pole of the amplitude, the amplitude factorizes into a product
of two on-shell amplitudes connected by a propagator. Suppose that
$i$,$j$ appear on one side of the propagator and $k$,$l$ appear
on the other side. In this case, the momentum in the propagator
shifts linearly as follows:\[
p_{f}\rightarrow\hat{p}_{f}(z)=p_{f}+zq{\normalcolor .}\]
The poles of this propagator are therefore obtained by solving a quadratic
equation. Now suppose that only particle $i$ appears on one side
of the propagator. Then
\[
p_{f}\rightarrow\hat{p}_{f}(z)=p_{f}+zq_{a}+z^{2}q_{b}\,,
\]
where $\left(q_{a}\right)_{\alpha\beta}=\lambda_{i(\alpha}\bar{\rho}_{\beta)}$
and $\left(q_{b}\right)_{\alpha\beta}=\frac{1}{4}\bar{\rho}_{\alpha}\bar{\rho}_{\beta}$.
Since $q_{b}^{2}=q_{a}\cdot q_{b}=0$, $\hat{p}_{f}(z)^{2}$ is once
again quadratic in $z$.

Following the argument
in section 3.1, it is not difficult to show that the recursion formula for
the four-line shift is
\[
A(z=0)=\sum_{f} \int d^{3}\eta\frac{1}{p_{f}^{2}} \left(  H (z_f^+,z_f^- ) A_L ( z_f^+ )A_R (z_f^+) + ( z_f^+ \leftrightarrow z_f^- )  \right) ,
\]
where $\sum_{f}$ is the sum over all possible channels, $z_{f}^{\pm}$
are the zeroes of $\hat{p}_{f}^{2}(z)$, and
\[ H ( x, y ) = \frac{ y}{x-y} \,.
\]
Note the similarity to the recursion formula for the two-line shift proposed in section 3.1. Although
the recursion formula for the four-line shift looks a bit simpler, the two-line shift still has several advantages. For example, the recursion formula for the two-line shift involves fewer channels than the recursion formula for the four-line shift.

\subsection*{Superconformally covariant form}

The BCFW recursion relation for $\CN=4$ sYM is known to take the simplest form in the super-twistor notation~\cite{acc0903,ms0903}.
By doing a similar manipulation in our case,
we will show (rather heuristically) that the recursion relation derived in section 3 can be written in a
superconformally covariant form.

\paragraph{Review of the 4d result}

Let us review the ideas of refs.~\cite{acc0903,ms0903}.
They begin with the usual BCFW recursion relation in terms
of amplitudes with the momentum conserving delta functions removed.
Schematically,
\be
A = \sum_f A_L[p_i(z^*); -P_L(z^*)] \frac{1}{P_L^2} A_R[p_j(z^*); -P_R(z^*)] \,. \label{4d-BCFW}
\ee
Here $z^*$ is the value of the BCFW deformation parameter $z$
satisfying $[P_L(z^*)]^2=0$. In translating eq.\eqref{4d-BCFW}
into the twistor language, the first step
is to insert the momentum conserving delta functions on both sides using the identity
\be
\d^4(P) = \d^4(P_L(z^*) + P_R(z^*) ) = \int d^4 p \,\d^4(P_L(z^*) - p) \d^4(P_R(z^*) + p) \,.
\ee
Using the notation $\CA = A \,\delta^4(P)$, we have
\be
\CA = \sum_f \int d^4 p \left(\CA_L[p_i(z^*);-p] \frac{1}{P_L^2} \CA_R[p_j(z^*);+p ]\right)\,.
\ee
The next step could be called ``undoing the BCFW residue calculus"
as it recovers the $z$-integral. Technically, we use the identity
\be
\frac{1}{P_L^2} &=& -{\rm sgn}([i|P_L| j\rangle)\int \frac{dz}{z}
\d(z [i|P_L|j\rangle + P_L^2)
= -{\rm sgn}([i|P_L| j\rangle)\int \frac{dz}{z} \d( P_L(z)^2) \,,
\label{undo-4d}
\ee
and use the delta functions to rewrite the formula such that
\be
\CA = -\sum_f \int\frac{dz}{z} \int d^4 p \d(p^2) {\rm sgn}([i|p| j\rangle) \CA_L[p_i(z);-p]  \CA_R[p_j(z);+p ] \,.
\label{4d-step2}
\ee
Reinstating the $z$-integral highlights the on-shell nature of the BCFW recursion relation in two ways.
First, it removes the off-shell propagator $(1/P_L^2)$.
Second, it restricts the integration domain of $p$ to on-shell
via $\d(p^2)$. The on-shell momentum integral $\int d^4p \,\d(p^2)$ transforms into the $\int d^2\l d^2\bar{\l}$ integral at a later stage,
which in turn translates into the twistor language.

In eq.\eqref{4d-step2}, the $z$-integral looks exactly
the same as the BCFW contour integral, although with a different integration
contour. It is not a coincidence; it follows from the identity,
\be
\oint \frac{dz}{2\pi i} \frac{1}{g(z)} = \int dz \d(g(z)) \,,
\ee
which holds at least formally for an arbitrary function $g(z)$.

\paragraph{Generalization to 3d}

It should be clear from the above review that
the first two steps $-$ inserting momentum conserving delta functions
and undoing the residue calculus $-$ can be straightforwardly
carried over to three dimensions. We find
\be
\CA = \sum_f \int\frac{dz}{z-1} \int d^3 p \d(p^2) X_{ij}(p)
\CA_L[p_i(z);-p]  \CA_R[p_j(z);+p ] \,,
\label{3d-step2}
\ee
with some function $X_{ij}(p)$ analogous to ${\rm sgn}([i|p| j\rangle)$ in eq.\eqref{undo-4d}. It must be a dimensionless function of $p$, $\l_i$ and $\l_j$, but its precise form is not of interest in this paper.

The next step is to make the replacement
\be
\int d^3p \, \d(p^2) \goto \int d^2 \l \,,
\ee
which follows easily from a change of variable such as
\be
p_{11} = \l_1 \l_1 \,, \;\;\; p_{22} = \l_2 \l_2 \,, \;\;\;
p_{12} = s \,\l_1 \l_2 \,.
\ee
Supplementing
the $\l$-integral with the $\eta$-integral,
we can write down the superconformally covariant version of the three-dimensional recursion relation:
\footnote{
A similar gluing of amplitudes without the $z$-deformation
was considered in ref.~\cite{blm}.
Strictly speaking, the superconformal covariance of
this $\Lambda$-integral representation relies on
the precise form of $X_{ij}(\Lambda)$.}
\be
\CA &=& \sum_f \int\frac{dz}{z-1} \int d^{2|3}\L \, X_{ij}(\L)
\CA_L[\L_i(z);\L]  \CA_R[\L_j(z);i\L ]
\nn \\
&=&  \sum_f \int d^{2|3}\L \, X_{ij}(\L) \int\frac{dz}{z-1} z^{M_{ij}} \left(\CA_L[\L_i;\L]  \CA_R[\L_j;i\L ] \right) \,.
\label{3d-step4}
\ee
On the second line, $M_{ij}$ denotes the SO$(2)$ deformation generator
written in terms of the $\L$-variables:
\be
M_{ij} =  i \left( \L_i \frac{\partial}{\partial \L_j} - \L_j \frac{\partial}{\partial \L_i} \right)  \,.
\ee

In four dimensions, after similar rearrangements and half Fourier transformations,
all the $z$ dependence are absorbed into
\be
\int \frac{dz}{z} e^{iz W_i\cdot Z_j} \propto {\rm sgn}(W_i \cdot Z_j)
\,,
\ee
where $W_i$ and $Z_j$ are four dimensional twistor variables
for the legs $i$ and $j$, respectively.
It is not clear how to further simplify the $z$-dependence in
eq.\eqref{3d-step4}.

\section{Computational details}

\subsection*{Large-$z$ behavior  \label{AppB}}
We demonstrate the cancelation of bad large $z$ behavior for the four-point diagrams in Fig.\ref{4ptcontact}.
Note that the hard scalar and spinor have the same SU$(4)$ R-index in these diagram. As a result,
the 4-point contact term involving $\epsilon^{ABCD}$ does not contribute.

The diagram in Fig.\ref{4ptcontact}(a) is given by
\[
g^2\left\langle\psi(\hat{p}_{1})\right|\sigma^{\mu} \bar{\Psi}(k_{1}) \frac{\epsilon_{\mu\nu\lambda}(-k_{2}-\hat{p}_{l})^{\lambda}}{\left(k_{2}+\hat{p}_{l}\right)^{2}}\left(\hat{p}_{l}-k_{2}\right)^{\nu}\Phi(k_2){\normalcolor .}\]
In the large-$z$ limit, $\hat{p}_{l}\rightarrow z^{2}q,\psi(\hat{p}_{1})\rightarrow z\lambda_q$, and the above
formula simplifies to\[
zg^{2}\frac{\left\langle q\right|\sigma^{\mu} \bar{\Psi}(k_{1}) \epsilon_{\mu\nu\lambda}q^{\nu}k_{2}^{\lambda}}{q\cdot k_{2}}\Phi(k_2)+\mathcal{O}\left(1/z\right){\normalcolor .}\]
Using eq.(\ref{epsilon}), it is not difficult to show that
 \begin{equation}
\left\langle q\right|\sigma^{\mu}\bar{\Psi}(k_{1})(\epsilon_{\mu\nu\rho}q^{\nu}k_{2}^{\rho})\Phi(k_2)=\left\langle q\right|\bar{\Psi}(k_{1})( k_{2}\cdot q )\Phi(k_2)\,.
\label{eq:1}
\end{equation}
Hence, we get
 \[
zg^{2}\left\langle q\right|\bar{\Psi}(k_{1})\Phi(k_2) +\mathcal{O}\left(1/z\right)\,.\]

The diagram in Fig.\ref{4ptcontact}(b) is simply given by
\[
-g^{2} \langle\psi(\hat{p}_{1}) |\bar{\Psi}(k_{1})\Phi(k_2) \,.
\]
In the large-$z$ limit, this becomes
\[
-zg^{2}\langle\psi(\hat{p}_{1}) |\bar{\Psi}(k_{1})\Phi(k_2) +\mathcal{O}\left(1/z\right){\normalcolor .}
\]
Thus we see that in the large-$z$ limit, the terms of $\mathcal{O}(z)$
cancel out.


\subsection*{The 8-point calculation } \label{details}

We present some of the complicated details in calculating the
8-point amplitude. We begin with a closer look at the orthonormal basis $\{e , \bar{e}\}$ defined in section \ref{8pt-generality}. The possible ambiguities in the orientation of the basis can be lifted by choosing the signs of  $ \varepsilon_1 , \varepsilon_2 , \varepsilon_3 $ in the following identities
\begin{align}
&(e_{1})_r (e_{2})_s - (e_{1})_s (e_{2})_r = \varepsilon_1 \frac{\langle r s \rangle }{\sqrt{-p_{0}^2}}, \qquad\;\;\;\; (\bar{e}_1)_{\bar{r}} (\bar{e}_2)_{\bar{s}} - (\bar{e}_1)_{\bar{s}} (\bar{e}_2)_{\bar{r}} = \varepsilon_1 \frac{-\langle \bar{r}\bar{s}\rangle}{\sqrt{-p_{0}^2}}, \nonumber
\\
&(e_{3})_r (e_{4})_s - (e_{3})_s (e_{4})_r  =  \frac{\varepsilon_2}2 \epsilon_{rspq}\frac{\langle p q \rangle}{\sqrt{-p_{0}^2}}, \quad  (\bar{e}_3)_{\bar{r}} (\bar{e}_4)_{\bar{s}} - (\bar{e}_3)_{\bar{s}} (\bar{e}_4)_{\bar{r}}=  \frac{\varepsilon_3}2 \epsilon_{\bar{r}\bar{s}\bar{p}\bar{q}} \frac{-\langle\bar{p}\bar{q}\rangle}{\sqrt{-p_{0}^2}}, \nonumber
\\
&(e_{3})_r (e_{3})_s + (e_{4})_r (e_{4})_s = \delta_{rs} + \frac{\langle r| p_{0} | s\rangle}{p_{0}^2}, \quad (\bar{e}_3)_{\bar{r}} (\bar{e}_3)_{\bar{s}} + (\bar{e}_4)_{\bar{r}} (\bar{e}_4)_{\bar{s}} = \delta_{\bar{r}\bar{s}} - \frac{\langle \bar{r}| p_{0}| \bar{s}\rangle}{p_{0}^2}, \nonumber
\\
&(e_1)_r (\bar{e}_1)_{\bar{s}} + (e_{2})_r (\bar{e}_2)_{\bar{s}} =  i \frac{ \langle r |p_{0} |\bar{s}\rangle}{p_{0}^2}, \qquad  \qquad (p_0 \equiv \sum_{ \bar{r} } p_{ \bar{r}}). \label{identities}
\end{align}
where $p_0 = p_{1234}$ for the factorization gauge, and $p_0 = p_{2468} $ for the cyclic gauge.
In this paper, we always work with the choice $ \varepsilon_1 = \varepsilon_2 = \varepsilon_3=+1$.

The constraint equations imposed by the bosonic delta functions are
\begin{align}
c_{\bar{r} s}   c_{ \bar p s} + \delta_{ \bar{r} \bar{p}} =0 , \quad c_{ \bar{r} s}   \lambda_{s} = -\lambda_{ \bar{r} } .  \label{bosonic-delta}
\end{align}
One particular solution for the equations is
\begin{align}
c= - i (\bar{e}_1 e_1^T +\bar{e}_2 e_2^T + \bar{e}_3 e_3^T +\bar{e}_4 e_4^T) . \label{particular-sol}
\end{align}
One can check that eq.\eqref{particular-sol} indeed solves eq.\eqref{bosonic-delta} using the completeness and orthonormality of the basis and the momentum conservation condition of $\lambda$'s.

We use the following pair of one-parameter family of solutions $\{c_+(\tau), c_- (\tau)\}$ ($\tau = e^{i \theta}$)
\begin{align}
&c_+ (\theta)= i (\bar{e}_1 e_1^T +\bar{e}_2 e_2^T + \bar{e}_3  \hat{e}_3^T (\theta)  + \bar{e}_4 \hat{e}_4(\theta)^T), \quad \left(
   \begin{array}{c}
     \hat{e}_3 (\theta) \\
     \hat{e}_4 (\theta) \\
   \end{array}
 \right) = \left(
             \begin{array}{cc}
               \cos \theta  & \sin \theta \\
               - \sin \theta  &  \cos \theta\\
             \end{array}
           \right) \left(
   \begin{array}{c}
     e_3  \\
     e_4  \\
   \end{array}
 \right), \nonumber
\\
&c_- (\theta)= i (\bar{e}_1 e_1^T +\bar{e}_2 e_2^T + \bar{e}_3  \hat{e}_3^T (\theta)  + \bar{e}_4 \hat{e}_4(\theta)^T), \quad \left(
   \begin{array}{c}
     \hat{e}_3 (\theta) \\
     \hat{e}_4 (\theta) \\
   \end{array}
 \right) = \left(
             \begin{array}{cc}
               \cos \theta  & \sin \theta \\
               - \sin \theta  &  \cos \theta\\
             \end{array}
           \right)  \left(
   \begin{array}{c}
     e_3  \\
     -e_4  \\
   \end{array}
 \right). \nonumber
\end{align}
To reduce the Grassmannian integral \eqref{8pt-master}
down to an ordinary contour integral in $\tau$, we introduce the following new variables ($4\times 4$ matrix) $(\omega)_{\bar{r} i}$.
\begin{align}
(c)_{\bar{r}p}  = - i (\omega \cdot e)_{\bar{r}p}, \quad \textrm{$4\times 4$ matrices } (e)_{ip}=(e_i)_p , ( \omega)_{ \bar r i} = ( \omega_{ \bar{r} } )_i .
\end{align}
In terms of these variables, the solutions are given by
\begin{align}
\omega^*_\pm (\theta) = \{ \bar{e}_1 , \bar{e}_2, \cos \theta \bar{e}_3 + \sin \theta \bar{e}_4 , \mp \sin \theta \bar{e}_3 \pm \cos \theta \bar{e}_4 \}.
\end{align}
Ignoring the fermionic delta function for a while,
the integral gets transformed to
\begin{align}
\mathcal{L}_8 (\Lambda)  & \stackrel{\rm bos.}{=} \int d^{16} c \frac{\delta^{10}(c\cdot c^T +I ) \delta^8(c\cdot \lambda + \bar{\lambda})}{M_1 M_2 M_3 M_4}, \nonumber
\\
&= J^4 \int d^{16}\omega \frac{\delta^{10}(\omega \cdot \omega^T -I ) \delta^8(  \omega \cdot e \cdot \lambda + i  \bar{\lambda})}{M_1 M_2 M_3 M_4}, \quad J = \det (e_1 , e_2 , e_3 ,e_4) = \pm 1. \nonumber
\\
&=  \frac{1}{(p_{0})^2} \int d^4(\omega)_{\bar{r}3} d^4 (\omega)_{\bar{r}4} \frac{\delta^{10}(\omega \cdot \omega^T -I )}{M_1 M_2 M_3 M_4}, \quad \textrm{integrating out 8 variables $(\omega)_{\bar{r}1}, (\omega)_{\bar{r}2}$}. \nonumber \end{align}
Now let the four barred indices be
\begin{align}
\bar{r}_1 , \bar{r}_2 , \bar{r}_3, \bar{r}_4 ,
\end{align}
 where $ \{ \bar{r}_1 , \bar{r}_2 , \bar{r}_3, \bar{r}_4 \}= \{ 1,2,3,4 \}$ for the factorization gauge. We now integrate out six variables $\omega_{\bar{r}_2 3}, \omega_{\bar{r}_2  4}, \omega_{\bar{r}_3 3}, \omega_{\bar{r}_3  4} , \omega_{\bar{r}_4 3}, \omega_{\bar{r}_4  4}$.
Then the bosonic integral becomes
\begin{align}
\mathcal{L}_8 (\Lambda)  & \stackrel{\rm bos.}{=}   \frac{1}{(p_{0})^2} \int d (\omega_{\bar{r}_1})_{ 3} d ( \omega_{ \bar{r}_1})_{ 4} \frac{\delta (\omega_{ \bar{r}_1 }\cdot \omega_{{\bar r}_1}-1)\delta(\omega_{\bar{r}_2} \cdot \omega_{ \bar{r}_3 }) \delta(\omega_{\bar{r}_3 } \cdot \omega_{\bar{r}_4 })\delta(\omega_{\bar{r}_4 } \cdot \omega_{\bar{r}_2 })}{M_1 M_2 M_3 M_4 \prod_{K=2,3,4}2(\omega_{\bar{r}_1 3}\omega_{\bar{r}_K 4}- \omega_{\bar{r}_1 4} \omega_{\bar{r}_K 3})}.
\end{align}
Using the identity
\begin{align}
\delta(\omega_{\bar{r}_2 } \cdot \omega_{{ \bar r}_3 })\delta(\omega_{\bar{r}_3} \cdot \omega_{ \bar{r}_4})\delta(\omega_{\bar{r}_4}\cdot \omega_{\bar{r}_2}) = \delta^3 (P )\langle \bar{r}_2 \bar{r}_3 \rangle \langle \bar{r}_3 \bar{r}_4 \rangle \langle \bar{r}_4 \bar{r}_2  \rangle,
\end{align}
the integral is further simplified to
\begin{align}
\mathcal{L}_8 ( \Lambda)  & \stackrel{bos.}{=}  \frac{ \delta^3(P) \langle \bar{r} \bar{r} \rangle^3 }{(p_{0}^2)^2} \int d ( \omega_{\bar{ r}_1})_{ 3} d ( \omega_{\bar{r}_1} )_{4} \frac{\delta({\omega}_{\bar{r}_1}  \cdot {\omega}_{\bar{r}_1 } -1)  }{M_{1}M_{2}M_{3}M_{4}\prod_{K =2,3,4}2(\omega_{ \bar{r}_1 3}\omega_{\bar{r}_K 4}-\omega_{ \bar{r}_1 4}\omega_{\bar{r}_K 3})} \nonumber
\\
&= \frac{ \delta^3(P) \langle \bar{r} \bar{r} \rangle^3 }{(p_{0}^2)^2} \int d\theta (\frac{d\omega_{\bar{r}_1 3}}{d\theta})\frac{1}{2\omega_{\bar{r}_1 4}}\frac{1}{M_{1}M_{2}M_{3}M_{4}\prod_{K=2,3,4}2(\omega_{ \bar{r}_1 3}\omega_{\bar{r}_K 4}-\omega_{\bar{r}_1 4}\omega_{\bar{r}_K 3})} \nonumber
\\
&= \frac{ \delta^3(P)  \langle \bar{r} \bar{r} \rangle^3 }{2(p_{0}^2)^2}
\left[ \int d \theta \frac{1}{M_{1}M_{2}M_{3}M_{4}\prod_{ K =2,3,4}2(\omega_{\bar{r}_1 3}\omega_{\bar{r}_K 4}-\omega_{ \bar{r}_1 4}\omega_{\bar{r}_K 3})}|_{\omega=\omega^*_+} \right. \nonumber
\\
& \qquad \qquad \qquad \qquad
\left. -\int d \theta \frac{1}{M_{1}M_{2}M_{3}M_{4}\prod_{K = 2,3,4}2(\omega_{ \bar{r}_1 3}\omega_{\bar{r}_K 4}-\omega_{ \bar{r}_1 4}\omega_{\bar{r}_K 3})}|_{\omega=\omega^*_-}\right]   \nonumber
\\
&=\frac{ \delta^3(P)  }{16\sqrt{-p_{0}^2} } \left[ \int d \theta \frac{1}{M_{1}M_{2}M_{3}M_{4}}|_{\omega=\omega^*_+} +\int d \theta \frac{1}{M_{1}M_{2}M_{3}M_{4}}|_{\omega=\omega^*_-}\right].
\end{align}
In the first three lines, we used the shorthand notation $\langle \bar{r} \bar{r} \rangle^3 \equiv \langle \bar{r}_2 \bar{r}_3 \rangle \langle \bar{r}_3 \bar{r}_4 \rangle \langle \bar{r}_4 \bar{r}_2  \rangle$.
In the last line, we used $\omega_{ \bar{r}_1 3}\omega_{\bar{r}_K 4}-\omega_{ \bar{r}_1 4}\omega_{\bar{r}_K 3}|_{\omega = \omega^*_\pm} = \pm ((\bar{e}_3)_{ \bar{r}_1 } (\tilde{e}_{4})_{\bar{r}_K } -(\tilde{e}_{4})_{\bar{r}_1 } (\tilde{e}_{3})_{\bar{r}_K })$ and the identity \eqref{identities}.
Including the fermionic delta function and changing the integral variable by $\tau= e^{i\theta}$, we obtain the final expression in eq.\eqref{8pt-master} by choosing some  appropriate contours.


\end{document}